\documentclass{article}

\usepackage[english]{babel}

\usepackage[letterpaper,top=2cm,bottom=2cm,left=3cm,right=3cm,marginparwidth=1.75cm]{geometry}

\usepackage{rotating}
\usepackage{natbib}
\usepackage{adjustbox}
\usepackage{amsmath}
\usepackage{graphicx}
\usepackage[colorlinks=true, allcolors=blue]{hyperref}
\usepackage{multirow}
\usepackage{amssymb}
\usepackage{float}
\usepackage{authblk}
\usepackage{longtable}
\usepackage[
singlelinecheck=false 
]{caption}
\usepackage{enumitem}
\usepackage{diagbox}
\usepackage{setspace}
\usepackage{slashbox}
\usepackage{booktabs,subcaption,amsfonts,dcolumn}
\usepackage{tablefootnote}
\newcolumntype{d}[1]{D..{#1}}

\usepackage[markup=underlined,dvipsnames,svgnames,x11names]{xcolor}
\usepackage{changes}

\newcommand{\bigzero}{\mbox{\normalfont\Large\bfseries 0}}

\definechangesauthor[color=BrickRed]{AU}
\usepackage{todonotes}
\setcommentmarkup{\todo[color={authorcolor!20},size=\scriptsize]{#3: #1}}

\providecommand{\keywords}[1]
{
  \small	
  \textbf{\text{Keywords:}} #1
}

\begin{document}

\title{Homogeneity Tests and Interval Estimations of Risk Differences for Stratified Bilateral and Unilateral Correlated Data}

\author{Shuyi Liang$^1$, Xin-Wei Huang$^1$, Yijing Xin$^2$, Kai-Tai Fang$^3$, and Chang-Xing Ma$^{1*}$}

\maketitle
$^1$Department of Biostatistics, University at Buffalo, New York, USA

$^2$The First Affiliated Hospital of Xiamen University, Fujian, China

$^3$Guangdong Provincial Key Laboratory of Interdisciplinary Research and Application for Data Science, Beijing Normal University—Hong Kong Baptist University United International College, Zhuhai 519087, China

*cxma@buffalo.edu

\begin{abstract}

In clinical trials studying paired parts of a subject with binary outcomes, it is expected to collect measurements bilaterally. However, there are cases where subjects contribute measurements for only one part. By utilizing combined data, it is possible to gain additional information compared to using bilateral or unilateral data alone. With the combined data, this article investigates homogeneity tests of risk differences with the presence of stratification effects and proposes interval estimations of a common risk difference if stratification does not introduce underlying dissimilarities. Under Dallal's model \citeyearpar{dallal1988paired}, we propose three test statistics and evaluate their performances regarding type I error controls and powers. Confidence intervals of a common risk difference with satisfactory coverage probabilities and interval length are constructed. Our simulation results show that the score test is the most robust and the profile likelihood confidence interval outperforms other methods proposed. Data from a study of acute otitis media is used to illustrate our proposed procedures.
\end{abstract}

\keywords{Dallal's model; homogeneity test; confidence interval; stratified design; unilateral and bilateral data; risk difference}

\section{Introduction}
In clinical trials, particularly in ophthalmologic and otology studies, data are commonly collected in pairs. Each subject may contribute measurements from both sides (bilateral) or just one side (unilateral), depending on the disease status. The response variables in these studies are typically binary outcomes (with or without disease). For example, participants with both eyes infected in a double-blind randomized trial are assigned to either a treatment or a control group. Throughout the duration of the study, the conditions of the eyes are observed and the endpoint of interest is whether the eye is cured or not. Measurements from two eyes of the same subject are usually correlated. Any statistical approaches that ignore this attribute will lead to misleading inferences according to Rosner \citeyearpar{rosner1982statistical}, Dallal \citeyearpar{dallal1988paired}, and Donner and Banting \citeyearpar{donner1988analysis}. Zhang and Ying evaluated \citeyearpar{zhang2018statistical} statistical applications in eye data analyses published in the British Journal of Ophthalmology (BJO) in 1995 and 2017. Their finding revealed that a majority of studies did not adjust for the intereye correlation.

Rosner \citeyearpar{rosner1982statistical} introduced a "constant R" model with an assumption that the conditional probability of a disease at one site given a disease at the other site is proportional to the prevalence of the corresponding group. This proportion is assumed constant among all groups. For bilateral data, Tang et al. \citeyearpar{tang2008testing} presented a set of eight statistical tests to assess the equality of prevalence between two groups. Ma et al. \citeyearpar{ma2015homogeneity} proposed three asymptotic test statistics for testing the homogeneity of multiple proportions given bilateral data. Tang et al. \citeyearpar{tang2006statistical}, Shan and Ma \citeyearpar{shan2014exact}, and Liu et al. \citeyearpar{liu2017exact} developed exact and approximate unconditional procedures for small sample designs or sparse data structures. Tang et al. \citeyearpar{tang2011asymptotic} constructed asymptotic confidence intervals for proportion difference specifically when there are two groups involved in the study. In practical scenarios, it is common to conduct the same study design on different populations, such as individuals of different ages, genders, ethnicities, or races. Combining data from multiple sources can enhance the statistical power of the analysis. However, it is crucial to thoroughly investigate and assess the assumption of the absence of confounding effects before merging the data. Qiu et al. \citeyearpar{qiu2019tests, qiu2019construction} studied homogeneity tests of risk differences, derived sample size formulas, and proposed confidence intervals in a stratified design. Donner \citeyearpar{donner1989statistical} presented an alternative approach to the "constant R" model assuming that the intra-cluster correlation for all individuals is constant. Focusing specifically on bilateral data, Ma and Liu \citeyearpar{ma2017testing} proposed asymptotic methods for testing the equality of multiple proportions. Liu et al. \citeyearpar{liu2017exact} employed exact methods for the equality test when the sample size is small. Confidence intervals of proportion differences in a two-arm randomized clinical trial were studied by Pei et al. \citeyearpar{pei2012confidence}. Yang et al. \citeyearpar{yang2021simultaneous} explored simultaneous confidence intervals for many-to-one comparisons of proportion differences. In the context of stratified studies, Shen et al. \citeyearpar{shen2018testing, shen2019common} developed statistical tests of the homogeneity of the difference of two proportions and interval estimations of a common risk difference. 

Under the combined data framework, Ma and Wang \citeyearpar{ma2021testing} investigated asymptotic procedures for equality of multiple proportions using the "constant R" model. Li et al. \citeyearpar{li2023testing} considered testing the common risk difference of proportions under stratified design with the same model. Qiu et al. \citeyearpar{qiu2022confidence} developed numerous confidence intervals for assessing the equivalence of two treatments using both asymptotic and bootstrap resampling techniques. Additionally, Qiu et al. \citeyearpar{qiu2021confidence} expanded their research by incorporating the impact of stratification effects. Pei et al. \citeyearpar{pei2008testing} examined the equality of cure rates between two treatment groups based on Donner's model \citeyearpar{donner1989statistical}. Ma and Wang \citeyearpar{ma2022testing} extended their work by considering scenarios involving more than 2 groups.

Dallal \citeyearpar{dallal1988paired} pointed out that the "constant R" model will give a poor fit if the characteristic is almost certain to occur bilaterally with widely varying group-specific prevalences and assumed that the conditional probability of a response at one site given a response at the other, denoted as $\gamma$, is independent of the prevalence of the corresponding group. In Dallal's work, three models were proposed based on different settings as the followings:

\[\begin{cases}
     \text{Model 1: homogeneous prevalence and the conditional probability $\gamma$ across different groups;}\\
      \text{Model 2: group-specific prevalences and homogeneous $\gamma$;}\\
      \text{Model 3: group-specific prevalences and $\gamma$'s;}\\
    \end{cases} \]\\
Dallal obtained the likelihood-ratio statistic for pairwise comparisons among the three models. In addition to the aforementioned models, the following case is worth considering when the primary focus is on evaluating the prevalences: 
\begin{itemize}
  \item Model 4: homogeneous prevalence and group-specific $\gamma$'s.
\end{itemize}
Under Dallal's model with bilateral data, Li et al. \citeyearpar{li2020statistical} developed asymptotic and exact test procedures for comparing Model 2 and Model 3. Chen et al. \citeyearpar{chen2022further} further studied four statistical tests including likelihood-ratio, score, Wald-type, and Rosner's tests for comparing Model 1 vs. Model 2 and Model 3 vs. Model 4, and concluded that the score test was the most robust method. For stratified designs, Sun et al. \citeyearpar{sun2022risk} explored tests of homogeneity of risk differences with only bilateral data presuming group-specific $\gamma$'s. However, there remains an compelling and unresolved area of research regarding the homogeneity testing of risk differences in combined bilateral and unilateral data frameworks, assuming a common conditional probability $\gamma$, while accounting for stratification effects.

In this paper, we present three statistical homogeneity tests and five confidence intervals for estimations of a common risk difference. In Section 2, we give a brief introduction to Dallal's model and describe the data structure. In Section 3, three test statistics, likelihood-ratio test, score test, and Wald-type test, are derived and their asymptotic behaviors are investigated as they are widely employed methods for handling multiple parameters and have demonstrated robustness in the literatures reviewed above. Confidence intervals are constructed using the three statistics. Simulation results are presented in Section 4 to compare these methods in terms of the type I error controls, powers, coverage probabilities, and mean interval lengths. All procedures are applied to a real example from an otology study for illustration in Section 5. Finally, we summarize the conclusions in Section 6.

\section{Dallal’s Model}

Assume that there are a total of $S$ stratum and each contains 2 groups. For the $s$th stratum ($s=1, 2, ..., S$), $N_s$ subjects contribute a pair of measurements and $M_s$ subjects contribute only one measurement of a paired organs. Let $n_{sgr}$ denote the number of subjects contributing bilateral data with $r$ response(s) in the $i$th group, where $g=1, 2$ and $r=0, 1, 2$. Similarly, let $m_{sgr^*}$ denote the number of subjects contributing data on only one of a paired organ with $r$ response(s) in the $i$th group, where $g=1, 2$ and $r^*=0, 1$. Let $n_{sg}$ and $m_{sg}$ be the total number of subjects in the $g$th group of the bilateral cohort and unilateral cohort, respectively. The data structure is presented in \hyperref[tab:datastructure]{Table \ref*{tab:datastructure}}. Clearly, \(N_s=\sum_{g=1}^{2} n_{sg}=\sum_{g=1}^{2} \sum_{r=0}^{2} n_{sgr}\) and \(M_s=\sum_{g=1}^{2} m_{sg}=\sum_{g=1}^{2} \sum_{r^*=0}^{1} m_{sgr^*}\). For the bilateral cohort, $\boldsymbol{n_{sg}}$ is used to refer to the random vector $(n_{sg0}, n_{sg1}, n_{sg2})^{T}$ following a multinomial distribution $Multi (n_{sg}; P_{bsg0}, P_{bsg1}, P_{bsg2})$, where $P_{bsgr}$ corresponds to the probability of a subject from the $ith$ group having $r$ response(s) $(r=0, 1, 2)$. For the unilateral cohort, $\boldsymbol{m_{sg}}$ is used to refer to the random vector $(m_{sg0}, m_{sg1})^{T}$ following a binomial distribution $B(m_{sg}; P_{usg0}, P_{usg1})$, where $P_{usgr^*}$ corresponds to the probability of a subject from the $ith$ group having $r^*$ response(s) $(r^*=0, 1)$.\\
\begin{table}[H]
\centering
\caption{Data Structure for the $s$th Stratum ($s=1, 2, ..., S$)}
\label{tab:datastructure}
\begin{tabular}{l|cccc|ccc}
\toprule
\multirow{3}{*}{group ($g$)} & \multicolumn{4}{c|}{Bilateral}                                                                                            & \multicolumn{3}{c}{Unilateral}                                                          \\ \cline{2-8} 
                             & \multicolumn{3}{c|}{Response ($r$)}                                                              & \multirow{2}{*}{Total} & \multicolumn{2}{c|}{Response ($r^*$)}                           & \multirow{2}{*}{Total} \\ \cline{2-4} \cline{6-7}
                             & \multicolumn{1}{c|}{0}         & \multicolumn{1}{c|}{1}         & \multicolumn{1}{c|}{2}         &                        & \multicolumn{1}{c|}{0}         & \multicolumn{1}{c|}{1}         &                        \\ \hline
1                            & \multicolumn{1}{c|}{$n_{s10}$} & \multicolumn{1}{c|}{$n_{s11}$} & \multicolumn{1}{c|}{$n_{s12}$} & $n_{s1}$               & \multicolumn{1}{c|}{$m_{s10}$} & \multicolumn{1}{c|}{$m_{s11}$} & $m_{s1}$               \\ \hline
2                            & \multicolumn{1}{c|}{$n_{s20}$} & \multicolumn{1}{c|}{$n_{s21}$} & \multicolumn{1}{c|}{$n_{s22}$} & $n_{s2}$               & \multicolumn{1}{c|}{$m_{s20}$} & \multicolumn{1}{c|}{$m_{s21}$} & $m_{s2}$               \\ \hline
Total                        & \multicolumn{1}{c|}{-}         & \multicolumn{1}{c|}{-}         & \multicolumn{1}{c|}{-}         & $N_s$                  & \multicolumn{1}{c|}{-}         & \multicolumn{1}{c|}{-}         & $M_s$                  \\ 
\bottomrule
\end{tabular}

\end{table}

Under Dallal's model, we assume the probability of observing a response on the $k$th site of the $j$th subject from the $g$th group within $s$th stratum is $\pi_{sg}$, where $j=1,2,...,n_{sg}$ or $j=1,2,...,m_{sg}$ depending on which cohort (bilateral or unilateral) the subject belongs to. Define $Z_{sgjk}=1 (k=1, 2)$ as the occurrence of such an event. The conditional probability of observing a response at one site given a response at the other site of the subject is described by a parameter $\gamma_{sg}$. Then we have
\[ Pr(Z_{sgjk}=1)=\pi_{sg},\; Pr(Z_{sgjk}=1|Z_{sgj(3-k)}=1)=\gamma_{sg} \;,\]
where $0\leq \pi_{sg}\leq1$ and $0\leq\gamma_{sg}\leq 1$. It is straightforward to show that
\[\begin{cases}
P_{bsg0}=1-2\pi_{sg}+\pi_{sg}\gamma_{sg},\; P_{bsg1}=2\pi_{sg}(1-\gamma_{sg}),\; P_{bsg2}=\pi_{sg}\gamma_{sg} \\
P_{usg0}=1-\pi_{sg},\; P_{usg1}=\pi_{sg}.
\end{cases}\]\\
Obviously, there are constraints on the parameter space. That is, $0\leq p_{bsg0},\;p_{bsg1},\;p_{bsg2},\; p_{usg0},\;p_{usg1}\leq 1$, which is equivalent to $0\leq \pi_{sg}\leq 1/(2-\gamma_{sg})$, $\gamma_{sg}\in [0, 1]$. 

In general circumstances, we are interested in testing if stratified designs will cause heterogeneous risk differences, denoted as $d_s=\pi_{s1}-\pi_{s2}$, across different strata under the condition that $\gamma_{s1}=\gamma_{s2}$ ($s=1, 2, ..., S$). Therefore, the null and the alternative hypotheses are as follows:

\[\begin{cases}
     \text{$H_0$: $d_1=d_2=...=d_S$ and $\gamma_{s1}=\gamma_{s2}$, where $s=1, 2, ..., S$}\\
      \text{$H_a$: $d_i\neq d_j$ for some $i\neq j$ and $\gamma_{s1}=\gamma_{s2}$, where $i,j,s \in \{1, 2, ..., S\}$}
    \end{cases} \]\\

For convenience, we will use the notation $\gamma_s$ to represent $\gamma_{sg}$ since testing the correlations is not relevant to the discussion of this paper.

\section{Methods}
\subsection{Maximum Likelihood Estimators (MLEs)}
\subsubsection{Unconstrained MLEs}
Assume independence between different strata or groups and let $\boldsymbol{\beta}=(\pi_{11}, \pi_{12}, \gamma_1,... ,\pi_{S1}, \pi_{S2}, \gamma_S)^T$. Under the alternative hypothesis $H_a$, the likelihood function is given by
\begin{flalign*}
    \mathcal{L}(\boldsymbol{\beta}|(\boldsymbol{n_{11}}, \boldsymbol{n_{12}}, \boldsymbol{m_{11}}, \boldsymbol{m_{12}}..., \boldsymbol{m_{S1}},  \boldsymbol{m_{S2}}))&=\prod_{s=1}^{S}\prod_{g=1}^{2}\frac{n_{sg}!}{n_{sg0}!n_{sg1}!n_{sg2}!}P_{bsg0}^{n_{sg0}}P_{bsg1}^{n_{sg1}}P_{bsg2}^{n_{sg2}}\times \frac{m_{sg}!}{m_{sg0}!m_{sg1}!}P_{usg0}^{m_{sg0}}P_{usg1}^{m_{sg1}},
\end{flalign*}
and the corresponding log-likelihood can be expressed as
\begin{flalign*}   
 l(\boldsymbol{\beta}|(\boldsymbol{n_{11}}, \boldsymbol{n_{12}}, \boldsymbol{m_{11}}, \boldsymbol{m_{12}}..., \boldsymbol{m_{S1}},  \boldsymbol{m_{S2}}))=\sum_{s=1}^{S}\sum_{g=1}^{2}&\Biggl\{m_{\textrm{sg0}} \,\log \left(1-\pi_{\textrm{sg}} \right)+n_{\textrm{sg2}} \,\log \left(\gamma_s \,\pi_{\textrm{sg}} \right) \\
 &+n_{\textrm{sg0}} \,\log \left(\gamma_s \,\pi_{\textrm{sg}} -2\,\pi_{\textrm{sg}} +1\right) \\
 &+n_{\textrm{sg1}} \,\log \left(2\,\pi_{\textrm{sg}} -2\,\gamma_s \,\pi_{\textrm{sg}} \right)+m_{\textrm{sg1}} \,\log \left(\pi_{\textrm{sg}} \right)\Biggl\}.
\end{flalign*}

The unconstrained MLEs of $\pi_{sg}$ and $\gamma_{s}$, denoted as $\hat{\pi}_{sg}$ and $\hat{\gamma}_s$, respectively, can be obtained by solving the following equations:
\[\frac{\partial l}{\partial \pi_{sg}}=\frac{m_{\textrm{sg1}} }{\pi_{\textrm{sg}} }+\frac{n_{\textrm{sg1}} }{\pi_{\textrm{sg}} }+\frac{n_{\textrm{sg2}} }{\pi_{\textrm{sg}} }+\frac{m_{\textrm{sg0}} }{\pi_{\textrm{sg}} -1}+\frac{n_{\textrm{sg0}} \,{\left(\gamma_s -2\right)}}{\gamma_s \,\pi_{\textrm{sg}} -2\,\pi_{\textrm{sg}} +1}=0 \] \\
\[\frac{\partial l}{\partial \gamma_s}=\frac{n_{\textrm{s12}} }{\gamma_s }+\frac{n_{\textrm{s22}} }{\gamma_s }+\frac{n_{\textrm{s11}} }{\gamma_s -1}+\frac{n_{\textrm{s21}} }{\gamma_s -1}+\frac{n_{\textrm{s10}} \,\pi_{\textrm{s1}} }{\gamma_s \,\pi_{\textrm{s1}} -2\,\pi_{\textrm{s1}} +1}+\frac{n_{\textrm{s20}} \,\pi_{\textrm{s2}} }{\gamma_s \,\pi_{\textrm{s2}} -2\,\pi_{\textrm{s2}} +1}=0, \]\\
where $s=1, 2, ..., S$ and $g=1, 2$. However, there is no closed form of solution for $\gamma_s$. Hence, iterative methods described in Ma et al. \citeyearpar{ma2015homogeneity} are applied. The $(t+1)$th update of $\pi_{sg}$ can be updated by solving the following second-order polynomial: 

\begin{flalign*}  
{\left(\gamma_s \,m_{\textrm{sg0}} -2\,m_{\textrm{sg1}} -2\,n_{\textrm{sg0}} -2\,n_{\textrm{sg1}} -2\,n_{\textrm{sg2}} -2\,m_{\textrm{sg0}} +\gamma_s \,m_{\textrm{sg1}} +\gamma_s \,n_{\textrm{sg0}} +\gamma_s \,n_{\textrm{sg1}} +\gamma_s \,n_{\textrm{sg2}} \right)}\,{\pi_{\textrm{sg}} }^2 \\
+{\left(m_{\textrm{sg0}} +3\,m_{\textrm{sg1}} +2\,n_{\textrm{sg0}} +3\,n_{\textrm{sg1}} +3\,n_{\textrm{sg2}} -\gamma_s \,m_{\textrm{sg1}} -\gamma_s \,n_{\textrm{sg0}} -\gamma_s \,n_{\textrm{sg1}} -\gamma_s \,n_{\textrm{sg2}} \right)}\,\pi_{\textrm{sg}} \\
-m_{\textrm{sg1}} -n_{\textrm{sg1}} -n_{\textrm{sg2}}=0,
\end{flalign*}
and $\gamma_s$ can be updated by the Newton-Raphson method:

\[
\gamma_s^{(t+1)}=\gamma_s^{(t)}-\left(\frac{\partial^2 l}{\partial \gamma_s^2} \right)^{-1}\frac{\partial l}{\partial \gamma_s}\Bigg|\textrm{$\pi_{s1}=\pi_{s1}^{(t)}, \pi_{s2}=\pi_{s2}^{(t)}, \gamma_s=\gamma_{s}^{(t)}$},
\]
where
\[\frac{\partial^2 l}{\partial \gamma_s^2}=
-\frac{n_{\textrm{s12}} }{{\gamma_s }^2 }-\frac{n_{\textrm{s22}} }{{\gamma_s }^2 }-\frac{n_{\textrm{s11}} }{{{\left(\gamma_s -1\right)}}^2 }-\frac{n_{\textrm{s21}} }{{{\left(\gamma_s -1\right)}}^2 }-\frac{n_{\textrm{s10}} \,{\pi_{\textrm{s1}} }^2 }{{{\left(\gamma_s \,\pi_{\textrm{s1}} -2\,\pi_{\textrm{s1}} +1\right)}}^2 }-\frac{n_{\textrm{s20}} \,{\pi_{\textrm{s2}} }^2 }{{{\left(\gamma_s \,\pi_{\textrm{s2}} -2\,\pi_{\textrm{s2}} +1\right)}}^2 }.
\]

\subsubsection{Constrained MLEs}
Assume $H_0$: $d_1=d_2=...=d_S=d$ is true where $d$ is unknown, then the constrained MLEs of $\pi_{s1}$, $d$, and $\gamma_s$, denoted as $\Tilde{\pi}_{s1}$, $\Tilde{d}$, and $\Tilde{\gamma}_s$, respectively, can be obtained by the following iterative procedure as described in Shen and Ma \citeyearpar{shen2018testing}:

(1) Initialize $\pi_{s1}^{(0)}=\hat{\pi}_{s1}$, $\gamma_s^{(0)}=\hat{\gamma}_s$, and $d^{(0)}=(1/S)\sum_{s=1}^{S}(\hat{\pi}_{s1}-\hat{\pi}_{s2})$, where $\hat{\pi}_{s1}$, $\hat{\pi}_{s2}$, and $\hat{\gamma}_s$ are the unconstrained MLEs;

(2) Update
\begin{align*}
\begin{bmatrix}
\pi_{s1}^{(t+1)} \\
\gamma_{s}^{(t+1)}          
\end{bmatrix}
=
\begin{bmatrix}
\pi_{s1}^{(t)} \\
\gamma_{s}^{(t)}          
\end{bmatrix}  - 
\begin{bmatrix}
\frac{\partial^2 l_s}{\partial \pi_{s1}^2} & \frac{\partial^2 l_s}{\partial \pi_{s1}\partial \gamma_s}\\
\frac{\partial^2 l_s}{\partial \pi_{s1}\partial \gamma_s} & \frac{\partial^2 l_s}{\partial \gamma_{s}^2}        
         \end{bmatrix}^{-1} \times
  \begin{bmatrix}
  \frac{\partial l_s}{\partial \pi_{s1}} \\
  \frac{\partial l_s}{\partial \gamma_{s}}
  \end{bmatrix}\Bigg|\textrm{$\pi_{s1}=\pi_{s1}^{(t)}, \gamma_s=\gamma_{s}^{(t)}, d=d^{(t)}$},
\end{align*}
where $l_s$ is the log-likelihood function of the $s$th stratum ($s=1, 2, ..., S$);

(3) Update
\begin{align*}
d^{(t+1)}=d^{(t)}-\left( \sum_{s=1}^{S}\frac{\partial^2 l_s}{\partial d^2} \right)^{-1}\times \left(\sum_{s=1}^{S}\frac{\partial l_s}{\partial d} \right)\Bigg|\textrm{$\pi_{s1}=\pi_{s1}^{(t)}, \gamma_s=\gamma_{s}^{(t)}, d=d^{(t)}$};
\end{align*}

(4) Repeat steps (2) and (3) until convergence.\\

Subsequently, the constrained MLE of $\pi_{s2}$ can be obtained by $\Tilde{\pi}_{s2}=\Tilde{\pi}_{s1}-\Tilde{d}$.

\subsubsection{Conditional MLEs}
Given a known common risk difference $d_0$, the log-likelihood function can be written by substituting $\pi_{s1}-d_0$ for $\pi_{s2}$. Letting the first derivative of the log-likelihood function with respect to $\gamma_{s}$ equal zero leads to solving a cubic equation for $\gamma_s$. There is no explicit form of solution for $\pi_{s1}$, therefore, an iterative procedure in a similar manner to the method above for the constrained MLEs is applied. The steps are as follows:

(1) Initialize $\pi_{s1}^{(0)}=\Tilde{\pi}_{s1}$ and $\gamma_s^{(0)}=\Tilde{\gamma}_s$, where $\Tilde{\pi}_{s1}$ and $\Tilde{\gamma}_s$ are the constrained MLEs;

(2) Update 
\[
\pi_{s1}^{(t+1)}=\pi_{s1}^{(t)}-\left(\frac{\partial^2 l}{\partial \pi_{s1}^2} \right)^{-1}\frac{\partial l}{\partial \pi_{s1}}\Bigg|\textrm{$\gamma_s=\gamma_{s}^{(t)}, \pi_{s1}=\pi_{s1}^{(t)}, d=d_0$},
\]

(3) Solve the cubic function for $\gamma_s$ resulting from 
\begin{flalign*}  
\frac{\partial l}{\partial \gamma_{s}} =&\frac{n_{\textrm{s12}} }{\gamma_s }+\frac{n_{\textrm{s22}} }{\gamma_s }+\frac{n_{\textrm{s11}} }{\gamma_s -1}+\frac{n_{\textrm{s21}} }{\gamma_s -1} \\
&+\frac{n_{\textrm{s10}} \,\pi_{\textrm{s1}}^{(t+1)} }{\gamma_s \,\pi_{\textrm{s1}}^{(t+1)} -2\,\pi_{\textrm{s1}}^{(t+1)} +1}-\frac{n_{\textrm{s20}} \,{\left(d_0-\pi_{\textrm{s1}}^{(t+1)} \right)}}{2\,d_0-2\,\pi_{\textrm{s1}}^{(t+1)} -d_0\,\gamma_s +\gamma_s \,\pi_{\textrm{s1}}^{(t+1)} +1}=0
\end{flalign*}  

(4) Repeat steps (2) and (3) until convergence.

\subsection{Statistical Tests}
\subsubsection{Likelihood-Ratio Test ($T_{LR}$)}
Let $\hat{\boldsymbol{\beta}}$ and $\Tilde{\boldsymbol{\beta}}$ be the MLEs of $\boldsymbol{\beta}$ under $H_a$ and $H_0$, respectively. Then the likelihood-ratio (LR) test is given by
\begin{flalign*}T_{LR}&=2(l(\hat{\boldsymbol{\beta}})-l(\Tilde{\boldsymbol{\beta}})) \; \\
&=2(l(\hat{\pi}_{11}, \hat{\pi}_{12}, \hat{\gamma}_1, ..., \hat{\pi}_{S1}, \hat{\pi}_{S2}, \hat{\gamma}_S)-l(\Tilde{\pi}_{11}, \Tilde{\pi}_{12}, \Tilde{\gamma}_1, ..., \Tilde{\pi}_{S1}, \Tilde{\pi}_{S2}, \Tilde{\gamma}_S))\; \\
&=2(l(\hat{\pi}_{11}, \hat{\pi}_{12}, \hat{\gamma}_1, ..., \hat{\pi}_{S1}, \hat{\pi}_{S2}, \hat{\gamma}_S)-l(\Tilde{\pi}_{11}, \Tilde{\pi}_{11}-\Tilde{d}, \Tilde{\gamma}_1, ..., \Tilde{\pi}_{S1}, \Tilde{\pi}_{S1}-\Tilde{d}, \Tilde{\gamma}_S)).\; 
\end{flalign*}
Under the regularity conditions, $T_{LR}$ follows asymptotically a $\mathcal{X}^2$ distribution with $S-1$ degrees of freedom as the sample size approaches infinity according to Wilks \citeyearpar{wilks1938large}.

\subsubsection{Score Test ($T_{SC}$)}
Denote $U(\boldsymbol{\beta})=(\frac{\partial{l}}{\partial{\pi_{11}}},\frac{\partial{l}}{\partial{\pi_{12}}},\frac{\partial{l}}{\partial{\gamma_{1}}},...,\frac{\partial{l}}{\partial{\pi_{S1}}},\frac{\partial{l}}{\partial{\pi_{S2}}},\frac{\partial{l}}{\partial{\gamma_{S}}})^T$. Under $H_0$, the score test statistic is defined as
\[T_{SC}=U(\boldsymbol{\beta})^T\mathcal{I}_a^{-1}(\boldsymbol{\beta})U(\boldsymbol{\beta})|\boldsymbol{\beta}=\Tilde{\boldsymbol{\beta}},\]
where $\mathcal{I}_a$ is the Fisher information matrix for $\boldsymbol{\beta}$ under the parameter settings when $H_a$ is true. The score test statistic is asymptotically distributed as a $\mathcal{X}^2$ distribution with $S-1$ degrees of freedom according to Rao \citeyearpar{rao1948large}. The expression of $\mathcal{I}_a$ is as follows:

\[\mathcal{I}_a(\boldsymbol{\beta})=
\begin{pmatrix}
  \begin{matrix}
  I_{1,11} & I_{1,12} & I_{1,13} \\
  I_{1,21} & I_{1,22} & I_{1,23} \\
  I_{1,31} & I_{1,32} & I_{1,33}
  \end{matrix}
  & \bigzero   & ... & \bigzero \\
 
  \bigzero &  \begin{matrix}
  I_{2,11} & I_{2,12} & I_{2,13} \\
  I_{2,21} & I_{2,22} & I_{2,23} \\
  I_{2,31} & I_{2,32} & I_{2,33}
  \end{matrix}   & ... & \bigzero \\
 &.&& \\
 &.&& \\
 &.&& \\
  \bigzero &  \bigzero  & ...  & \begin{matrix}
  I_{S,11} & I_{S,12} & I_{S,13} \\
  I_{S,21} & I_{S,22} & I_{S,23} \\
  I_{S,31} & I_{S,32} & I_{S,33}
  \end{matrix} 
\end{pmatrix}_{3S\times 3S}
\]
where 
\begin{flalign*}
&I_{s,11}=E(-\frac{\partial{}^2l}{\partial{\pi_{s1}}^2})=-\frac{m_{s1} +2\,n_{s1} -\gamma_s \,n_{s1} -2\,m_{s1} \,\pi_{\textrm{s1}} -2\,n_{s1} \,\pi_{\textrm{s1}} +\gamma_s \,m_{s1} \,\pi_{\textrm{s1}} +\gamma_s \,n_{s1} \,\pi_{\textrm{s1}} }{\pi_{\textrm{s1}} \,{\left(\pi_{\textrm{s1}} -1\right)}\,{\left(\gamma_s \,\pi_{\textrm{s1}} -2\,\pi_{\textrm{s1}} +1\right)}}\;, \\
&I_{s,12}=I_{s,21}=E(-\frac{\partial{}^2l}{\partial{\pi_{s1}} \partial{\pi_{s2}}})=0\;, \\
&I_{s,13}=I_{s,31}=E(-\frac{\partial{}^2l}{\partial{\pi_{s1}}\partial{\gamma_s}})=-\frac{n_{s1}}{\gamma_s \,\pi_{\textrm{s1}} -2\,\pi_{s1} +1}\;,\\
&I_{s,22}=E(-\frac{\partial{}^2l}{\partial{\pi_{s2}}^2})=-\frac{m_{s2} +2\,n_{s2} -\gamma_s \,n_{s2} -2\,m_{s2} \,\pi_{\textrm{s2}} -2\,n_{s2} \,\pi_{\textrm{s2}} +\gamma_s \,m_{s2} \,\pi_{\textrm{s2}} +\gamma_s \,n_{s2} \,\pi_{\textrm{s2}} }{\pi_{\textrm{s2}} \,{\left(\pi_{\textrm{s2}} -1\right)}\,{\left(\gamma_s \,\pi_{\textrm{s2}} -2\,\pi_{\textrm{s2}} +1\right)}}\;, \\
&I_{s,23}=I_{s,32}=E(-\frac{\partial{}^2l}{\partial{\pi_{s2}}\partial{\gamma_s}})=-\frac{n_{s2} }{\gamma_s \,\pi_{\textrm{s2}} -2\,\pi_{\textrm{s2}} +1}\;, \\
&I_{s,33}=E(-\frac{\partial{}^2l}{\partial{\gamma_{s}}^2})=\frac{n_{s1} \,\pi_{\textrm{s1}} }{\gamma_s }+\frac{n_{s2} \,\pi_{\textrm{s2}} }{\gamma_s }+\frac{n_{s1} \,{\pi_{\textrm{s1}} }^2 }{\gamma_s \,\pi_{\textrm{s1}} -2\,\pi_{\textrm{s1}} +1}+\frac{n_{s2} \,{\pi_{\textrm{s2}} }^2 }{\gamma_s \,\pi_{\textrm{s2}} -2\,\pi_{\textrm{s2}} +1}-\frac{2\,n_{s1} \,\pi_{\textrm{s1}} }{\gamma_s -1}-\frac{2\,n_{s2} \,\pi_{\textrm{s2}} }{\gamma_s -1}\;, \\
&\textrm{and s=1, 2, 3, ..., S.}
\end{flalign*}
Furthermore, the score test statistic can be simplified as
\begin{flalign*}
T_{SC}=\sum_{s=1}^{S} \Biggl\{ (\frac{\partial{l}}{\partial{\pi_{s1}}},\frac{\partial{l}}{\partial{\pi_{s2}}},\frac{\partial{l}}{\partial{\gamma_{s}}})
  \begin{pmatrix}
  I_{s,11} & 0 & I_{s,13} \\
  0 & I_{s,22} & I_{s,23} \\
  I_{s,31} & I_{s,32} & I_{s,33}
  \end{pmatrix}
  (\frac{\partial{l}}{\partial{\pi_{s1}}},\frac{\partial{l}}{\partial{\pi_{s2}}},\frac{\partial{l}}{\partial{\gamma_{s}}})^T \Biggr\}\Bigg|\boldsymbol{\beta}=\Tilde{\boldsymbol{\beta}}
\end{flalign*}

\subsubsection{Wald-Type Test ($T_W$)}
The null hypothesis $H_0$: $d_1=d_2=...=d_S$ is equivalent to $C\boldsymbol{\beta}=0$ where $\boldsymbol{\beta}=(\pi_{11}, \pi_{12}, \gamma_1,... ,\pi_{S1}, \pi_{S2}, \gamma_S)^T$ and
\[C=
\begin{bmatrix}
1 & -1 & 0 & -1& 1 & 0 & 0 & 0 & 0 & 0 &0 & 0 & ... &0 \\
0 &  0 & 0 & 1& -1 & 0 & -1& 1 & 0 & 0 &0 & 0 & ... &0 \\
0 &  0 & 0 & 0& 0 & 0 &  1& -1 & 0 & -1 &1 & 0 & ... &0 \\
&   &  & &  &  &  & ... &  &  & &  &  & \\
0 &  0 & 0 & 0& 0 & 0 &  ...& 0 & 1 & -1 &0 & -1 & 1 &0
\end{bmatrix}_{(S-1)\times(3S)}.\]
Wald-type test statistic ($T_W$) for testing $H_0$ can be expressed as 
\[T_{W}=(C\boldsymbol{\beta})^T(C\mathcal{I}_a(\boldsymbol{\beta})^{-1}C^T)^{-1}C\boldsymbol{\beta}|\boldsymbol{\beta}=\hat{\boldsymbol{\beta}},\]
and it follows asymptotically a $\mathcal{X}^2$ distribution with $S-1$ degrees of freedom according to Wald \citeyearpar{wald1943tests}. It is easy to show the matrix $C\mathcal{I}_a(\boldsymbol{\beta})^{-1}C^T$ is a tridiagonal matrix. That is,

\[
C\mathcal{I}_a(\boldsymbol{\beta})^{-1}C^T=
\begin{bmatrix}
a_1 & b_1 & & & &\\
b_1 & a_2 & b_2 & & &\\
   &  b_2 & a_3 &  & &\\
   & & & ... & &\\
   & & & &a_{S-2} &b_{S-2}\\
  & & & &b_{S-2} &a_{S-1}
\end{bmatrix}_{(S-1)\times (S-1)},
\]
where $a_s=inv_{s,11} -2\,inv_{s,12} +inv_{s,22} +inv_{s+1,11} -2\,inv_{s+1,12} +inv_{s+1,22}$ ($s=1, 2, ..., S-1$), $b_s=2\,inv_{s+1,12} -inv_{s+1,11} -inv_{s+1,22}$ ($s=1, 2, ..., S-2$), and

\[
\begin{bmatrix}
inv_{s,11}&inv_{s,12} &inv_{s,13}\\
inv_{s,21}&inv_{s,22} &inv_{s,23}\\
inv_{s,31}&inv_{s,32} &inv_{s,33}\\
\end{bmatrix}
=
\begin{bmatrix}
I_{s,11}&I_{s,12} &I_{s,13}\\
I_{s,21}&I_{s,22} &I_{s,23}\\
I_{s,31}&I_{s,32} &I_{s,33}\\
\end{bmatrix}^{-1} \textrm{for s=1, 2, ..., S.}
\]
Denote the inverse matrix of $C\mathcal{I}_a(\boldsymbol{\beta})^{-1}C^T$ as $I_{CIC}^{-1}$. The formula for calculating this inverse matrix, as derived by Da Fonseca \citeyearpar{da2007eigenvalues}, Usmani \citeyearpar{usmani1994inversion}, Mallik \citeyearpar{mallik2001inverse}, and K{\i}l{\i}{\c{c}} \citeyearpar{kilicc2008explicit}, is as follows:
\[
\textrm{$(I_{CIC}^{-1})_{ij}$}=
\begin{cases}
(-1)^{i+j}b_i...b_j\theta_{i-1}\phi_{j+1}/\theta_n  \textrm{, if $i<j$}\\
\theta_{i-1}\phi_{j+1}/\theta_n  \textrm{, if $i=j$}\\
(-1)^{i+j}b_j...b_{i-1}\theta_{j-1}\phi_{i+1}/\theta_n  \textrm{, if $i>j$}
\end{cases}
\]\\
where $(I_{CIC}^{-1})_{ij}$ represents the element in $i$th row and $j$th column of the inverse matrix and 
\[
\begin{cases}
\theta_i=a_i\theta_{i-1}-b_{i-1}c_{i-1}\theta_{i-2}\textrm{, i=2, 3, ..., S-1}\\
\theta_0=1 \\
\theta_1=a_1 \\
\phi_i=a_i\phi_{i+1}=b_ic_i\phi_{i+2}\textrm{, i=S-2, ..., 1}\\
\phi_{S}=1\\
\phi_{S-1}=a_{S-1}.
\end{cases}
\]\\

\subsection{Confidence Intervals (CIs)}
Given a homogeneity test result failing to reject $H_0: d_1=d_2=...=d_S$ or prior knowledge of the existence of a common risk difference, the main target now is to estimate this quantity. We propose five confidence interval estimations and compare their empirical coverage probabilities and mean interval widths.   

\subsubsection{Unconstrained Wald-Type CI}
The unconstrained Wald-type confidence interval is constructed by considering $S$ strata independently. For each stratum, let $\boldsymbol{\beta}_s=(\pi_{s1}, \pi_{s2}, \gamma_s)^T$, then the Fisher information matrix of $\boldsymbol{\beta}_s$ can be expressed as
\[\mathcal{I}_s(\boldsymbol{\beta}_s)=\begin{bmatrix}
I_{s,11} & I_{s,12} & I_{s,13} \\
I_{s,21} & I_{s,22} & I_{s,23} \\
I_{s,31} & I_{s,32} & I_{s,33} 
\end{bmatrix} \textrm{for s=1, 2, ..., S.}\]\\
Denote $C_u=(1, -1, 0)$ and the common risk difference can be written as $d=C_u\boldsymbol{\beta}_s$. The quantity $$\frac{\sum_{s=1}^{S} w_sC_u\hat{\boldsymbol{\beta}}_s}{\sqrt{\sum_{s=1}^{S}w_s^2C_u\mathcal{I}_s(\hat{\boldsymbol{\beta}}_s)^{-1}C_u^T}}$$ is asymptotically distributed as the standard normal distribution. Two types of weights $w_s$'s are considered: (1) uniform weights $w_s=1/S$ and (2) sample-based weights $w_s=(N_s+M_s)/N_{all}$, where $N_{all}=\sum_{s=1}^{S} (N_s+M_s)$. Therefore, the $100(1-\alpha)\%$ CI of $d$ is given by
\begin{flalign*}
\Biggl[max\left(-1, (\sum_{s=1}^{S} w_sC_u\hat{\boldsymbol{\beta}}_s)-Z_{1-\alpha/2}\sqrt{\sum_{s=1}^{S}w_s^2C_u\mathcal{I}_s(\hat{\boldsymbol{\beta}}_s)^{-1}C_u^T}\right),\\
min\left(1, (\sum_{s=1}^{S} w_sC_u\hat{\boldsymbol{\beta}}_s)+Z_{1-\alpha/2}\sqrt{\sum_{s=1}^{S}w_s^2C_u\mathcal{I}_s(\hat{\boldsymbol{\beta}}_s)^{-1}C_u^T}\right)
\Biggr],
\end{flalign*}
where $\hat{\boldsymbol{\beta}}_s$ is the unconstrained MLE of $\boldsymbol{\beta}_s$, $Z_{1-\alpha/2}$ is the $(1-\alpha/2)$ quantile of the standard normal distribution, and $\alpha$ is a pre-defined type I error rate.

\subsubsection{Constrained Wald-Type CI}
\label{sec:con_wald_CI}
The constrained Wald-type CI is derived from the parameter space with a common risk difference. Denote $\boldsymbol{\beta^*}=(d, \pi_{11}, \gamma_1, ..., \pi_{S1}, \gamma_S)^{T}$ and the Fisher information matrix of $\boldsymbol{\beta^*}$ as\\
\[\mathcal{I}_0(\boldsymbol{\beta}^*)=
\begin{bmatrix}

I^*_{d} & 
I^*_{11} & I^*_{12}& I^*_{21} & I^*_{22}  & ... & I^*_{S1} & I^*_{S2}\\
I^*_{11} & I^*_{1,11} & I^*_{1,12} & 0 & 0 & ... & 0 & 0 \\
I^*_{12} & I^*_{1,21} & I^*_{1,22} & 0 & 0 & ... & 0 & 0 \\

I^*_{21} & 0 & 0  & I^*_{2,11} & I^*_{2,12} & ... & 0 & 0 \\
I^*_{22} & 0 & 0 & I^*_{2,21} & I^*_{2,22} & ... & 0 & 0 \\
&&&&&...&&\\
I^*_{S1} &0&0&0&0&...& I^*_{S,11} & I^*_{S,12}\\
I^*_{S2} &0&0&0&0&...& I^*_{S,21} & I^*_{S,22}\\

\end{bmatrix} 
\]\\
where
\begin{flalign*}
I^*_{d}&=E(-\frac{\partial{}^2l}{\partial{d}^2})=\sum_{s=1}^{S} \biggl\{ \frac{m_{\textrm{s2}} }{d-\pi_{\textrm{s1}} +1}-\frac{m_{\textrm{s2}} }{d-\pi_{\textrm{s1}} }+\frac{n_{\textrm{s2}} \,{\left(2\,\pi_{\textrm{s1}} -2\,d+2\,\gamma_s \,{\left(d-\pi_{\textrm{s1}} \right)}\right)}}{{{\left(d-\pi_{\textrm{s1}} \right)}}^2 } \\
&-\frac{\gamma_s \,n_{\textrm{s2}} }{d-\pi_{\textrm{s1}} }+\frac{n_{\textrm{s2}} \,{{\left(\gamma_s -2\right)}}^2 \,{\left(2\,d-2\,\pi_{\textrm{s1}} -\gamma_s \,{\left(d-\pi_{\textrm{s1}} \right)}+1\right)}}{{{\left(2\,d-2\,\pi_{\textrm{s1}} -d\,\gamma_s +\gamma_s \,\pi_{\textrm{s1}} +1\right)}}^2 } \biggr\}\;, \\
I^*_{s1}&=E(-\frac{\partial{}^2l}{\partial d\partial \pi_{s1}})=\frac{m_{\textrm{s2}} }{d-\pi_{\textrm{s1}} }-\frac{m_{\textrm{s2}} }{d-\pi_{\textrm{s1}} +1}-\frac{n_{\textrm{s2}} \,{\left(2\,\pi_{\textrm{s1}} -2\,d+2\,\gamma_s \,{\left(d-\pi_{\textrm{s1}} \right)}\right)}}{{{\left(d-\pi_{\textrm{s1}} \right)}}^2 } \\
&+\frac{\gamma_s \,n_{\textrm{s2}} }{d-\pi_{\textrm{s1}} }-\frac{n_{\textrm{s2}} \,{{\left(\gamma_s -2\right)}}^2 \,{\left(2\,d-2\,\pi_{\textrm{s1}} -\gamma_s \,{\left(d-\pi_{\textrm{s1}} \right)}+1\right)}}{{{\left(2\,d-2\,\pi_{\textrm{s1}} -d\,\gamma_s +\gamma_s \,\pi_{\textrm{s1}} +1\right)}}^2 }\;, \\
I^*_{s2}&=E(-\frac{\partial{}^2l}{\partial d\partial \gamma_{s}})=\frac{n_{\textrm{s2}} }{2\,d-2\,\pi_{\textrm{s1}} -d\,\gamma_s +\gamma_s \,\pi_{\textrm{s1}} +1}\;,  \\
I^*_{s,11}&=E(-\frac{\partial{}^2l}{\partial \pi_{s1}^2})=\frac{m_{\textrm{s1}} }{\pi_{\textrm{s1}} }-\frac{m_{\textrm{s2}} }{d-\pi_{\textrm{s1}} }+\frac{m_{\textrm{s2}} }{d-\pi_{\textrm{s1}} +1}-\frac{m_{\textrm{s1}} }{\pi_{\textrm{s1}} -1}+\frac{\gamma_s \,n_{\textrm{s1}} }{\pi_{\textrm{s1}} }-\frac{\gamma_s \,n_{\textrm{s2}} }{d-\pi_{\textrm{s1}} }-\frac{2\,n_{\textrm{s1}} \,{\left(\gamma_s -1\right)}}{\pi_{\textrm{s1}} }\\
&+\frac{2\,n_{\textrm{s2}} \,{\left(\gamma_s -1\right)}}{d-\pi_{\textrm{s1}} }+\frac{n_{\textrm{s2}} \,{{\left(\gamma_s -2\right)}}^2 }{2\,d-2\,\pi_{\textrm{s1}} -d\,\gamma_s +\gamma_s \,\pi_{\textrm{s1}} +1}+\frac{n_{\textrm{s1}} \,{{\left(\gamma_s -2\right)}}^2 }{\gamma_s \,\pi_{\textrm{s1}} -2\,\pi_{\textrm{s1}} +1}\;,\\
I^*_{s,12}&=I^*_{s,21}=E(-\frac{\partial{}^2l}{\partial \pi_{s1} \partial \gamma_{s}})\;, \\
&=-\frac{n_{\textrm{s1}} +n_{\textrm{s2}} +2\,d\,n_{\textrm{s1}} -2\,n_{\textrm{s1}} \,\pi_{\textrm{s1}} -2\,n_{\textrm{s2}} \,\pi_{\textrm{s1}} -d\,\gamma_s \,n_{\textrm{s1}} +\gamma_s \,n_{\textrm{s1}} \,\pi_{\textrm{s1}} +\gamma_s \,n_{\textrm{s2}} \,\pi_{\textrm{s1}} }{{\left(\gamma_s \,\pi_{\textrm{s1}} -2\,\pi_{\textrm{s1}} +1\right)}\,{\left(2\,d-2\,\pi_{\textrm{s1}} -d\,\gamma_s +\gamma_s \,\pi_{\textrm{s1}} +1\right)}}\; \\
I^*_{s,22}&=E(-\frac{\partial{}^2l}{\partial \gamma_{s}^2})=\frac{n_{\textrm{s1}} \,\pi_{\textrm{s1}} }{\gamma_s }-\frac{n_{\textrm{s2}} \,{\left(d-\pi_{\textrm{s1}} \right)}}{\gamma_s }+\frac{n_{\textrm{s1}} \,{\pi_{\textrm{s1}} }^2 }{\gamma_s \,\pi_{\textrm{s1}} -2\,\pi_{\textrm{s1}} +1}-\frac{2\,n_{\textrm{s1}} \,\pi_{\textrm{s1}} }{\gamma_s -1}\; \\
&+\frac{n_{\textrm{s2}} \,{{\left(d-\pi_{\textrm{s1}} \right)}}^2 }{2\,d-2\,\pi_{\textrm{s1}} -d\,\gamma_s +\gamma_s \,\pi_{\textrm{s1}} +1}+\frac{2\,n_{\textrm{s2}} \,{\left(d-\pi_{\textrm{s1}} \right)}}{\gamma_s -1}\; \textrm{, s=1, 2, ..., S.}
\end{flalign*}\\
Hence, the $(1-\alpha)\%$ CI is given by 
\begin{flalign*}
\Biggl[max\left(-1, C_c\Tilde{\boldsymbol{\beta}}^*-Z_{1-\alpha/2}\sqrt{C_c\mathcal{I}_0(\Tilde{\boldsymbol{\beta}}^*)^{-1}C_c^T}\right),\\
min\left(1,  C_c\Tilde{\boldsymbol{\beta}}^*+Z_{1-\alpha/2}\sqrt{C_c\mathcal{I}_0(\Tilde{\boldsymbol{\beta}}^*)^{-1}C_c^T}\right)
\Biggr],
\end{flalign*}\\
where $C_c=(1, 0, 0, ..., 0)$ of size $S$ and $\Tilde{\boldsymbol{\beta}}^*$ is the constrained MLE of $\boldsymbol{\beta}^*$. Note that $C_c\mathcal{I}_0(\Tilde{\boldsymbol{\beta}}^*)^{-1}C_c^T$ is the element at the first row and first column of $\mathcal{I}_0(\Tilde{\boldsymbol{\beta}}^*)^{-1}$ and $C_c\Tilde{\boldsymbol{\beta}}^*=\Tilde{d}$. For the purpose of parallel computation, $C_c\mathcal{I}_0(\Tilde{\boldsymbol{\beta}}^*)^{-1}C_c^T$ can be written as 
\begin{flalign*}
C_c\mathcal{I}_0(\Tilde{\boldsymbol{\beta}}^*)^{-1}C_c^T=\frac{A}{B}
\end{flalign*}
where
\begin{flalign*}
A&=\prod_{s=1}^{S} \Biggl(I^*_{s,11}I^*_{s,22}-I^*_{s,12}I^*_{s,21} \Biggr), \\
B&=I^*_{d}\Biggl\{\prod_{s=1}^{S} \Biggl(I^*_{s,11}I^*_{s,22}-I^*_{s,12}I^*_{s,21} \Biggr)\Biggr\}+\sum_{s=1}^{S} \Biggl\{ \Biggl[
-I^*_{s1}(I^*_{s1}I^*_{s,22}-I^*_{s2}I^*_{s,12})\; \\
&+I^*_{s2}(I^*_{s1}I^*_{s,21}-I^*_{s2}I^*_{s,11}) \Biggr] \times \Biggl[ \prod_{j=1, j \neq s}^{S} \Biggl(I^*_{j,11}I^*_{j,22}-I^*_{j,12}I^*_{j,21} \Biggr) \Biggr]\Biggr\}.
\end{flalign*}
The quantity $B$ is the determinant of $\mathcal{I}_0(\Tilde{\boldsymbol{\beta}}^*)^{-1}$ and is calculated by the Laplace expansion on the first row.

\subsubsection{Profile Likelihood CI}
Under the assumption of a common risk difference, we define the null and alternative hypotheses as $H_{00}: d_1=d_2=...=d_S=d_0$ and $H_0: d_1=d_2=...=d_S$, where $d_0$ is a known quantity. Then the likelihood ratio test statistic $2(l(\Tilde{\pi}_{11}, \Tilde{\pi}_{11}-\Tilde{d}, \Tilde{\gamma}_1, ..., \Tilde{\pi}_{S1}, \Tilde{\pi}_{S1}-\Tilde{d}, \Tilde{\gamma}_S)-l({\pi}_{11}^*, {\pi}_{11}^*-d_0, {\gamma}_1^*, ..., {\pi}_{S1}^*, {\pi}_{S1}^*-d_0, {\gamma}_S^*))$ follows asymptotically a $\mathcal{X}^2$ distribution with one degree of freedom, where $\pi_{s1}^*$ and $\gamma_s^*$ are the conditional MLEs of $\pi_{s1}$ and $\gamma_s$, respectively, given $d=d_0$ ($s=1, 2, ..., S$). The upper bound of the profile likelihood confidence interval can be obtained by the following algorithm modified from the bisection method by Yang et al. \citeyearpar{yang2021simultaneous}:\\
(1) Set the initial value $d^{(0)}=\Tilde{d}$, where $\Tilde{d}$ is the constrained MLE of $d$. Let flag = 1 and stepsize = 0.1;\\
(2) Update $d^{(t+1)}=d^{(t)}$ + flag $\times$ stepsize and compute the constrained MLEs $\pi_{s1}^{(t+1)}$ and $\gamma_{s}^{(t+1)}$ given $d^{(t+1)}$ for $s=1, 2, .., S$. Compare the likelihood-ratio test statistic $2(l(\Tilde{\pi}_{11}, \Tilde{\pi}_{11}-\Tilde{d}, \Tilde{\gamma}_1, ..., \Tilde{\pi}_{S1}, \Tilde{\pi}_{S1}-\Tilde{d}, \Tilde{\gamma}_S)-l({\pi}_{11}^{(t+1)}, {\pi}_{11}^{(t+1)}-d^{(t+1)}, {\gamma}_1^{(t+1)}, ..., {\pi}_{S1}^{(t+1)}, {\pi}_{S1}^{(t+1)}-d^{(t+1)}, {\gamma}_S^{(t+1)}))$ with $\mathcal{X}^2_{1, 1-\alpha}$, the $(1-\alpha)$ quantile of a $\mathcal{X}^2$ distribution with one degree of freedom; \\
(3) If $2(l(\Tilde{\pi}_{11}, \Tilde{\pi}_{11}-\Tilde{d}, \Tilde{\gamma}_1, ..., \Tilde{\pi}_{S1}, \Tilde{\pi}_{S1}-\Tilde{d}, \Tilde{\gamma}_S)-l({\pi}_{11}^{(t+1)}, {\pi}_{11}^{(t+1)}-d^{(t+1)}, {\gamma}_1^{(t+1)}, ..., {\pi}_{S1}^{(t+1)}, {\pi}_{S1}^{(t+1)}-d^{(t+1)}, {\gamma}_S^{(t+1)})) \geq \mathcal{X}^2_{1, 1-\alpha}$, update the search direction by letting flag = -1 and change stepsize = stepsize $\times 1/\pi$, then return to step (2). Otherwise, let flag = 1 and return to step (2);\\
(4) Repeat steps (2) and (3) until convergence, i.e., the stepsize is sufficiently small (e.g., $10^{-4}$).

The lower bound of the profile likelihood CI can be found in a similar way. Simplify initialize flag = -1 in step (1) and then update it to 1 if the statistic $\geq \mathcal{X}^2_{1, 1-\alpha}$ or -1 if the statistic $<\mathcal{X}^2_{1, 1-\alpha}$

\subsubsection{Score CI}
With the existence of a common risk difference $d$, denote $U^*(\boldsymbol{\beta}^*)=(\frac{\partial{l}}{\partial{d}},\frac{\partial{l}}{\partial{\pi_{11}}},\frac{\partial{l}}{\partial{\gamma_{1}}},...,\frac{\partial{l}}{\partial{\pi_{S1}}},\frac{\partial{l}}{\partial{\gamma_{S}}})^T$, which is derived under $H_0: d_1=d_2=...=d_S$. The upper limit of the score CI can be found via the following procedures:\\
(1) Set the initial value of $d$ as $\Tilde{d}$, where $\Tilde{d}$ is the constrained MLE of $d$. Let flag = 1 and stepsize = 0.1;\\
(2) Update  $d^{(t+1)}=d^{(t)}$ + flag $\times$ stepsize and compute the conditional MLEs $\pi_{s1}^{(t+1)}$ and $\gamma_{s}^{(t+1)}$ given $d^{(t+1)}$ for $s=1, 2, .., S$. Compute the score test statistic $(T^*_{SC})^{(t+1)}=U^*(\boldsymbol{\beta}^*)^T\mathcal{I}_0^{-1}(\boldsymbol{\beta}^*)U^*(\boldsymbol{\beta}^*)|\boldsymbol{\beta}^*=(\boldsymbol{\beta}^*)^{(t+1)}$, where $(\beta^*)^{(t+1)}=(d^{(t+1)}, \pi_{11}^{(t+1)}, \gamma_1^{(t+1)}, ..., \pi_{S1}^{(t+1)}, \gamma_S^{(t+1)})$ and $\mathcal{I}_0(\boldsymbol{\beta}^*)$ is the Fisher information matrix of $\boldsymbol{\beta}^*$ defined in Section \ref{sec:con_wald_CI};\\
(3) If $(T^*_{SC})^{(t+1)} \geq \mathcal{X}^2_{1, 1-\alpha}$, update the search direction by letting flag = -1 and change stepsize = stepsize $\times 1/\pi$, then return to step (2). Otherwise, let flag = 1 and return to step (2);\\
(4) Repeat steps (2) and (3) until convergence, i.e., the stepsize is sufficiently small (e.g., $10^{-4}$).

The lower limit of the score CI can be found by initializing flag = -1 in step (1) and then update it to 1 if $(T^*_{SC})^{(t+1)} \geq \mathcal{X}^2_{1, 1-\alpha}$ or -1 if $(T^*_{SC})^{(t+1)} <\mathcal{X}^2_{1, 1-\alpha}$. It is easy to show that the the conditional MLEs $\pi_{s1}^{*}$ and $\gamma_{s}^{*}$ given $d^{*}$ satisfy
\[
\begin{cases}
\frac{\partial{l}}{\partial{\pi_{s1}}}|_{\pi_{s1}=\pi_{11}^{*}, \gamma_s=\gamma_s^{*}, d=d^{*} }=0 \textrm{, }\\
\frac{\partial{l}}{\partial{\gamma_{s}}}|_{\pi_{s1}=\pi_{11}^{*}, \gamma_s=\gamma_s^{*}, d=d^{*} }=0  \textrm{, s=1, 2, ..., S. }
\end{cases}
\]\\
Hence, the score test statistic can be further simplified as 
\[
T^*_{SC}=\frac{\partial{l}}{\partial{d}}(I_0(\boldsymbol{\beta}^*))^{-1}_{(1, 1)}\frac{\partial{l}}{\partial{d}},
\]\\
where $(I_0(\boldsymbol{\beta}^*))^{-1}_{(1, 1)}$ is the element at the upper-left corner of the inverse matrix of $I_0(\boldsymbol{\beta}^*)$.

\section{Monte Carlo Simulation Studies}
\begin{figure}[H]
        \caption{Notched box plots of empirical type I errors for $n_{s1}=25$ and $m_{s1}=15$}
        \label{fig:tien25}
     \centering
\includegraphics[width=\textwidth]{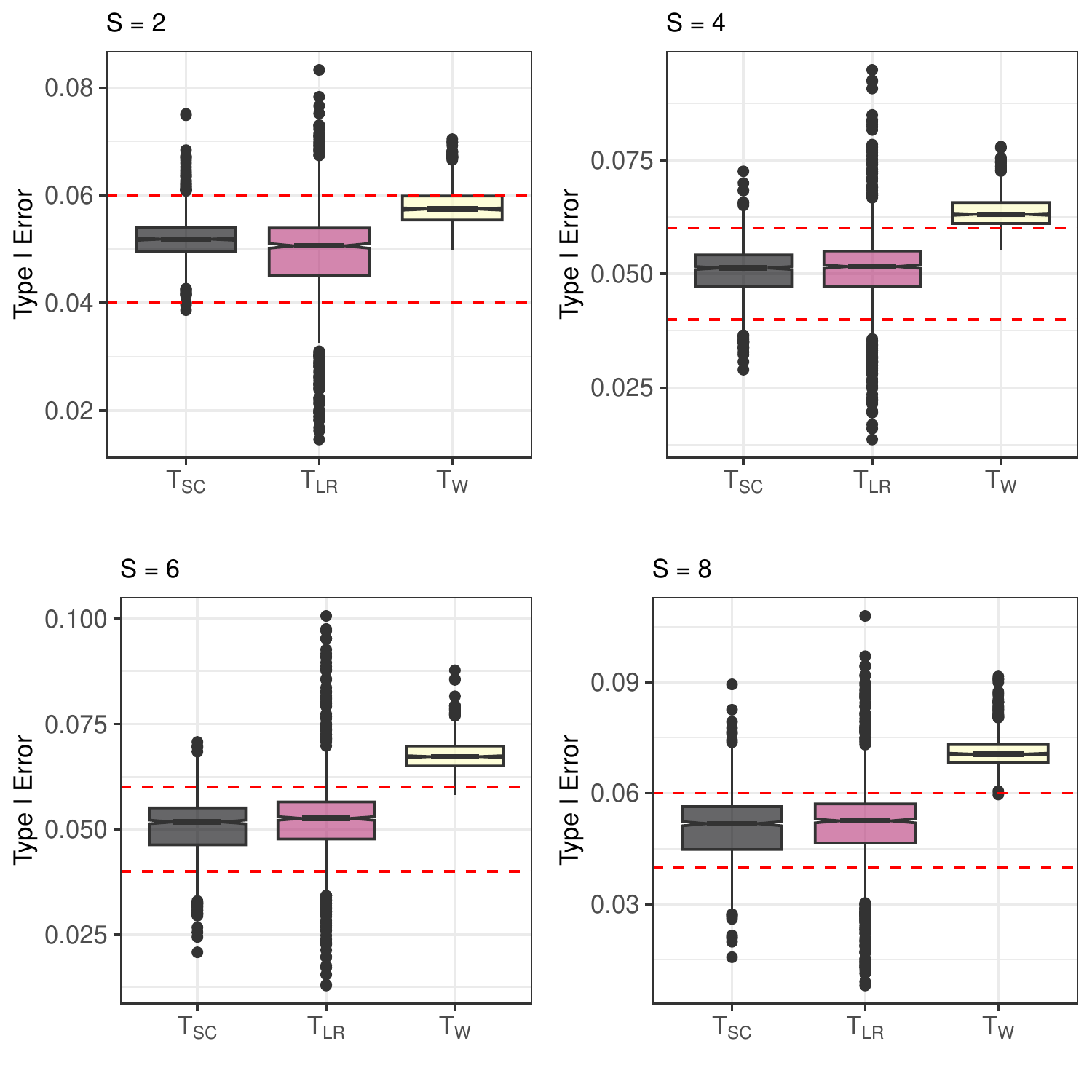}
\end{figure}

\begin{figure}[H]
        \caption{Notched box plots of empirical type I errors for $n_{s1}=100$ and $m_{s1}=90$}
        \label{fig:tien100}
     \centering
\includegraphics[width=\textwidth]{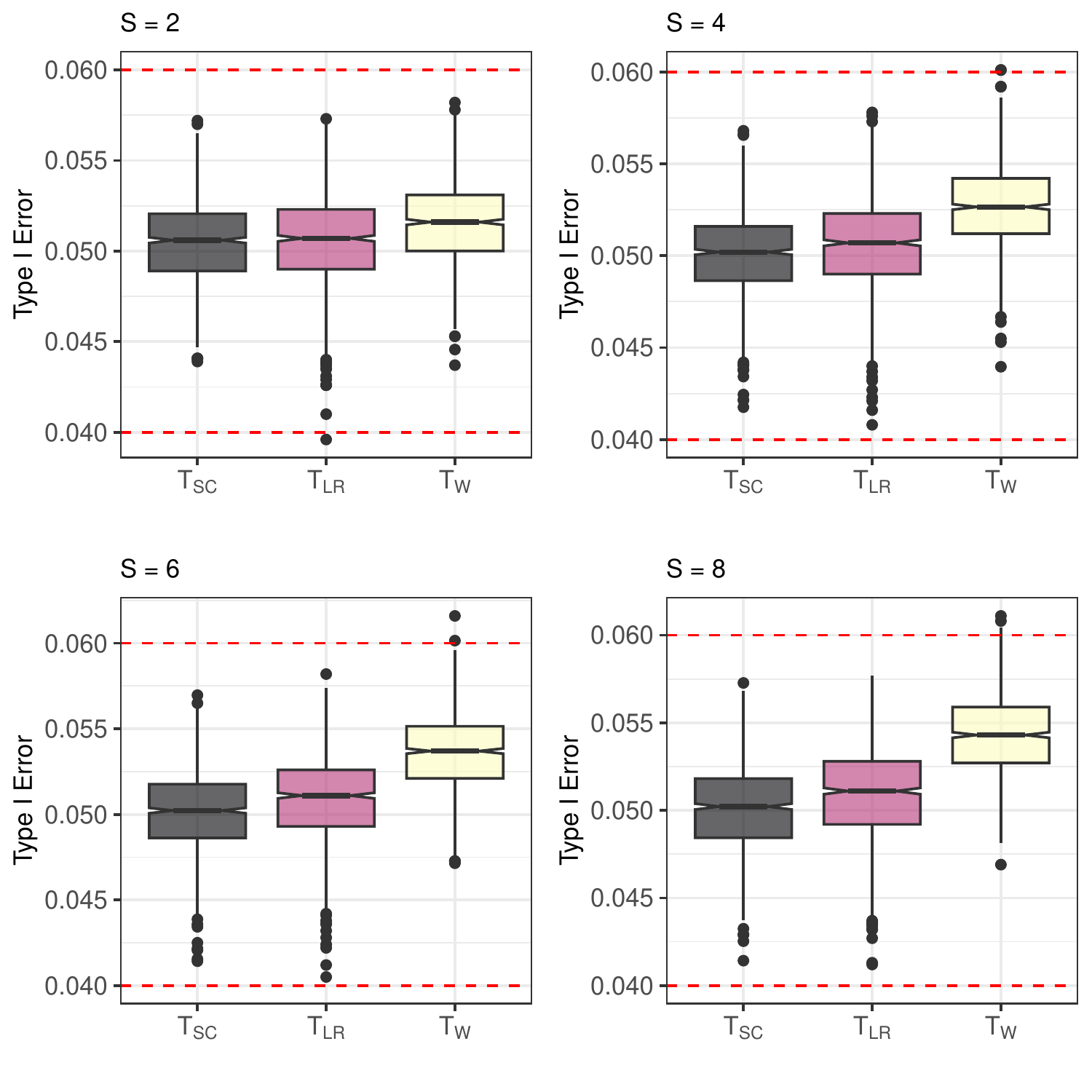}
\end{figure}

In this section, we investigate the performances of the proposed test methods and confidence interval estimations discussed in previous sections under different parameter configurations. First, we investigate the behaviors of the type I error rates of these test procedures with various combinations of sample size, $d_s$, $\pi_{s1}$, and $\gamma_s$. Notations of parameter configurations are presented in \hyperref[tab:tieconf2]{Table \ref*{tab:tieconf2}} to \hyperref[tab:tieconf8]{Table \ref*{tab:tieconf8}}. The restriction of $p_{bsg0}$, $p_{bsg1}$, $p_{bsg2}$, $p_{usg0}$, $p_{usg1} \in$ (0.1, 1) is added. And the lower bound 0.1 is considered for the purpose of limiting cases in which some $n_{sgr}=0$ $(r=0, 1, 2)$ or $m_{sgr}=0$ $(r=0, 1)$, since these rarely happen in real-world analyses and can hinder the convergence of the proposed algorithms. Some remedies for these situations include recruiting more participants to avoid observing zero cells and adding a small number to all table cells, e.g., $10^{-4}$. We assume $n_{s1}=n_{s2}=n_{s'1}=n_{s'2}$ and $m_{s1}=m_{s2}=m_{s'1}=m_{s'2}$ for $s, s'\in \{1, 2, ..., S\}$. In each combination, a random sample under the null hypothesis $H_0: d_1=d_2=...=d_S$ is generated and the three test statistics can be calculated. Given a nominal level $\alpha$ = 0.05, the decision rule is to reject $H_0$ if the test statistics is greater than $\mathcal{X}_{S-1,1-\alpha}^2$. Here $\mathcal{X}_{S-1,1-\alpha}^2$ is the $100(1-\alpha)$ percentile of the chi-square distribution with $S-1$ degrees of freedom. We repeat the above procedure until there are 10,000 replicates. Then the empirical type I error rates can be computed as the number of rejections/10,000. The results are presented in \hyperref[tab:tie2]{Table \ref*{tab:tie2}} to \hyperref[tab:tie8]{Table \ref*{tab:tie8}}. Although \hyperref[tab:tieconf2]{Table \ref*{tab:tieconf2}} to \hyperref[tab:tieconf8]{Table \ref*{tab:tieconf8}} contain a wide range of parameter settings, it is possible that some scenarios are not considered. Therefore, we generate 1,000 random sets of ($d_s, \pi_{s1}, \gamma_s$) for each combination of stratum number $S$ and sample size ($n_{s1}, m_{s1}$) while constraining $p_{bsg0}$, $p_{bsg1}$, $p_{bsg2}$, $p_{usg0}$, $p_{usg1} \in$ (0.1, 1). For each random set, 10,000 simulations are conducted to investigate the empirical type I error rates. \hyperref[fig:tien25]{Figure \ref*{fig:tien25}} to \hyperref[fig:tien100]{Figure \ref*{fig:tien100}} display notched box plots of empirical type I errors of all methods based on the 1,000 sets of parameters. Following Tang et al. \citeyearpar{tang2008testing}, a test is said to be liberal if the ratio of the empirical type I error rate to the nominal type I error rate is greater than 1.2 (i.e., $>$6\% for $\alpha=5\%$), conservative if the ratio is less than 0.8 (i.e., $<4\%$), and robust if the ratio is between 0.8 and 1.2.

When the sample size and the number of strata are both small, the three tests yield comparable results while the Wald-type test produces liberal type I errors under several parameter settings, as indicated in \hyperref[tab:tie2]{Table \ref*{tab:tie2}}. From \hyperref[tab:tie4]{Table \ref*{tab:tie4}} to \hyperref[tab:tie8]{Table \ref*{tab:tie8}}, the Wald-type test exhibits inflated type I error rates and becomes extremely liberal when $n_{s1}$ and $m_{s1}$ are small and the number of strata is large, i.e., S = 8. Under the 1,000 random parameter settings, the Wald-type test also demonstrates a tendency for liberal type I errors when the sample size is small and the number of strata is large based on \hyperref[fig:tien25]{Figure \ref*{fig:tien25}}. Additionally, the likelihood-ratio test produces a higher proportion of cases that lie outside the interval $(0.04,0.06)$ when compared with the score test, particularly in situations where the sample size is small. In general, the score test $T_{SC}$ is the most robust among the three methods since its satisfactory type I error rates for any configurations. All tests generate closer results when the sample size becomes larger regardless of the number of strata.

\begin{singlespace}
\begin{table}[H]
\captionsetup[subtable]{justification=centering}
\caption{Parameter configurations for type I error rates (2 strata)}
\label{tab:tieconf2}
  \begin{subtable}[t]{0.48\textwidth}
        \centering
\begin{tabular}[t]{l|c}
\toprule

  Notation   & $d_1=d_2$     \\ \hline
$E^{T, 2}_{d,1}$ & 0            \\ 
$E^{T, 2}_{d,2}$ & 0.05 \\ 
$E^{T, 2}_{d,3}$ & 0.1     \\ 
$E^{T, 2}_{d,4}$ & 0.15 \\ 

\bottomrule
\end{tabular}
\caption{}
\end{subtable}
\hfill
  \begin{subtable}[t]{0.48\textwidth}
        \centering
\begin{tabular}[t]{l|c}
\toprule
 Notation  & $(\gamma_1, \gamma_2)$     \\ \hline
$E^{T, 2}_{\gamma,1}$ & (0.4, 0.4)            \\ 
$E^{T, 2}_{\gamma,2}$ & (0.5, 0.5)  \\ 
$E^{T, 2}_{\gamma,3}$ & (0.6, 0.6)      \\ 
$E^{T, 2}_{\gamma,4}$ & (0.7, 0.7)  \\ 
$E^{T, 2}_{\gamma,5}$ & (0.4, 0.5)  \\ 
$E^{T, 2}_{\gamma,6}$ & (0.5, 0.6)  \\ 
$E^{T, 2}_{\gamma,7}$ & (0.6, 0.7)  \\

\bottomrule
\end{tabular}
\caption{}
\end{subtable}

\bigskip 

  \begin{subtable}[t]{0.48\textwidth}
        \centering
\begin{tabular}[t]{l|c}
\toprule

 Notation  & $(\pi_{11}, \pi_{21})$     \\ \hline
$E^{T, 2}_{\pi,1}$ & (0.5, 0.5,)            \\ 
$E^{T, 2}_{\pi,2}$ & (0.45, 0.45)  \\ 
$E^{T, 2}_{\pi,3}$ & (0.45, 0.5)      \\ 
$E^{T, 2}_{\pi,4}$ & (0.5, 0.45)  \\ 

\bottomrule
\end{tabular}
\caption{}
\end{subtable}
\hspace{\fill}
\begin{subtable}[t]{0.48\textwidth}
\begin{tabular}[t]{l|c}
\toprule

\bottomrule
\end{tabular}

\end{subtable}

\end{table}
\end{singlespace}

\begin{singlespace}
\begin{table}[H]
\captionsetup[subtable]{justification=centering}
\caption{Parameter configurations for type I error rates (4 strata)}
\label{tab:tieconf4}
  \begin{subtable}[t]{0.48\textwidth}
        \centering
\begin{tabular}[t]{l|c}
\toprule

  Notation   & $d_1=d_2=d_3=d_4$     \\ \hline
$E^{T, 4}_{d,1}$ & 0            \\ 
$E^{T, 4}_{d,2}$ & 0.05 \\ 
$E^{T, 4}_{d,3}$ & 0.1     \\ 
$E^{T, 4}_{d,4}$ & 0.15 \\ 

\bottomrule
\end{tabular}
\caption{}
\end{subtable}
\hfill
  \begin{subtable}[t]{0.48\textwidth}
        \centering
\begin{tabular}[t]{l|c}
\toprule
 Notation  & $(\gamma_1, \gamma_2, \gamma_3, \gamma_4)$     \\ \hline
$E^{T, 4}_{\gamma,1}$ & (0.4, 0.4, 0.4, 0.4)            \\ 
$E^{T, 4}_{\gamma,2}$ & (0.5, 0.5, 0.5, 0.5)  \\ 
$E^{T, 4}_{\gamma,3}$ & (0.6, 0.6, 0.6, 0.6)      \\ 
$E^{T, 4}_{\gamma,4}$ & (0.7, 0.7, 0.7, 0.7)  \\ 
$E^{T, 4}_{\gamma,5}$ & (0.4, 0.4, 0.5, 0.5)  \\ 
$E^{T, 4}_{\gamma,6}$ & (0.4, 0.4, 0.6, 0.6)  \\ 
$E^{T, 4}_{\gamma,7}$ & (0.4, 0.4, 0.7, 0.7)  \\ 
$E^{T, 4}_{\gamma,8}$ & (0.5, 0.5, 0.6, 0.6)  \\ 
$E^{T, 4}_{\gamma,9}$ & (0.5, 0.5, 0.7, 0.7)  \\ 
$E^{T, 4}_{\gamma,10}$ & (0.6, 0.6, 0.7, 0.7)  \\ 

\bottomrule
\end{tabular}
\caption{}
\end{subtable}

\bigskip 

  \begin{subtable}[t]{0.48\textwidth}
        \centering
\begin{tabular}[t]{l|c}
\toprule

 Notation  & $(\pi_{11}, \pi_{21}, \pi_{31}, \pi_{41})$     \\ \hline
$E^{T, 4}_{\pi,1}$ & (0.5, 0.5, 0.5, 0.5)            \\ 
$E^{T, 4}_{\pi,2}$ & (0.45, 0.45, 0.45, 0.45)  \\ 
$E^{T, 4}_{\pi,3}$ & (0.45, 0.45, 0.5, 0.5)      \\ 
$E^{T, 4}_{\pi,4}$ & (0.45, 0.5, 0.45, 0.5)  \\ 

\bottomrule
\end{tabular}
\caption{}
\end{subtable}
\hspace{\fill}
\begin{subtable}[t]{0.48\textwidth}
\begin{tabular}[t]{l|c}
\toprule

\bottomrule
\end{tabular}

\end{subtable}

\end{table}
\end{singlespace}

\begin{singlespace}
\begin{table}[H]
\captionsetup[subtable]{justification=centering}
\caption{Parameter configurations for type I error rates (6 strata)}
\label{tab:tieconf6}
\begin{subtable}[t]{0.48\textwidth}

\begin{tabular}[t]{l|c}
\toprule

  Notation   & $d_1=d_2=d_3=d_4=d_5=d_6$     \\ \hline
$E^{T, 6}_{d,1}$ & 0            \\ 
$E^{T, 6}_{d,2}$ & 0.05 \\ 
$E^{T, 6}_{d,3}$ & 0.1     \\ 
$E^{T, 6}_{d,4}$ & 0.15 \\ 

\bottomrule
\end{tabular}
\caption{}
\end{subtable}
\hspace{\fill}
\begin{subtable}[t]{0.48\textwidth}

\begin{tabular}[t]{l|c}
\toprule
 Notation  & $(\gamma_1, \gamma_2, \gamma_3, \gamma_4, \gamma_5, \gamma_6)$     \\ \hline
$E^{T, 6}_{\gamma,1}$ & (0.4, 0.4, 0.4, 0.4, 0.4, 0.4)            \\ 
$E^{T, 6}_{\gamma,2}$ & (0.5, 0.5, 0.5, 0.5, 0.5, 0.5)  \\ 
$E^{T, 6}_{\gamma,3}$ & (0.6, 0.6, 0.6, 0.6, 0.6, 0.6)      \\ 
$E^{T, 6}_{\gamma,4}$ & (0.7, 0.7, 0.7, 0.7, 0.7, 0.7)  \\ 
$E^{T, 6}_{\gamma,5}$ & (0.4, 0.4, 0.5, 0.5, 0.6, 0.6)  \\ 
$E^{T, 6}_{\gamma,6}$ & (0.5, 0.5, 0.6, 0.6, 0.7, 0.7)  \\ 
$E^{T, 6}_{\gamma,7}$ & (0.4, 0.4, 0.4, 0.5, 0.5, 0.5)  \\ 
$E^{T, 6}_{\gamma,8}$ & (0.5, 0.5, 0.5, 0.6, 0.6, 0.6)  \\ 
$E^{T, 6}_{\gamma,9}$ & (0.6, 0.6, 0.6, 0.7, 0.7, 0.7)  \\ 
$E^{T, 6}_{\gamma,10}$ & (0.4, 0.4, 0.4, 0.6, 0.6, 0.6)  \\ 
$E^{T, 6}_{\gamma,11}$ & (0.5, 0.5, 0.5, 0.7, 0.7, 0.7)  \\ 
$E^{T, 6}_{\gamma,12}$ & (0.4, 0.4, 0.4, 0.7, 0.7, 0.7)  \\ 

\bottomrule
\end{tabular}
\caption{}
\end{subtable}

\bigskip 

\begin{subtable}[t]{0.48\textwidth}
\begin{tabular}[t]{l|c}
\toprule

 Notation  & $(\pi_{11}, \pi_{21}, \pi_{31}, \pi_{41}, \pi_{51}, \pi_{61})$     \\ \hline
$E^{T, 6}_{\pi,1}$ & (0.5, 0.5, 0.5, 0.5, 0.5, 0.5)            \\ 
$E^{T, 6}_{\pi,2}$ & (0.45, 0.45, 0.45, 0.45, 0.45, 0.45)  \\ 
$E^{T, 6}_{\pi,3}$ & (0.45, 0.45, 0.5, 0.5, 0.45, 0.45)      \\ 
$E^{T, 6}_{\pi,4}$ & (0.45, 0.5, 0.45, 0.5, 0.45, 0.5)  \\ 

\bottomrule
\end{tabular}
\caption{}
\end{subtable}
\hspace{\fill}
\begin{subtable}[t]{0.48\textwidth}
\begin{tabular}[t]{l|c}
\toprule

\bottomrule
\end{tabular}

\end{subtable}

\end{table}
\end{singlespace}

\begin{singlespace}
\begin{table}[H]
\captionsetup[subtable]{justification=centering}
\caption{Parameter configurations for type I error rates (8 strata)}
\label{tab:tieconf8}
\begin{subtable}{1\linewidth}
\centering
\begin{tabular}[t]{l|c}
\toprule

  Notation   & $d_1=d_2=d_3=d_4=d_5=d_6=d_7=d_8$     \\ \hline
$E^{T, 8}_{d,1}$ & 0            \\ 
$E^{T, 8}_{d,2}$ & 0.05 \\ 
$E^{T, 8}_{d,3}$ & 0.1     \\ 
$E^{T, 8}_{d,4}$ & 0.15 \\ 

\bottomrule
\end{tabular}
\captionsetup{justification=centering}
\caption{}
\end{subtable}

\bigskip

\begin{subtable}{1\linewidth}
\centering
\begin{tabular}[t]{l|c}
\toprule
 Notation  & $(\gamma_1, \gamma_2, \gamma_3, \gamma_4, \gamma_5, \gamma_6, \gamma_7, \gamma_8)$     \\ \hline
$E^{T, 8}_{\gamma,1}$ & (0.4, 0.4, 0.4, 0.4, 0.4, 0.4, 0.4, 0.4)   \\ 
$E^{T, 8}_{\gamma,2}$ & (0.5, 0.5, 0.5, 0.5, 0.5, 0.5, 0.5, 0.5)  \\ 
$E^{T, 8}_{\gamma,3}$ & (0.6, 0.6, 0.6, 0.6, 0.6, 0.6, 0.6, 0.6)      \\ 
$E^{T, 8}_{\gamma,4}$ & (0.7, 0.7, 0.7, 0.7, 0.7, 0.7, 0.7, 0.7)  \\ 
$E^{T, 8}_{\gamma,5}$ & (0.4, 0.4, 0.5, 0.5, 0.6, 0.6, 0.7, 0.7)  \\ 
$E^{T, 8}_{\gamma,6}$ & (0.4, 0.4, 0.4, 0.4, 0.5, 0.5, 0.5, 0.5)  \\ 
$E^{T, 8}_{\gamma,7}$ & (0.4, 0.4, 0.4, 0.4, 0.6, 0.6, 0.6, 0.6)  \\ 
$E^{T, 8}_{\gamma,8}$ & (0.4, 0.4, 0.4, 0.4, 0.7, 0.7, 0.7, 0.7)  \\ 
$E^{T, 8}_{\gamma,9}$ & (0.5, 0.5, 0.5, 0.5, 0.6, 0.6, 0.6, 0.6)  \\ 
$E^{T, 8}_{\gamma,10}$ & (0.5, 0.5, 0.5, 0.5, 0.7, 0.7, 0.7, 0.7)  \\ 
$E^{T, 8}_{\gamma,11}$ & (0.6, 0.6, 0.6, 0.6, 0.7, 0.7, 0.7, 0.7)  \\

\bottomrule
\end{tabular}
\captionsetup{justification=centering}
\caption{}
\end{subtable}

\bigskip 

\begin{subtable}{1\linewidth}
\centering
\begin{tabular}[t]{l|c}
\toprule

 Notation  & $(\pi_{11}, \pi_{21}, \pi_{31}, \pi_{41}, \pi_{51}, \pi_{61}, \pi_{71}, \pi_{81})$     \\ \hline
$E^{T, 8}_{\pi,1}$ & (0.5, 0.5, 0.5, 0.5, 0.5, 0.5, 0.5, 0.5)            \\ 
$E^{T, 8}_{\pi,2}$ & (0.45, 0.45, 0.45, 0.45, 0.45, 0.45, 0.45, 0.45)  \\ 
$E^{T, 8}_{\pi,3}$ & (0.45, 0.45, 0.5, 0.5, 0.45, 0.45, 0.5, 0.5)      \\ 
$E^{T, 8}_{\pi,4}$ & (0.45, 0.5, 0.45, 0.5, 0.45, 0.5, 0.45, 0.5)  \\ 
$E^{T, 8}_{\pi,5}$ & (0.45, 0.45, 0.45, 0.5, 0.5, 0.45, 0.45, 0.45)  \\

\bottomrule
\end{tabular}
\captionsetup{justification=centering}
\caption{}
\end{subtable}

\end{table}
\end{singlespace}

\begin{singlespace}
\begin{center}
\addtolength{\tabcolsep}{-0.1em}
\begin{longtable}[H]{c c c | c c c | c c c | c c c | c c c}
\caption{Type I error rates (percent) of various procedures under $H_0$: $d_1=d_2=...=d_S$ at $\alpha$=0.05 based on 10,000 replicates (2 strata) } \label{tab:tie2} \\

\hline
 &  &  & \multicolumn{3}{c|}{$n_{s1}=25$} & \multicolumn{3}{c|}{$n_{s1}=60$} & \multicolumn{3}{c|}{$n_{s1}=80$} & \multicolumn{3}{c} {$n_{s1}=100$} \\ 
  &  &  & \multicolumn{3}{c|}{$m_{s1}=15$} & \multicolumn{3}{c|}{$m_{s1}=30$} & \multicolumn{3}{c|}{$m_{s1}=40$} & \multicolumn{3}{c} {$m_{s1}=90$} \\ \cline{4-6} \cline{7-9} \cline{10-12} \cline{13-15}
$d$ & $\gamma$ & $\pi$ & $T_{SC}$ & $T_{LR}$ & $T_W$ & $T_{SC}$ & $T_{LR}$ & $T_W$ & $T_{SC}$ & $T_{LR}$ & $T_W$  & $T_{SC}$ & $T_{LR}$ & $T_W$ \\ \hline

\endfirsthead

\multicolumn{15}{l}%
{{ \tablename\ \thetable{}: (Continued)}} \\
\hline
 &  &  & \multicolumn{3}{c|}{$n_{s1}=25$} & \multicolumn{3}{c|}{$n_{s1}=60$} & \multicolumn{3}{c|}{$n_{s1}=80$} & \multicolumn{3}{c} {$n_{s1}=100$} \\ 
  &  &  & \multicolumn{3}{c|}{$m_{s1}=15$} & \multicolumn{3}{c|}{$m_{s1}=30$} & \multicolumn{3}{c|}{$m_{s1}=40$} & \multicolumn{3}{c} {$m_{s1}=90$} \\ \cline{4-6} \cline{7-9} \cline{10-12} \cline{13-15}
$d$ & $\gamma$ & $\pi$ & $T_{SC}$ & $T_{LR}$ & $T_W$ & $T_{SC}$ & $T_{LR}$ & $T_W$ & $T_{SC}$ & $T_{LR}$ & $T_W$  & $T_{SC}$ & $T_{LR}$ & $T_W$ \\ \hline
\endhead

\hline \multicolumn{15}{c}{{Continued on next page}} \\ \hline
\endfoot

\hline \hline
\endlastfoot

$E^{T, 2}_{d,1}$ & $E^{T, 2}_{\gamma,1}$ & $E^{T, 2}_{\pi,1}$ & 5.10 & 5.02 & 5.65 & 5.14 & 5.16 & 5.27 & 4.92 & 5.07 & 5.16 & 4.99 & 5.11 & 5.17 \\ 
        ~ & ~ & $E^{T, 2}_{\pi,2}$ & 4.91 & 5.03 & 5.40 & 5.33 & 5.34 & 5.42 & 5.00 & 5.09 & 5.23 & 5.19 & 5.15 & 5.24 \\ 
        ~ & ~ & $E^{T, 2}_{\pi,3}$ & 4.93 & 5.05 & 5.81 & 4.98 & 5.08 & 5.30 & 5.04 & 5.11 & 5.21 & 4.98 & 5.02 & 5.06 \\ 
        ~ & ~ & $E^{T, 2}_{\pi,4}$ & 5.07 & 5.17 & 5.63 & 5.56 & 5.61 & 5.74 & 4.93 & 5.02 & 5.11 & 5.71 & 5.81 & 5.84 \\ 
        ~ & $E^{T, 2}_{\gamma,2}$ & $E^{T, 2}_{\pi,1}$ & 4.79 & 5.03 & 5.57 & 5.34 & 5.35 & 5.38 & 4.87 & 4.93 & 5.00 & 4.95 & 4.98 & 5.04 \\ 
        ~ & ~ & $E^{T, 2}_{\pi,2}$ & 4.90 & 5.12 & 5.39 & 5.09 & 5.09 & 5.21 & 5.12 & 5.14 & 5.24 & 5.13 & 5.18 & 5.22 \\ 
        ~ & ~ & $E^{T, 2}_{\pi,3}$ & 5.14 & 5.38 & 5.73 & 5.13 & 5.20 & 5.35 & 4.62 & 4.67 & 4.75 & 5.09 & 5.16 & 5.22 \\ 
        ~ & ~ & $E^{T, 2}_{\pi,4}$ & 5.39 & 5.51 & 5.89 & 5.08 & 5.20 & 5.33 & 5.05 & 5.03 & 5.06 & 5.34 & 5.37 & 5.43 \\
\end{longtable}
\end{center}
\end{singlespace}

\begin{singlespace}
\begin{center}
\addtolength{\tabcolsep}{-0.1em}
\begin{longtable}[H]{c c c | c c c | c c c | c c c | c c c}
\caption{Type I error rates (percent) of various procedures under $H_0$: $d_1=d_2=...=d_S$ at $\alpha$=0.05 based on 10,000 replicates (4 strata) } \label{tab:tie4} \\

\hline
 &  &  & \multicolumn{3}{c|}{$n_{s1}=25$} & \multicolumn{3}{c|}{$n_{s1}=60$} & \multicolumn{3}{c|}{$n_{s1}=80$} & \multicolumn{3}{c} {$n_{s1}=100$} \\ 
  &  &  & \multicolumn{3}{c|}{$m_{s1}=15$} & \multicolumn{3}{c|}{$m_{s1}=30$} & \multicolumn{3}{c|}{$m_{s1}=40$} & \multicolumn{3}{c} {$m_{s1}=90$} \\ \cline{4-6} \cline{7-9} \cline{10-12} \cline{13-15}
$d$ & $\gamma$ & $\pi$ & $T_{SC}$ & $T_{LR}$ & $T_W$ & $T_{SC}$ & $T_{LR}$ & $T_W$ & $T_{SC}$ & $T_{LR}$ & $T_W$  & $T_{SC}$ & $T_{LR}$ & $T_W$ \\ \hline

\endfirsthead

\multicolumn{15}{l}%
{{ \tablename\ \thetable{}: (Continued)}} \\
\hline
 &  &  & \multicolumn{3}{c|}{$n_{s1}=25$} & \multicolumn{3}{c|}{$n_{s1}=60$} & \multicolumn{3}{c|}{$n_{s1}=80$} & \multicolumn{3}{c} {$n_{s1}=100$} \\ 
  &  &  & \multicolumn{3}{c|}{$m_{s1}=15$} & \multicolumn{3}{c|}{$m_{s1}=30$} & \multicolumn{3}{c|}{$m_{s1}=40$} & \multicolumn{3}{c} {$m_{s1}=90$} \\ \cline{4-6} \cline{7-9} \cline{10-12} \cline{13-15}
$d$ & $\gamma$ & $\pi$ & $T_{SC}$ & $T_{LR}$ & $T_W$ & $T_{SC}$ & $T_{LR}$ & $T_W$ & $T_{SC}$ & $T_{LR}$ & $T_W$  & $T_{SC}$ & $T_{LR}$ & $T_W$ \\ \hline
\endhead

\hline \multicolumn{15}{c}{{Continued on next page}} \\ \hline
\endfoot

\hline \hline
\endlastfoot

 $E^{T, 4}_{d,1}$ & $E^{T, 4}_{\gamma,1}$ & $E^{T, 4}_{\pi,1}$ & 4.34 & 5.00 & 5.60 & 5.34 & 5.69 & 5.99 & 5.15 & 5.31 & 5.53 & 4.53 & 4.71 & 4.90 \\ 
        ~ & ~ & $E^{T, 4}_{\pi,2}$ & 4.76 & 5.21 & \textbf{6.05} & 5.01 & 5.19 & 5.52 & 5.00 & 5.09 & 5.27 & 5.15 & 5.20 & 5.29 \\ 
        ~ & $E^{T, 4}_{\gamma,2}$ & $E^{T, 4}_{\pi,1}$ & 4.66 & 4.93 & 5.83 & 5.01 & 5.26 & 5.58 & 4.97 & 4.98 & 5.22 & 4.65 & 4.81 & 4.98 \\ 
        ~ & ~ & $E^{T, 4}_{\pi,2}$ & 5.10 & 5.37 & \textbf{6.04} & 5.21 & 5.32 & 5.63 & 5.13 & 5.20 & 5.42 & 4.86 & 4.89 & 5.10 \\ 
        ~ & $E^{T, 4}_{\gamma,3}$ & $E^{T, 4}_{\pi,1}$ & 5.68 & \textbf{6.01} & \textbf{6.77} & 5.00 & 5.22 & 5.55 & 5.08 & 5.18 & 5.39 & 4.94 & 4.98 & 5.23 \\ 
        ~ & ~ & $E^{T, 4}_{\pi,2}$ & 5.38 & 5.55 & \textbf{6.05} & 5.17 & 5.33 & 5.65 & 4.94 & 5.00 & 5.11 & 5.31 & 5.36 & 5.49 \\ 
        ~ & $E^{T, 4}_{\gamma,4}$ & $E^{T, 4}_{\pi,1}$ & 5.00 & 5.32 & \textbf{6.18} & 5.26 & 5.41 & 5.80 & 5.22 & 5.26 & 5.49 & 4.78 & 4.81 & 4.95 \\ 
        ~ & ~ & $E^{T, 4}_{\pi,2}$ & 5.16 & 5.51 & \textbf{6.31} & 4.81 & 4.89 & 5.08 & 5.08 & 5.20 & 5.35 & 5.16 & 5.22 & 5.31 \\
\end{longtable}
\end{center}
\end{singlespace}

\begin{singlespace}
\begin{center}
\addtolength{\tabcolsep}{-0.1em}
\begin{longtable}[H]{c c c | c c c | c c c | c c c | c c c}
\caption{Type I error rates (percent) of various procedures under $H_0$: $d_1=d_2=...=d_S$ at $\alpha$=0.05 based on 10,000 replicates (6 strata) } \label{tab:tie6} \\

\hline
 &  &  & \multicolumn{3}{c|}{$n_{s1}=25$} & \multicolumn{3}{c|}{$n_{s1}=60$} & \multicolumn{3}{c|}{$n_{s1}=80$} & \multicolumn{3}{c} {$n_{s1}=100$} \\ 
  &  &  & \multicolumn{3}{c|}{$m_{s1}=15$} & \multicolumn{3}{c|}{$m_{s1}=30$} & \multicolumn{3}{c|}{$m_{s1}=40$} & \multicolumn{3}{c} {$m_{s1}=90$} \\ \cline{4-6} \cline{7-9} \cline{10-12} \cline{13-15}
$d$ & $\gamma$ & $\pi$ & $T_{SC}$ & $T_{LR}$ & $T_W$ & $T_{SC}$ & $T_{LR}$ & $T_W$ & $T_{SC}$ & $T_{LR}$ & $T_W$  & $T_{SC}$ & $T_{LR}$ & $T_W$ \\ \hline

\endfirsthead

\multicolumn{15}{l}%
{{ \tablename\ \thetable{}: (Continued)}} \\
\hline
 &  &  & \multicolumn{3}{c|}{$n_{s1}=25$} & \multicolumn{3}{c|}{$n_{s1}=60$} & \multicolumn{3}{c|}{$n_{s1}=80$} & \multicolumn{3}{c} {$n_{s1}=100$} \\ 
  &  &  & \multicolumn{3}{c|}{$m_{s1}=15$} & \multicolumn{3}{c|}{$m_{s1}=30$} & \multicolumn{3}{c|}{$m_{s1}=40$} & \multicolumn{3}{c} {$m_{s1}=90$} \\ \cline{4-6} \cline{7-9} \cline{10-12} \cline{13-15}
$d$ & $\gamma$ & $\pi$ & $T_{SC}$ & $T_{LR}$ & $T_W$ & $T_{SC}$ & $T_{LR}$ & $T_W$ & $T_{SC}$ & $T_{LR}$ & $T_W$  & $T_{SC}$ & $T_{LR}$ & $T_W$ \\ \hline
\endhead

\hline \multicolumn{15}{c}{{Continued on next page}} \\ \hline
\endfoot

\hline \hline
\endlastfoot

        $E^{T, 6}_{d,1}$ & $E^{T, 6}_{\gamma,1}$ & $E^{T, 6}_{\pi,1}$ & 4.83 & 5.74 & \textbf{6.82} & 5.37 & 5.69 & \textbf{6.06} & 4.89 & 5.26 & 5.52 & 5.14 & 5.33 & 5.64 \\ 
        ~ & ~ & $E^{T, 6}_{\pi,2}$ & 5.29 & 5.76 & \textbf{7.01} & 4.82 & 5.05 & 5.56 & 4.94 & 5.13 & 5.44 & 4.96 & 5.02 & 5.18 \\ 
        ~ & $E^{T, 6}_{\gamma,2}$ & $E^{T, 6}_{\pi,1}$ & 5.24 & 5.70 & \textbf{6.84} & 5.20 & 5.45 & 5.88 & 5.16 & 5.40 & 5.73 & 5.21 & 5.31 & 5.41 \\ 
        ~ & ~ & $E^{T, 6}_{\pi,2}$ & 5.11 & 5.50 & \textbf{6.67} & 5.07 & 5.22 & 5.67 & 4.96 & 5.11 & 5.43 & 5.31 & 5.30 & 5.53 \\ 
        ~ & $E^{T, 6}_{\gamma,3}$ & $E^{T, 6}_{\pi,1}$ & 4.86 & 5.32 & \textbf{6.27} & 4.98 & 5.27 & 5.67 & 5.24 & 5.31 & 5.66 & 5.02 & 5.11 & 5.37 \\ 
        ~ & ~ & $E^{T, 6}_{\pi,2}$ & 5.17 & 5.48 & \textbf{6.35} & 5.39 & 5.46 & 5.84 & 5.16 & 5.17 & 5.40 & 4.83 & 4.92 & 5.17 \\ 
        ~ & $E^{T, 6}_{\gamma,4}$ & $E^{T, 6}_{\pi,1}$ & 4.94 & 5.34 & \textbf{6.24} & 5.19 & 5.25 & 5.75 & 5.15 & 5.23 & 5.47 & 5.49 & 5.57 & 5.80 \\ 
        ~ & ~ & $E^{T, 6}_{\pi,2}$ & 5.11 & 5.27 & \textbf{6.23} & 5.07 & 5.05 & 5.46 & 5.28 & 5.39 & 5.61 & 4.95 & 5.11 & 5.25 \\
\end{longtable}
\end{center}
\end{singlespace}

\begin{singlespace}
\begin{center}
\addtolength{\tabcolsep}{-0.1em}
\begin{longtable}[H]{c c c | c c c | c c c | c c c | c c c}
\caption{Type I error rates (percent) of various procedures under $H_0$: $d_1=d_2=...=d_S$ at $\alpha$=0.05 based on 10,000 replicates (8 strata) } \label{tab:tie8} \\

\hline
 &  &  & \multicolumn{3}{c|}{$n_{s1}=25$} & \multicolumn{3}{c|}{$n_{s1}=60$} & \multicolumn{3}{c|}{$n_{s1}=80$} & \multicolumn{3}{c} {$n_{s1}=100$} \\ 
  &  &  & \multicolumn{3}{c|}{$m_{s1}=15$} & \multicolumn{3}{c|}{$m_{s1}=30$} & \multicolumn{3}{c|}{$m_{s1}=40$} & \multicolumn{3}{c} {$m_{s1}=90$} \\ \cline{4-6} \cline{7-9} \cline{10-12} \cline{13-15}
$d$ & $\gamma$ & $\pi$ & $T_{SC}$ & $T_{LR}$ & $T_W$ & $T_{SC}$ & $T_{LR}$ & $T_W$ & $T_{SC}$ & $T_{LR}$ & $T_W$  & $T_{SC}$ & $T_{LR}$ & $T_W$ \\  \hline

\endfirsthead

\multicolumn{15}{l}%
{{ \tablename\ \thetable{}: (Continued)}} \\
\hline
 &  &  & \multicolumn{3}{c|}{$n_{s1}=25$} & \multicolumn{3}{c|}{$n_{s1}=60$} & \multicolumn{3}{c|}{$n_{s1}=80$} & \multicolumn{3}{c} {$n_{s1}=100$} \\ 
  &  &  & \multicolumn{3}{c|}{$m_{s1}=15$} & \multicolumn{3}{c|}{$m_{s1}=30$} & \multicolumn{3}{c|}{$m_{s1}=40$} & \multicolumn{3}{c} {$m_{s1}=90$} \\ \cline{4-6} \cline{7-9} \cline{10-12} \cline{13-15}
$d$ & $\gamma$ & $\pi$ & $T_{SC}$ & $T_{LR}$ & $T_W$ & $T_{SC}$ & $T_{LR}$ & $T_W$ & $T_{SC}$ & $T_{LR}$ & $T_W$  & $T_{SC}$ & $T_{LR}$ & $T_W$ \\  \hline
\endhead

\hline \multicolumn{15}{c}{{Continued on next page}} \\ \hline
\endfoot

\hline \hline
\endlastfoot

        $E^{T, 8}_{d,1}$ & $E^{T, 8}_{\gamma,1}$ & $E^{T, 8}_{\pi,1}$ & 4.84 & \textbf{6.01} & \textbf{6.79} & 5.49 & 5.80 & \textbf{6.30} & 5.15 & 5.49 & 5.80 & 5.01 & 5.24 & 5.51 \\ 
        ~ & ~ & $E^{T, 8}_{\pi,2}$ & 5.22 & 5.71 & \textbf{7.02} & 4.89 & 5.27 & 5.73 & 5.07 & 5.34 & 5.80 & 4.98 & 5.12 & 5.38 \\ 
        ~ & $E^{T, 8}_{\gamma,2}$ & $E^{T, 8}_{\pi,1}$ & 5.36 & 5.97 & \textbf{7.20} & 5.30 & 5.65 & \textbf{6.08} & 5.09 & 5.37 & 5.83 & 5.19 & 5.32 & 5.63 \\ 
        ~ & ~ & $E^{T, 8}_{\pi,2}$ & 5.08 & 5.57 & \textbf{6.86} & 5.34 & 5.50 & \textbf{6.04} & 5.02 & 5.22 & 5.57 & 4.78 & 4.88 & 5.15 \\ 
        ~ & $E^{T, 8}_{\gamma,3}$ & $E^{T, 8}_{\pi,1}$ & 5.38 & 5.99 & \textbf{7.31} & 5.31 & 5.58 & \textbf{6.00} & 5.03 & 5.18 & 5.55 & 4.86 & 5.02 & 5.30 \\ 
        ~ & ~ & $E^{T, 8}_{\pi,2}$ & 5.10 & 5.43 & \textbf{6.74} & 4.98 & 5.09 & 5.51 & 5.03 & 5.12 & 5.54 & 5.50 & 5.61 & 5.85 \\ 
        ~ & $E^{T, 8}_{\gamma,4}$ & $E^{T, 8}_{\pi,1}$ & 5.47 & 5.99 & \textbf{7.23} & 4.79 & 4.99 & 5.48 & 5.54 & 5.75 & \textbf{6.03} & 5.02 & 5.07 & 5.25 \\ 
        ~ & ~ & $E^{T, 8}_{\pi,2}$ & 5.38 & 5.79 & \textbf{6.99} & 5.14 & 5.22 & 5.67 & 4.97 & 5.00 & 5.38 & 5.35 & 5.40 & 5.57 \\
\end{longtable}
\end{center}
\end{singlespace}

\begin{figure}[H]
        \caption{Notched box plots of empirical type I errors for $n_{s1}=60$ and $m_{s1}=30$}
        \label{fig:tien60}
     \centering
\includegraphics[width=\textwidth]{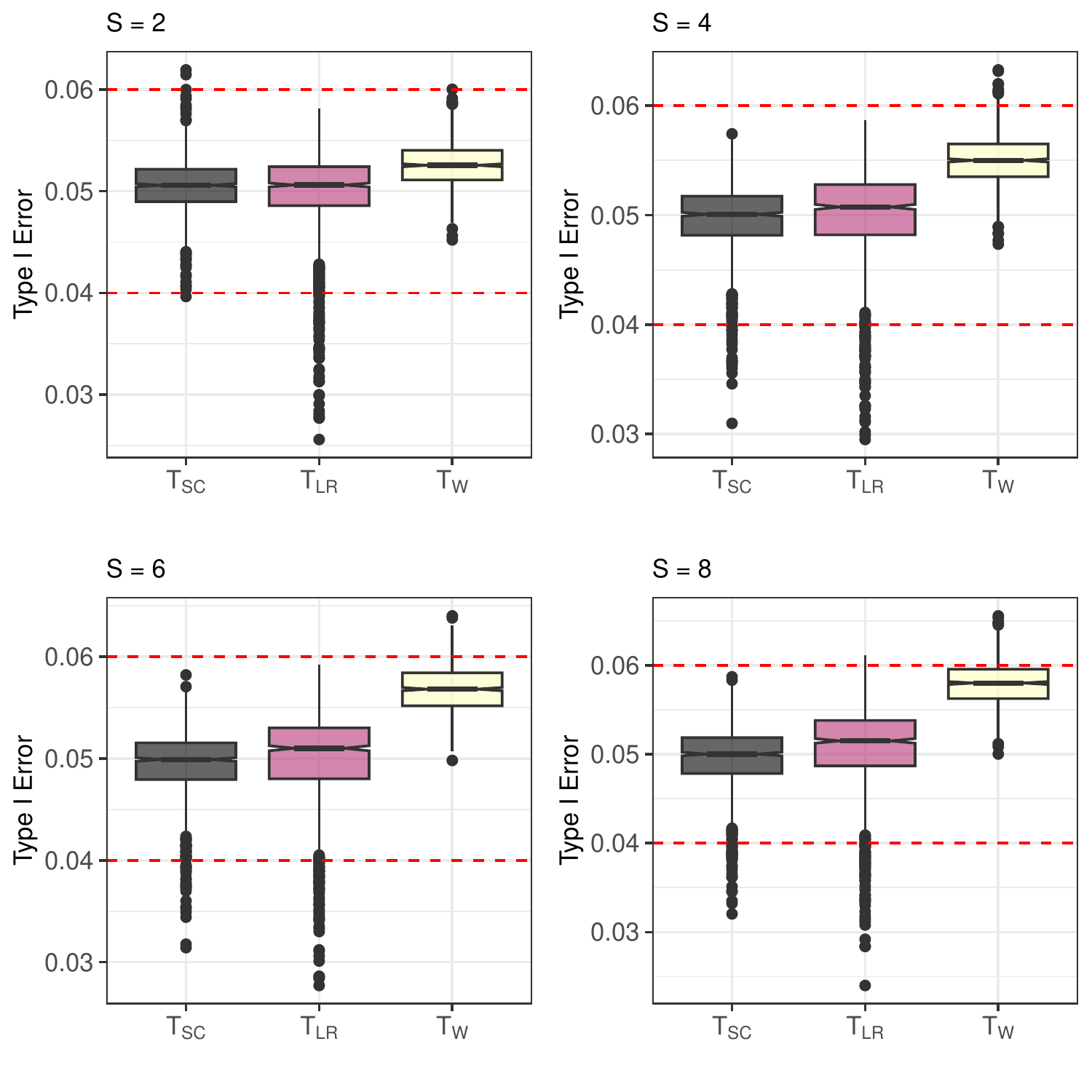}
\end{figure}

\begin{figure}[H]
        \caption{Notched box plots of empirical type I errors for $n_{s1}=80$ and $m_{s1}=40$}
        \label{fig:tien80}
     \centering
\includegraphics[width=\textwidth]{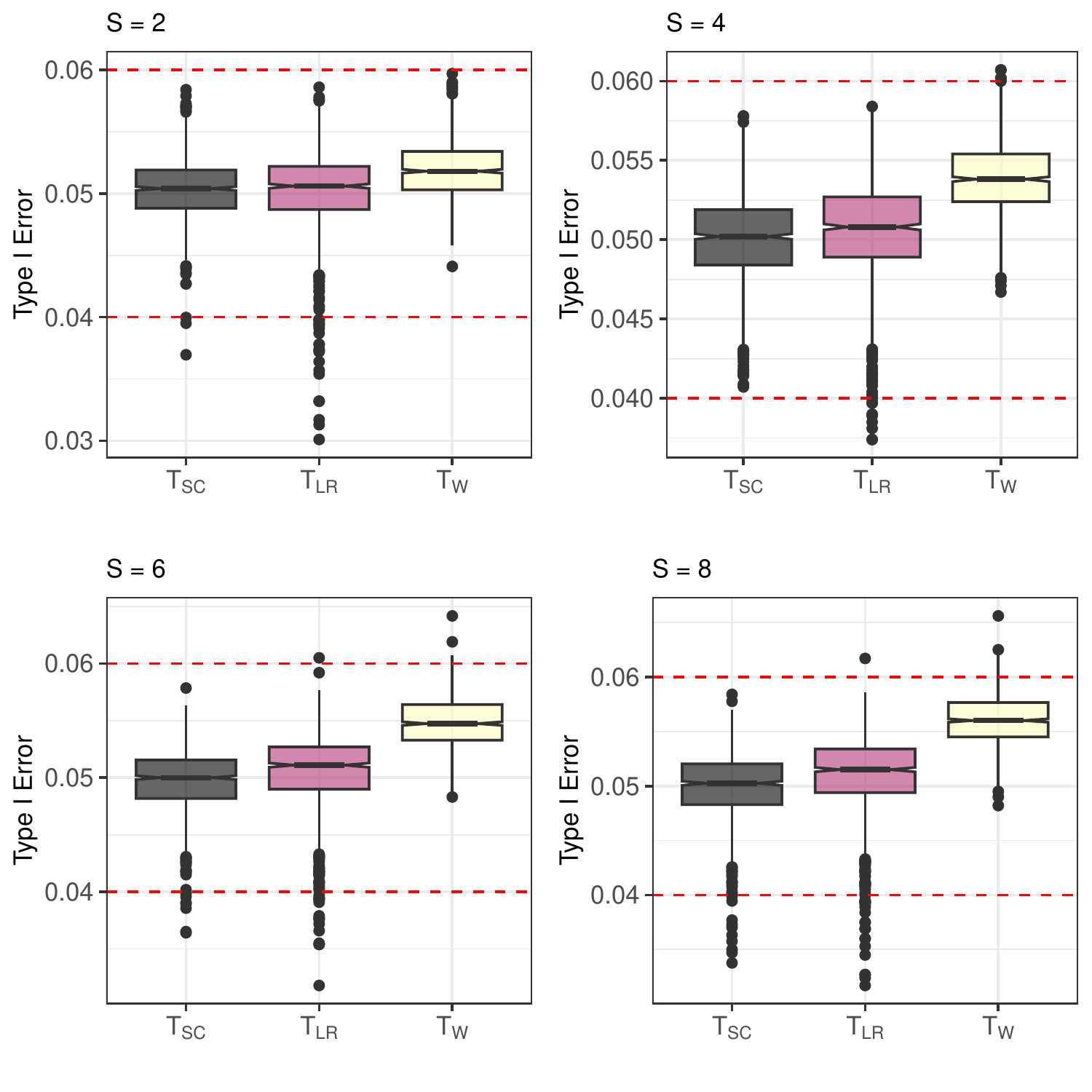}
\end{figure}

Next, we evaluate the power performances of the proposed methods under different combinations of $d_s$, $\pi_{s1}$, and $\gamma_s$. Notations of parameter configurations are presented in \hyperref[tab:powerconf2]{Table \ref*{tab:powerconf2}} to \hyperref[tab:powerconf8]{Table \ref*{tab:powerconf8}}. We still assume $n_{s1}=n_{s2}=n_{s'1}=n_{s'2}$ and $m_{s1}=m_{s2}=m_{s'1}=m_{s'2}$ for $s, s'\in \{1, 2, ..., S\}$. \hyperref[tab:power2]{Table \ref*{tab:power2}} to \hyperref[tab:power8]{Table \ref*{tab:power8}} show the empirical powers associated with $T_{SC}$, $T_{LR}$, and $T_W$. In general, both the score and the likelihood-ratio test generate comparable powers while the Wald-type test tends to produce much higher powers, especially in cases where the sample size is small. This is because the Wald-type test produces inflated type I error rates as observed from simulations for investigating type I errors. The powers of the three proposed methods improve and get closer as the sample size increases or as $H_0$ and $H_a$ are more different than similar. Overall, the score test is recommended as a result of reasonable powers with guaranteed type I error rate controls.

\begin{singlespace}
\begin{table}[H]
\captionsetup[subtable]{justification=centering}
\caption{Parameter configurations for powers (2 strata)}
\label{tab:powerconf2}
  \begin{subtable}[t]{0.48\textwidth}
        \centering
\begin{tabular}[t]{l|c}
\toprule

  Notation   & $(d_1, d_2)$     \\ \hline
$E^{P, 2}_{d,1}$ & (0, 0.05)            \\ 
$E^{P, 2}_{d,2}$ & (0, 0.1) \\ 
$E^{P, 2}_{d,3}$ & (0, 0.15)     \\ 
$E^{P, 2}_{d,4}$ & (0.05,0.1) \\ 
$E^{P, 2}_{d,5}$ & (0.05,0.15) \\ 
$E^{P, 2}_{d,6}$ & (0.1,0.15) \\ 

\bottomrule
\end{tabular}
\caption{}
\end{subtable}
\hfill
  \begin{subtable}[t]{0.48\textwidth}
        \centering
\begin{tabular}[t]{l|c}
\toprule
 Notation  & $(\gamma_1, \gamma_2)$     \\ \hline
$E^{P, 2}_{\gamma,1}$ & (0.4, 0.4)            \\ 
$E^{P, 2}_{\gamma,2}$ & (0.5, 0.5)  \\ 
$E^{P, 2}_{\gamma,3}$ & (0.6, 0.6)      \\ 
$E^{P, 2}_{\gamma,4}$ & (0.7, 0.7)  \\ 
$E^{P, 2}_{\gamma,5}$ & (0.4, 0.5)  \\ 
$E^{P, 2}_{\gamma,6}$ & (0.5, 0.6)  \\ 
$E^{P, 2}_{\gamma,7}$ & (0.6, 0.7)  \\ 

\bottomrule
\end{tabular}
\caption{}
\end{subtable}

\bigskip 

  \begin{subtable}[t]{0.48\textwidth}
        \centering
\begin{tabular}[t]{l|c}
\toprule

 Notation  & $(\pi_{11}, \pi_{21})$     \\ \hline
$E^{P, 2}_{\pi,1}$ & (0.5, 0.5)            \\ 
$E^{P, 2}_{\pi,2}$ & (0.45, 0.45)  \\ 
$E^{P, 2}_{\pi,3}$ & (0.5, 0.45)  \\ 

\bottomrule
\end{tabular}
\caption{}
\end{subtable}
\hspace{\fill}
\begin{subtable}[t]{0.48\textwidth}
\begin{tabular}[t]{l|c}
\toprule

\bottomrule
\end{tabular}

\end{subtable}

\end{table}
\end{singlespace}

\begin{singlespace}
\begin{table}[H]
\captionsetup[subtable]{justification=centering}
\caption{Parameter configurations for powers (4 strata)}
\label{tab:powerconf4}
  \begin{subtable}[t]{0.48\textwidth}
        \centering
\begin{tabular}[t]{l|c}
\toprule

  Notation   & $(d_1, d_2, d_3, d_4)$     \\ \hline
$E^{P, 4}_{d,1}$ & (0, 0.05, 0, 0.05)            \\ 
$E^{P, 4}_{d,2}$ & (0, 0.1, 0, 0.1) \\ 
$E^{P, 4}_{d,3}$ & (0, 0.15, 0, 0.15)     \\ 
$E^{P, 4}_{d,4}$ & (0, 0.05, 0.1, 0.15) \\ 

\bottomrule
\end{tabular}
\caption{}
\end{subtable}
\hfill
  \begin{subtable}[t]{0.48\textwidth}
        \centering
\begin{tabular}[t]{l|c}
\toprule
 Notation  & $(\gamma_1, \gamma_2, \gamma_3, \gamma_4)$     \\ \hline
$E^{P, 4}_{\gamma,1}$ & (0.4, 0.4, 0.4, 0.4)            \\ 
$E^{P, 4}_{\gamma,2}$ & (0.5, 0.5, 0.5, 0.5)  \\ 
$E^{P, 4}_{\gamma,3}$ & (0.6, 0.6, 0.6, 0.6)      \\ 
$E^{P, 4}_{\gamma,4}$ & (0.7, 0.7, 0.7, 0.7)  \\ 
$E^{P, 4}_{\gamma,5}$ & (0.4, 0.4, 0.5, 0.5)  \\ 
$E^{P, 4}_{\gamma,6}$ & (0.4, 0.4, 0.6, 0.6)  \\ 
$E^{P, 4}_{\gamma,7}$ & (0.4, 0.4, 0.7, 0.7)  \\ 
$E^{P, 4}_{\gamma,8}$ & (0.5, 0.5, 0.6, 0.6)  \\ 
$E^{P, 4}_{\gamma,9}$ & (0.5, 0.5, 0.7, 0.7)  \\ 
$E^{P, 4}_{\gamma,10}$ & (0.6, 0.6, 0.7, 0.7)  \\ 
$E^{P, 4}_{\gamma,11}$ & (0.4, 0.5, 0.6, 0.7)  \\ 

\bottomrule
\end{tabular}
\caption{}
\end{subtable}

\bigskip 

  \begin{subtable}[t]{0.48\textwidth}
        \centering
\begin{tabular}[t]{l|c}
\toprule

 Notation  & $(\pi_{11}, \pi_{21}, \pi_{31}, \pi_{41})$     \\ \hline
$E^{P, 4}_{\pi,1}$ & (0.5, 0.5, 0.5, 0.5)            \\ 
$E^{P, 4}_{\pi,2}$ & (0.45, 0.45, 0.45, 0.45)  \\ 
$E^{P, 4}_{\pi,3}$ & (0.5, 0.45, 0.5, 0.45)  \\ 

\bottomrule
\end{tabular}
\caption{}
\end{subtable}
\hspace{\fill}
\begin{subtable}[t]{0.48\textwidth}
\begin{tabular}[t]{l|c}
\toprule

\bottomrule
\end{tabular}

\end{subtable}

\end{table}
\end{singlespace}

\begin{singlespace}
\begin{table}[H]
\captionsetup[subtable]{justification=centering}
\caption{Parameter configurations for powers (6 strata)}
\label{tab:powerconf6}
\begin{subtable}[t]{0.48\textwidth}

\begin{tabular}[t]{l|c}
\toprule

  Notation   & $(d_1, d_2, d_3, d_4, d_5, d_6)$     \\ \hline
$E^{P, 6}_{d,1}$ & (0, 0.05, 0, 0.05, 0, 0.05)            \\ 
$E^{P, 6}_{d,2}$ & (0, 0.1, 0, 0.1, 0, 0.1) \\ 
$E^{P, 6}_{d,3}$ & (0, 0.15, 0, 0.15, 0, 0.15)     \\ 
$E^{P, 6}_{d,4}$ & (0, 0.05, 0.1, 0.15, 0.1, 0.05) \\ 

\bottomrule
\end{tabular}
\caption{}
\end{subtable}
\hspace{\fill}
\begin{subtable}[t]{0.48\textwidth}

\begin{tabular}[t]{l|c}
\toprule
 Notation  & $(\gamma_1, \gamma_2, \gamma_3, \gamma_4, \gamma_5, \gamma_6)$     \\ \hline
$E^{P, 6}_{\gamma,1}$ & (0.4, 0.4, 0.4, 0.4, 0.4, 0.4)            \\ 
$E^{P, 6}_{\gamma,2}$ & (0.5, 0.5, 0.5, 0.5, 0.5, 0.5)  \\ 
$E^{P, 6}_{\gamma,3}$ & (0.6, 0.6, 0.6, 0.6, 0.6, 0.6)      \\ 
$E^{P, 6}_{\gamma,4}$ & (0.7, 0.7, 0.7, 0.7, 0.7, 0.7)  \\ 
$E^{P, 6}_{\gamma,5}$ & (0.4, 0.4, 0.5, 0.5, 0.6, 0.6)  \\ 
$E^{P, 6}_{\gamma,6}$ & (0.5, 0.5, 0.6, 0.6, 0.7, 0.7)  \\ 
$E^{P, 6}_{\gamma,7}$ & (0.4, 0.5, 0.6, 0.7, 0.4, 0.5)  \\ 
$E^{P, 6}_{\gamma,8}$ & (0.4, 0.5, 0.6, 0.7, 0.6, 0.7)  \\ 

\bottomrule
\end{tabular}
\caption{}
\end{subtable}

\bigskip 

\begin{subtable}[t]{0.48\textwidth}
\begin{tabular}[t]{l|c}
\toprule

 Notation  & $(\pi_{11}, \pi_{21}, \pi_{31}, \pi_{41}, \pi_{51}, \pi_{61})$     \\ \hline
$E^{P, 6}_{\pi,1}$ & (0.5, 0.5, 0.5, 0.5, 0.5, 0.5)            \\ 
$E^{P, 6}_{\pi,2}$ & (0.45, 0.45, 0.45, 0.45, 0.45, 0.45)  \\ 
$E^{P, 6}_{\pi,3}$ & (0.5, 0.5, 0.5, 0.45, 0.45, 0.45)  \\ 

\bottomrule
\end{tabular}
\caption{}
\end{subtable}
\hspace{\fill}
\begin{subtable}[t]{0.48\textwidth}
\begin{tabular}[t]{l|c}
\toprule

\bottomrule
\end{tabular}

\end{subtable}

\end{table}
\end{singlespace}

\begin{singlespace}
\begin{table}[H]
\captionsetup[subtable]{justification=centering}
\caption{Parameter configurations for powers (8 strata)}
\label{tab:powerconf8}
\begin{subtable}{1\linewidth}
\centering
\begin{tabular}[t]{l|c}
\toprule

  Notation   & $(d_1, d_2, d_3, d_4, d_5, d_6, d_7, d_8)$     \\ \hline
$E^{P, 8}_{d,1}$ & (0, 0.05, 0, 0.05, 0, 0.05, 0, 0.05)            \\ 
$E^{P, 8}_{d,2}$ & (0, 0.1, 0, 0.1, 0, 0.1, 0, 0.1) \\ 
$E^{P, 8}_{d,3}$ & (0, 0.15, 0, 0.15, 0, 0.15, 0, 0.15)     \\ 
$E^{P, 8}_{d,4}$ & (0, 0.05, 0.1, 0.15, 0.15, 0.1, 0.05, 0) \\ 

\bottomrule
\end{tabular}
\captionsetup{justification=centering}
\caption{}
\end{subtable}

\bigskip

\begin{subtable}{1\linewidth}
\centering
\begin{tabular}[t]{l|c}
\toprule
 Notation  & $(\gamma_1, \gamma_2, \gamma_3, \gamma_4, \gamma_5, \gamma_6, \gamma_7, \gamma_8)$     \\ \hline
$E^{T, 8}_{\gamma,1}$ & (0.4, 0.4, 0.4, 0.4, 0.4, 0.4, 0.4, 0.4)            \\ 
$E^{P, 8}_{\gamma,2}$ & (0.5, 0.5, 0.5, 0.5, 0.5, 0.5, 0.5, 0.5)  \\ 
$E^{P, 8}_{\gamma,3}$ & (0.6, 0.6, 0.6, 0.6, 0.6, 0.6, 0.6, 0.6)      \\ 
$E^{P, 8}_{\gamma,4}$ & (0.7, 0.7, 0.7, 0.7, 0.7, 0.7, 0.7, 0.7)  \\ 
$E^{P, 8}_{\gamma,5}$ & (0.4, 0.4, 0.4, 0.4, 0.5, 0.5, 0.5, 0.5)  \\ 
$E^{P, 8}_{\gamma,6}$ & (0.4, 0.4, 0.4, 0.4, 0.6, 0.6, 0.6, 0.6)  \\ 
$E^{P, 8}_{\gamma,7}$ & (0.4, 0.4, 0.4, 0.4, 0.7, 0.7, 0.7, 0.7)  \\ 
$E^{P, 8}_{\gamma,8}$ & (0.5, 0.5, 0.5, 0.5, 0.6, 0.6, 0.6, 0.6)  \\ 
$E^{P, 8}_{\gamma,9}$ & (0.5, 0.5, 0.5, 0.5, 0.7, 0.7, 0.7, 0.7)  \\ 
$E^{P, 8}_{\gamma,10}$ & (0.6, 0.6, 0.6, 0.6, 0.7, 0.7, 0.7, 0.7)  \\ 
$E^{P, 8}_{\gamma,11}$ & (0.4, 0.4, 0.5, 0.5, 0.6, 0.6, 0.7, 0.7)  \\

\bottomrule
\end{tabular}
\captionsetup{justification=centering}
\caption{}
\end{subtable}

\bigskip 

\begin{subtable}{1\linewidth}
\centering
\begin{tabular}[t]{l|c}
\toprule

 Notation  & $(\pi_{11}, \pi_{21}, \pi_{31}, \pi_{41}, \pi_{51}, \pi_{61}, \pi_{71}, \pi_{81})$     \\ \hline
$E^{P, 8}_{\pi,1}$ & (0.5, 0.5, 0.5, 0.5, 0.5, 0.5, 0.5, 0.5)    \\ 
$E^{P, 8}_{\pi,2}$ & (0.45, 0.45, 0.45, 0.45, 0.45, 0.45, 0.45, 0.45)  \\ 
$E^{P, 8}_{\pi,3}$ & (0.5, 0.5, 0.5, 0.5, 0.45, 0.45, 0.45, 0.45)  \\

\bottomrule
\end{tabular}
\captionsetup{justification=centering}
\caption{}
\end{subtable}

\end{table}
\end{singlespace}

\begin{singlespace}
\begin{center}
\addtolength{\tabcolsep}{-0.25em}
\begin{longtable}[H]{c c c | c c c | c c c | c c c | c c c}
\caption{The empirical powers (percent) of various procedures at $\alpha$=0.05 based on 10,000 replicates (2 strata) } \label{tab:power2} \\

\hline
 &  &  & \multicolumn{3}{c|}{$n_{s1}=25$} & \multicolumn{3}{c|}{$n_{s1}=60$} & \multicolumn{3}{c|}{$n_{s1}=80$} & \multicolumn{3}{c} {$n_{s1}=100$} \\ 
  &  &  & \multicolumn{3}{c|}{$m_{s1}=15$} & \multicolumn{3}{c|}{$m_{s1}=30$} & \multicolumn{3}{c|}{$m_{s1}=40$} & \multicolumn{3}{c} {$m_{s1}=90$} \\ \cline{4-6} \cline{7-9} \cline{10-12} \cline{13-15}
$d$ & $\gamma$ & $\pi$ & $T_{SC}$ & $T_{LR}$ & $T_W$ & $T_{SC}$ & $T_{LR}$ & $T_W$ & $T_{SC}$ & $T_{LR}$ & $T_W$  & $T_{SC}$ & $T_{LR}$ & $T_W$ \\  \hline

\endfirsthead

\multicolumn{15}{l}%
{{ \tablename\ \thetable{}: (Continued)}} \\
\hline
 &  &  & \multicolumn{3}{c|}{$n_{s1}=25$} & \multicolumn{3}{c|}{$n_{s1}=60$} & \multicolumn{3}{c|}{$n_{s1}=80$} & \multicolumn{3}{c} {$n_{s1}=100$} \\ 
  &  &  & \multicolumn{3}{c|}{$m_{s1}=15$} & \multicolumn{3}{c|}{$m_{s1}=30$} & \multicolumn{3}{c|}{$m_{s1}=40$} & \multicolumn{3}{c} {$m_{s1}=90$} \\ \cline{4-6} \cline{7-9} \cline{10-12} \cline{13-15}
$d$ & $\gamma$ & $\pi$ & $T_{SC}$ & $T_{LR}$ & $T_W$ & $T_{SC}$ & $T_{LR}$ & $T_W$ & $T_{SC}$ & $T_{LR}$ & $T_W$  & $T_{SC}$ & $T_{LR}$ & $T_W$ \\  \hline
\endhead

\hline \multicolumn{15}{c}{{Continued on next page}} \\ \hline
\endfoot

\hline \hline
\endlastfoot

        $E^{P, 2}_{d,1}$ & $E^{P, 2}_{\gamma,1}$ & $E^{P, 2}_{\pi,1}$ &  8.05 &  8.19 &  8.82 & 12.51 & 12.68 & 12.89 & 15.73 & 15.89 & 16.06 & 18.65 & 18.80 & 18.88 \\ 
        ~ & ~ & $E^{P, 2}_{\pi,2}$ &  7.98 &  8.28 &  8.61 & 11.75 & 11.78 & 11.98 & 13.37 & 13.48 & 13.60 & 16.76 & 16.94 & 17.06 \\ 
        ~ & ~ & $E^{P, 2}_{\pi,3}$ &  8.15 &  8.45 &  8.61 & 11.67 & 11.84 & 12.12 & 14.02 & 14.16 & 14.40 & 18.03 & 18.19 & 18.47 \\ 
        ~ & $E^{P, 2}_{\gamma,2}$ & $E^{P, 2}_{\pi,1}$ &  7.92 &  8.15 &  8.67 & 11.09 & 11.17 & 11.36 & 13.09 & 13.15 & 13.27 & 16.37 & 16.51 & 16.61 \\ 
        ~ & ~ & $E^{P, 2}_{\pi,2}$ &  7.37 &  7.43 &  7.81 & 10.70 & 10.68 & 10.82 & 12.61 & 12.72 & 12.87 & 14.86 & 14.90 & 15.13 \\ 
        ~ & ~ & $E^{P, 2}_{\pi,3}$ &  7.50 &  7.78 &  8.06 & 10.92 & 11.06 & 11.33 & 13.16 & 13.21 & 13.38 & 15.41 & 15.49 & 15.60 \\ 
        ~ & $E^{P, 2}_{\gamma,3}$ & $E^{P, 2}_{\pi,1}$ &  7.46 &  7.65 &  7.94 & 10.28 & 10.31 & 10.48 & 11.25 & 11.33 & 11.54 & 14.44 & 14.51 & 14.61 \\ 
        ~ & ~ & $E^{P, 2}_{\pi,2}$ &  7.20 &  7.28 &  7.66 &  9.61 &  9.75 &  9.95 & 11.04 & 11.10 & 11.26 & 13.35 & 13.45 & 13.55 \\
\end{longtable}
\end{center}
\end{singlespace}

\begin{singlespace}
\begin{center}
\addtolength{\tabcolsep}{-0.25em}
\begin{longtable}[H]{c c c | c c c | c c c | c c c | c c c}
\caption{The empirical powers (percent) of various procedures at $\alpha$=0.05 based on 10,000 replicates (4 strata) } \label{tab:power4} \\

\hline
 &  &  & \multicolumn{3}{c|}{$n_{s1}=25$} & \multicolumn{3}{c|}{$n_{s1}=60$} & \multicolumn{3}{c|}{$n_{s1}=80$} & \multicolumn{3}{c} {$n_{s1}=100$} \\ 
  &  &  & \multicolumn{3}{c|}{$m_{s1}=15$} & \multicolumn{3}{c|}{$m_{s1}=30$} & \multicolumn{3}{c|}{$m_{s1}=40$} & \multicolumn{3}{c} {$m_{s1}=90$} \\ \cline{4-6} \cline{7-9} \cline{10-12} \cline{13-15}
$d$ & $\gamma$ & $\pi$ & $T_{SC}$ & $T_{LR}$ & $T_W$ & $T_{SC}$ & $T_{LR}$ & $T_W$ & $T_{SC}$ & $T_{LR}$ & $T_W$  & $T_{SC}$ & $T_{LR}$ & $T_W$ \\  \hline

\endfirsthead

\multicolumn{15}{l}%
{{ \tablename\ \thetable{}: (Continued)}} \\
\hline
 &  &  & \multicolumn{3}{c|}{$n_{s1}=25$} & \multicolumn{3}{c|}{$n_{s1}=60$} & \multicolumn{3}{c|}{$n_{s1}=80$} & \multicolumn{3}{c} {$n_{s1}=100$} \\ 
  &  &  & \multicolumn{3}{c|}{$m_{s1}=15$} & \multicolumn{3}{c|}{$m_{s1}=30$} & \multicolumn{3}{c|}{$m_{s1}=40$} & \multicolumn{3}{c} {$m_{s1}=90$} \\ \cline{4-6} \cline{7-9} \cline{10-12} \cline{13-15}
$d$ & $\gamma$ & $\pi$ & $T_{SC}$ & $T_{LR}$ & $T_W$ & $T_{SC}$ & $T_{LR}$ & $T_W$ & $T_{SC}$ & $T_{LR}$ & $T_W$  & $T_{SC}$ & $T_{LR}$ & $T_W$ \\  \hline
\endhead

\hline \multicolumn{15}{c}{{Continued on next page}} \\ \hline
\endfoot

\hline \hline
\endlastfoot

        $E^{P, 4}_{d,1}$ & $E^{P, 4}_{\gamma,1}$ & $E^{P, 4}_{\pi,1}$ &  8.86 &  9.31 & 10.47 & 13.35 & 13.82 & 14.38 & 16.96 & 17.40 & 17.88 & 21.37 & 21.82 & 22.45 \\ 
        ~ & ~ & $E^{P, 4}_{\pi,2}$ &  8.07 &  8.51 &  9.60 & 11.99 & 12.20 & 12.70 & 15.09 & 15.26 & 15.76 & 19.94 & 20.25 & 20.55 \\ 
        ~ & ~ & $E^{P, 4}_{\pi,3}$ &  8.35 &  9.15 & 10.03 & 13.24 & 13.61 & 14.30 & 15.44 & 15.82 & 16.35 & 20.55 & 20.68 & 21.12 \\ 
        ~ & $E^{P, 4}_{\gamma,2}$ & $E^{P, 4}_{\pi,1}$ &  7.73 &  8.37 &  9.24 & 11.83 & 12.00 & 12.58 & 14.27 & 14.53 & 14.99 & 18.40 & 18.61 & 18.93 \\ 
        ~ & ~ & $E^{P, 4}_{\pi,2}$ &  7.67 &  8.09 &  8.94 & 11.55 & 11.79 & 12.27 & 12.98 & 13.19 & 13.54 & 17.30 & 17.47 & 17.82 \\ 
        ~ & ~ & $E^{P, 4}_{\pi,3}$ &  8.07 &  8.46 &  9.43 & 11.45 & 11.59 & 12.06 & 13.11 & 13.28 & 13.70 & 17.45 & 17.58 & 18.06 \\ 
        ~ & $E^{P, 4}_{\gamma,3}$ & $E^{P, 4}_{\pi,1}$ &  7.52 &  7.92 &  8.76 & 10.37 & 10.62 & 11.08 & 12.56 & 12.77 & 13.15 & 15.20 & 15.28 & 15.59 \\ 
        ~ & ~ & $E^{P, 4}_{\pi,2}$ &  7.33 &  7.56 &  8.37 &  9.84 &  9.90 & 10.35 & 11.69 & 11.90 & 12.19 & 15.83 & 15.99 & 16.29 \\
\end{longtable}
\end{center}
\end{singlespace}

\begin{singlespace}
\begin{center}
\addtolength{\tabcolsep}{-0.25em}
\begin{longtable}[H]{c c c | c c c | c c c | c c c | c c c}
\caption{The empirical powers (percent) of various procedures at $\alpha$=0.05 based on 10,000 replicates (6 strata) } \label{tab:power6} \\

\hline
 &  &  & \multicolumn{3}{c|}{$n_{s1}=25$} & \multicolumn{3}{c|}{$n_{s1}=60$} & \multicolumn{3}{c|}{$n_{s1}=80$} & \multicolumn{3}{c} {$n_{s1}=100$} \\ 
  &  &  & \multicolumn{3}{c|}{$m_{s1}=15$} & \multicolumn{3}{c|}{$m_{s1}=30$} & \multicolumn{3}{c|}{$m_{s1}=40$} & \multicolumn{3}{c} {$m_{s1}=90$} \\ \cline{4-6} \cline{7-9} \cline{10-12} \cline{13-15}
$d$ & $\gamma$ & $\pi$ & $T_{SC}$ & $T_{LR}$ & $T_W$ & $T_{SC}$ & $T_{LR}$ & $T_W$ & $T_{SC}$ & $T_{LR}$ & $T_W$  & $T_{SC}$ & $T_{LR}$ & $T_W$ \\  \hline

\endfirsthead

\multicolumn{15}{l}%
{{ \tablename\ \thetable{}: (Continued)}} \\
\hline
 &  &  & \multicolumn{3}{c|}{$n_{s1}=25$} & \multicolumn{3}{c|}{$n_{s1}=60$} & \multicolumn{3}{c|}{$n_{s1}=80$} & \multicolumn{3}{c} {$n_{s1}=100$} \\ 
  &  &  & \multicolumn{3}{c|}{$m_{s1}=15$} & \multicolumn{3}{c|}{$m_{s1}=30$} & \multicolumn{3}{c|}{$m_{s1}=40$} & \multicolumn{3}{c} {$m_{s1}=90$} \\ \cline{4-6} \cline{7-9} \cline{10-12} \cline{13-15}
$d$ & $\gamma$ & $\pi$ & $T_{SC}$ & $T_{LR}$ & $T_W$ & $T_{SC}$ & $T_{LR}$ & $T_W$ & $T_{SC}$ & $T_{LR}$ & $T_W$  & $T_{SC}$ & $T_{LR}$ & $T_W$ \\  \hline
\endhead

\hline \multicolumn{15}{c}{{Continued on next page}} \\ \hline
\endfoot

\hline \hline
\endlastfoot

        $E^{P, 6}_{d,1}$ & $E^{P, 6}_{\gamma,1}$ & $E^{P, 6}_{\pi,1}$ &  9.08 & 10.14 & 11.67 & 14.84 & 15.53 & 16.44 & 19.26 & 19.94 & 20.76 & 26.14 & 26.51 & 27.13 \\ 
        ~ & ~ & $E^{P, 6}_{\pi,2}$ &  8.55 &  9.27 & 10.78 & 12.78 & 13.22 & 14.14 & 17.09 & 17.43 & 18.10 & 22.10 & 22.46 & 23.05 \\ 
        ~ & ~ & $E^{P, 6}_{\pi,3}$ &  8.97 &  9.84 & 11.14 & 14.54 & 15.16 & 15.92 & 18.43 & 18.91 & 19.50 & 24.58 & 25.01 & 25.58 \\ 
        ~ & $E^{P, 6}_{\gamma,2}$ & $E^{P, 6}_{\pi,1}$ &  8.60 &  9.25 & 10.81 & 12.74 & 13.10 & 13.91 & 15.89 & 16.37 & 16.95 & 20.75 & 20.90 & 21.36 \\ 
        ~ & ~ & $E^{P, 6}_{\pi,2}$ &  7.43 &  7.97 &  9.22 & 11.38 & 11.66 & 12.26 & 14.79 & 14.96 & 15.47 & 18.98 & 19.16 & 19.80 \\ 
        ~ & ~ & $E^{P, 6}_{\pi,3}$ &  8.31 &  8.96 & 10.24 & 12.62 & 12.86 & 13.62 & 15.14 & 15.55 & 16.25 & 20.41 & 20.63 & 21.10 \\ 
        ~ & $E^{P, 6}_{\gamma,3}$ & $E^{P, 6}_{\pi,1}$ &  7.45 &  8.06 &  9.45 & 10.92 & 11.17 & 11.68 & 13.99 & 14.25 & 14.78 & 18.19 & 18.43 & 18.95 \\ 
        ~ & ~ & $E^{P, 6}_{\pi,2}$ &  7.37 &  7.84 &  9.23 & 11.10 & 11.36 & 11.90 & 13.67 & 13.91 & 14.43 & 17.34 & 17.42 & 17.76 \\
\end{longtable}
\end{center}
\end{singlespace}

\begin{singlespace}
\begin{center}
\addtolength{\tabcolsep}{-0.25em}
\begin{longtable}[H]{c c c | c c c | c c c | c c c | c c c}
\caption{The empirical powers (percent) of various procedures at $\alpha$=0.05 based on 10,000 replicates (8 strata) } \label{tab:power8} \\

\hline
 &  &  & \multicolumn{3}{c|}{$n_{s1}=25$} & \multicolumn{3}{c|}{$n_{s1}=60$} & \multicolumn{3}{c|}{$n_{s1}=80$} & \multicolumn{3}{c} {$n_{s1}=100$} \\ 
  &  &  & \multicolumn{3}{c|}{$m_{s1}=15$} & \multicolumn{3}{c|}{$m_{s1}=30$} & \multicolumn{3}{c|}{$m_{s1}=40$} & \multicolumn{3}{c} {$m_{s1}=90$} \\ \cline{4-6} \cline{7-9} \cline{10-12} \cline{13-15}
$d$ & $\gamma$ & $\pi$ & $T_{SC}$ & $T_{LR}$ & $T_W$ & $T_{SC}$ & $T_{LR}$ & $T_W$ & $T_{SC}$ & $T_{LR}$ & $T_W$  & $T_{SC}$ & $T_{LR}$ & $T_W$ \\ \hline

\endfirsthead

\multicolumn{15}{l}%
{{ \tablename\ \thetable{}: (Continued)}} \\
\hline
 &  &  & \multicolumn{3}{c|}{$n_{s1}=25$} & \multicolumn{3}{c|}{$n_{s1}=60$} & \multicolumn{3}{c|}{$n_{s1}=80$} & \multicolumn{3}{c} {$n_{s1}=100$} \\ 
  &  &  & \multicolumn{3}{c|}{$m_{s1}=15$} & \multicolumn{3}{c|}{$m_{s1}=30$} & \multicolumn{3}{c|}{$m_{s1}=40$} & \multicolumn{3}{c} {$m_{s1}=90$} \\ \cline{4-6} \cline{7-9} \cline{10-12} \cline{13-15}
$d$ & $\gamma$ & $\pi$ & $T_{SC}$ & $T_{LR}$ & $T_W$ & $T_{SC}$ & $T_{LR}$ & $T_W$ & $T_{SC}$ & $T_{LR}$ & $T_W$  & $T_{SC}$ & $T_{LR}$ & $T_W$ \\ \hline
\endhead

\hline \multicolumn{15}{c}{{Continued on next page}} \\ \hline
\endfoot

\hline \hline
\endlastfoot

        $E^{P, 8}_{d,1}$ & $E^{P, 8}_{\gamma,1}$ & $E^{P, 8}_{\pi,1}$ &  9.79 & 10.99 & 12.54 & 16.54 & 17.50 & 18.55 & 21.48 & 22.27 & 23.36 & 29.56 & 30.04 & 30.94 \\ 
        ~ & ~ & $E^{P, 8}_{\pi,2}$ &  9.14 & 10.14 & 12.10 & 14.69 & 15.21 & 16.10 & 18.52 & 18.90 & 19.75 & 25.29 & 25.56 & 26.38 \\ 
        ~ & ~ & $E^{P, 8}_{\pi,3}$ &  9.25 & 10.37 & 12.07 & 15.75 & 16.31 & 17.36 & 20.03 & 20.61 & 21.52 & 26.67 & 27.15 & 27.82 \\ 
        ~ & $E^{P, 8}_{\gamma,2}$ & $E^{P, 8}_{\pi,1}$ &  8.31 &  9.27 & 11.12 & 13.87 & 14.35 & 15.26 & 17.78 & 18.21 & 18.88 & 23.20 & 23.67 & 24.38 \\ 
        ~ & ~ & $E^{P, 8}_{\pi,2}$ &  8.52 &  9.17 & 10.73 & 12.92 & 13.32 & 14.18 & 15.82 & 16.12 & 16.85 & 21.43 & 21.67 & 22.22 \\ 
        ~ & ~ & $E^{P, 8}_{\pi,3}$ &  8.59 &  9.57 & 11.14 & 13.29 & 13.71 & 14.70 & 16.81 & 17.04 & 17.73 & 24.12 & 24.38 & 25.10 \\ 
        ~ & $E^{P, 8}_{\gamma,3}$ & $E^{P, 8}_{\pi,1}$ &  8.08 &  8.64 & 10.37 & 12.38 & 12.70 & 13.60 & 15.21 & 15.49 & 16.16 & 20.13 & 20.45 & 20.87 \\ 
        ~ & ~ & $E^{P, 8}_{\pi,2}$ &  7.29 &  7.84 &  9.23 & 11.42 & 11.58 & 12.29 & 14.87 & 15.01 & 15.60 & 18.84 & 18.92 & 19.33 \\
\end{longtable}
\end{center}
\end{singlespace}

\begin{singlespace}
\begin{table}[H]
\captionsetup[subtable]{justification=centering}
\caption{Parameter configurations for CIs (2 strata)}
\label{tab:ciconf2}
\begin{subtable}[t]{0.48\textwidth}
\centering
\begin{tabular}[t]{l|c|c}
\toprule
  Notation   & $(n_{11}, n_{21})$  & $(m_{11}, m_{21})$  \\ \hline
$E^{C, 2}_{N,1}$ & (35, 40)   & (15, 20)     \\ 
$E^{C, 2}_{N,2}$ & (60, 80) & (30, 40) \\ 
$E^{C, 2}_{N,3}$ & (90, 100) &  (80, 90) \\ 

\bottomrule
\end{tabular}                                                

\caption{\footnotesize Note: $n_{s1}=n_{s2}$, $m_{s1}=m_{s2}$, $S=1, 2.$}

\end{subtable}
\hspace{\fill}
\begin{subtable}[t]{0.48\textwidth}
\centering
\begin{tabular}[t]{l|c}
\toprule
  Notation   & $d_1=d_2=d_0$     \\ \hline
$E^{C, 2}_{d,1}$ & 0            \\ 
$E^{C, 2}_{d,2}$ & 0.1     \\ 
$E^{C, 2}_{d,3}$ & 0.15 \\ 

\bottomrule
\end{tabular}
\caption{\footnotesize }

\end{subtable}

\bigskip 

\begin{subtable}[t]{0.48\textwidth}
\centering
\begin{tabular}[t]{l|c}
\toprule
 Notation  & $(\gamma_1, \gamma_2)$     \\ \hline
$E^{C, 2}_{\gamma,1}$ & (0.4, 0.5)            \\ 
$E^{C, 2}_{\gamma,2}$ & (0.6, 0.7)  \\

\bottomrule
\end{tabular}
\caption{\footnotesize }

\end{subtable}
\hspace{\fill}
\begin{subtable}[t]{0.48\textwidth}
\centering
\begin{tabular}[t]{l|c}
\toprule
 Notation  & $(\pi_{11}, \pi_{21})$     \\ \hline
$E^{C, 2}_{\pi,1}$ & (0.45, 0.45)            \\ 
$E^{C, 2}_{\pi,2}$ & (0.5, 0.5)  \\

\bottomrule
\end{tabular}
\caption{\footnotesize }

\end{subtable}
\end{table}
\end{singlespace}

\begin{singlespace}
\begin{table}[H]
\captionsetup[subtable]{justification=centering}
\caption{Parameter configurations for CIs (4 strata)}
\label{tab:ciconf4}
\begin{subtable}[t]{0.48\textwidth}
\centering
\begin{tabular}[t]{l|c|c}
\toprule
  Notation   & $(n_{11}, n_{21}, n_{31}, n_{41})$  & $(m_{11}, m_{21}, m_{31}, m_{41})$  \\ \hline
$E^{C, 4}_{N,1}$ & (35, 35, 40, 40)   & (15, 15, 20, 20)     \\ 
$E^{C, 4}_{N,2}$ & (60, 60, 80, 80) & (30, 30, 40, 40) \\ 
$E^{C, 4}_{N,3}$ & (90, 90, 100, 100) &  (80, 80, 90, 90) \\ 

\bottomrule
\end{tabular}                                                

\caption{\footnotesize Note: $n_{s1}=n_{s2}$, $m_{s1}=m_{s2}$, $S=1, 2, 3, 4.$}

\end{subtable}
\hspace{\fill}
\begin{subtable}[t]{0.48\textwidth}
\centering
\begin{tabular}[t]{l|c}
\toprule
  Notation   & $d_1=d_2=d_3=d_4=d_0$     \\ \hline
$E^{C, 4}_{d,1}$ & 0            \\ 
$E^{C, 4}_{d,2}$ & 0.1     \\ 
$E^{C, 4}_{d,3}$ & 0.15 \\ 

\bottomrule
\end{tabular}
\caption{\footnotesize }

\end{subtable}

\bigskip 

\begin{subtable}[t]{0.48\textwidth}
\centering
\begin{tabular}[t]{l|c}
\toprule
 Notation  & $(\gamma_1, \gamma_2, \gamma_3, \gamma_4)$     \\ \hline
$E^{C, 4}_{\gamma,1}$ & (0.4, 0.4, 0.5, 0.5)            \\ 
$E^{C, 4}_{\gamma,2}$ & (0.4, 0.5, 0.6, 0.7)  \\

\bottomrule
\end{tabular}
\caption{\footnotesize }

\end{subtable}
\hspace{\fill}
\begin{subtable}[t]{0.48\textwidth}
\centering
\begin{tabular}[t]{l|c}
\toprule
 Notation  & $(\pi_{11}, \pi_{21}, \pi_{31}, \pi_{41})$     \\ \hline
$E^{C, 4}_{\pi,1}$ & (0.45, 0.45, 0.5, 0.5)            \\ 
$E^{C, 4}_{\pi,2}$ & (0.45, 0.5, 0.45, 0.5)  \\

\bottomrule
\end{tabular}
\caption{\footnotesize }

\end{subtable}
\end{table}
\end{singlespace}

\begin{singlespace}
\begin{table}[H]
\captionsetup[subtable]{justification=centering}
\caption{Parameter configurations for CIs (6 strata)}
\label{tab:ciconf6}
\begin{subtable}{1\linewidth}
\centering
\begin{tabular}[t]{l|c|c}
\toprule
  Notation   & $(n_{11}, n_{21}, n_{31}, n_{41}, n_{51}, n_{61})$  & $(m_{11}, m_{21}, m_{31}, m_{41}, m_{51}, m_{61})$  \\ \hline
$E^{C, 6}_{N,1}$ & (35, 35, 35, 40, 40, 40)   & (25, 25, 25, 30, 30, 30)     \\ 
$E^{C, 6}_{N,2}$ & (40, 40, 60, 60, 80, 80) & (30, 30, 50, 50, 70, 70) \\ 
$E^{C, 6}_{N,3}$ & (80, 80, 90, 90, 100, 100) &  (70, 70, 80, 80, 90, 90) \\ 

\bottomrule
\end{tabular}                                                

\caption{\footnotesize Note: $n_{s1}=n_{s2}$, $m_{s1}=m_{s2}$, $S=1, ..., 6.$}

\end{subtable}
\bigskip
\\
\begin{subtable}{1\linewidth}
\centering
\begin{tabular}[t]{l|c}
\toprule
  Notation   & $d_1=d_2=...=d_6=d_0$     \\ \hline
$E^{C, 6}_{d,1}$ & 0            \\ 
$E^{C, 6}_{d,2}$ & 0.1     \\ 
$E^{C, 6}_{d,3}$ & 0.15 \\ 

\bottomrule
\end{tabular}
\caption{\footnotesize }

\end{subtable}

\bigskip

\begin{subtable}{1\linewidth}
\centering
\begin{tabular}[t]{l|c}
\toprule
 Notation  & $(\gamma_1, \gamma_2, \gamma_3, \gamma_4, \gamma_5, \gamma_6)$     \\ \hline
$E^{C, 6}_{\gamma,1}$ & (0.4, 0.4, 0.5, 0.5, 0.6, 0.6)            \\ 
$E^{C, 6}_{\gamma,2}$ & (0.5, 0.5, 0.6, 0.6, 0.7, 0.7)  \\

\bottomrule
\end{tabular}
\caption{\footnotesize }

\end{subtable}
\bigskip
\\
\begin{subtable}{1\linewidth}
\centering
\begin{tabular}[t]{l|c}
\toprule
 Notation  & $(\pi_{11}, \pi_{21}, \pi_{31}, \pi_{41}, \pi_{51}, \pi_{61})$     \\ \hline
$E^{C, 6}_{\pi,1}$ & (0.45, 0.45, 0.5, 0.5, 0.45, 0.45)            \\ 
$E^{C, 6}_{\pi,2}$ & (0.45, 0.5, 0.45, 0.5, 0.45, 0.5)  \\

\bottomrule
\end{tabular}
\caption{\footnotesize }

\end{subtable}
\end{table}
\end{singlespace}

\begin{figure}[H]
        \caption{Notched box plots of empirical coverage probabilities for sample case $E_{N, 1}^{C, 4}$, $E_{N, 1}^{C, 6}$, and $E_{N, 1}^{C, 8}$}
        \label{fig:ciprob1}
     \centering

         \includegraphics[width=\textwidth]{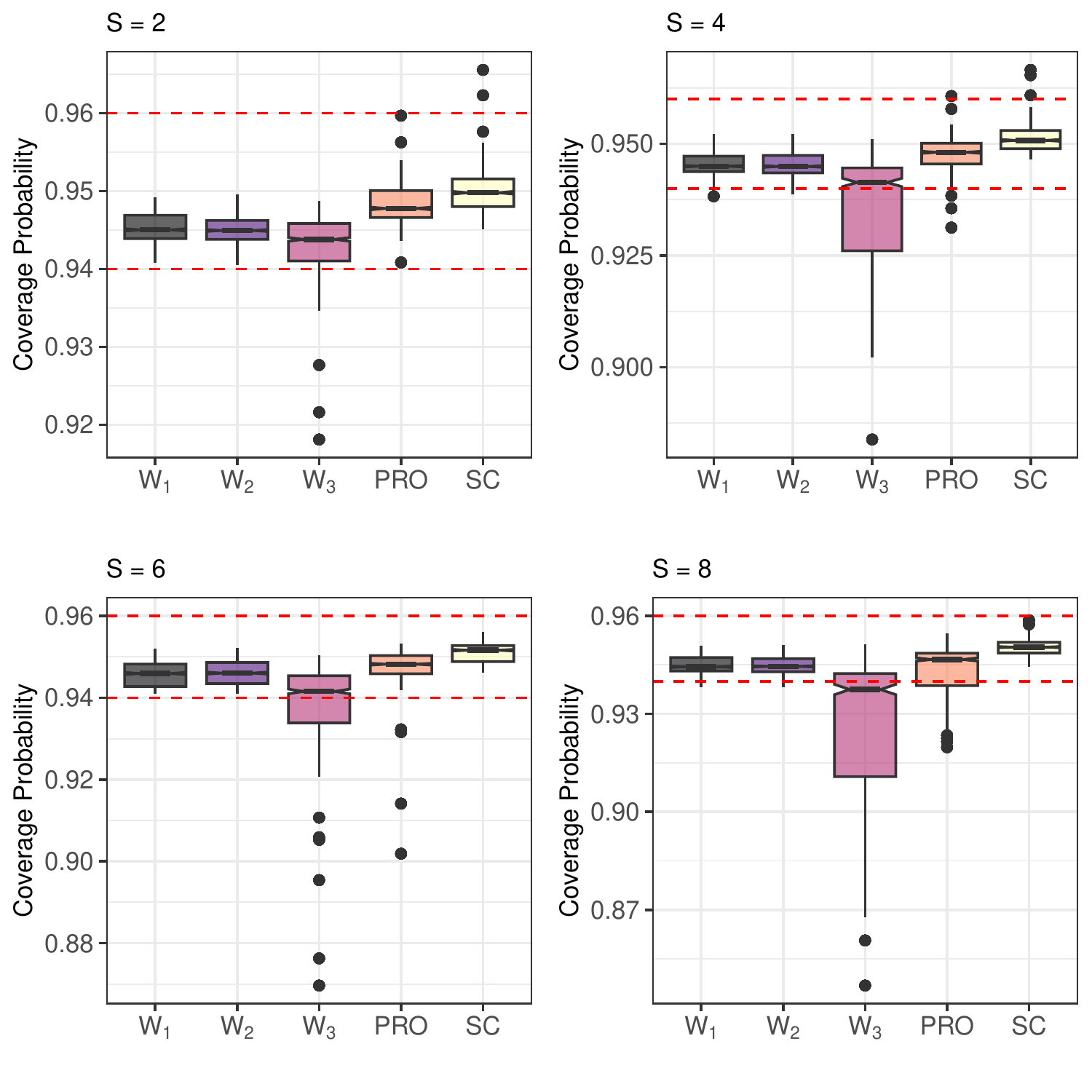}

\end{figure}

\begin{figure}[H]
        \caption{Notched box plots of mean lengths for sample case $E_{N, 3}^{C, 4}$, $E_{N, 3}^{C, 6}$, and $E_{N, 3}^{C, 8}$}
        \label{fig:cilen3}
     \centering

         \includegraphics[width=\textwidth]{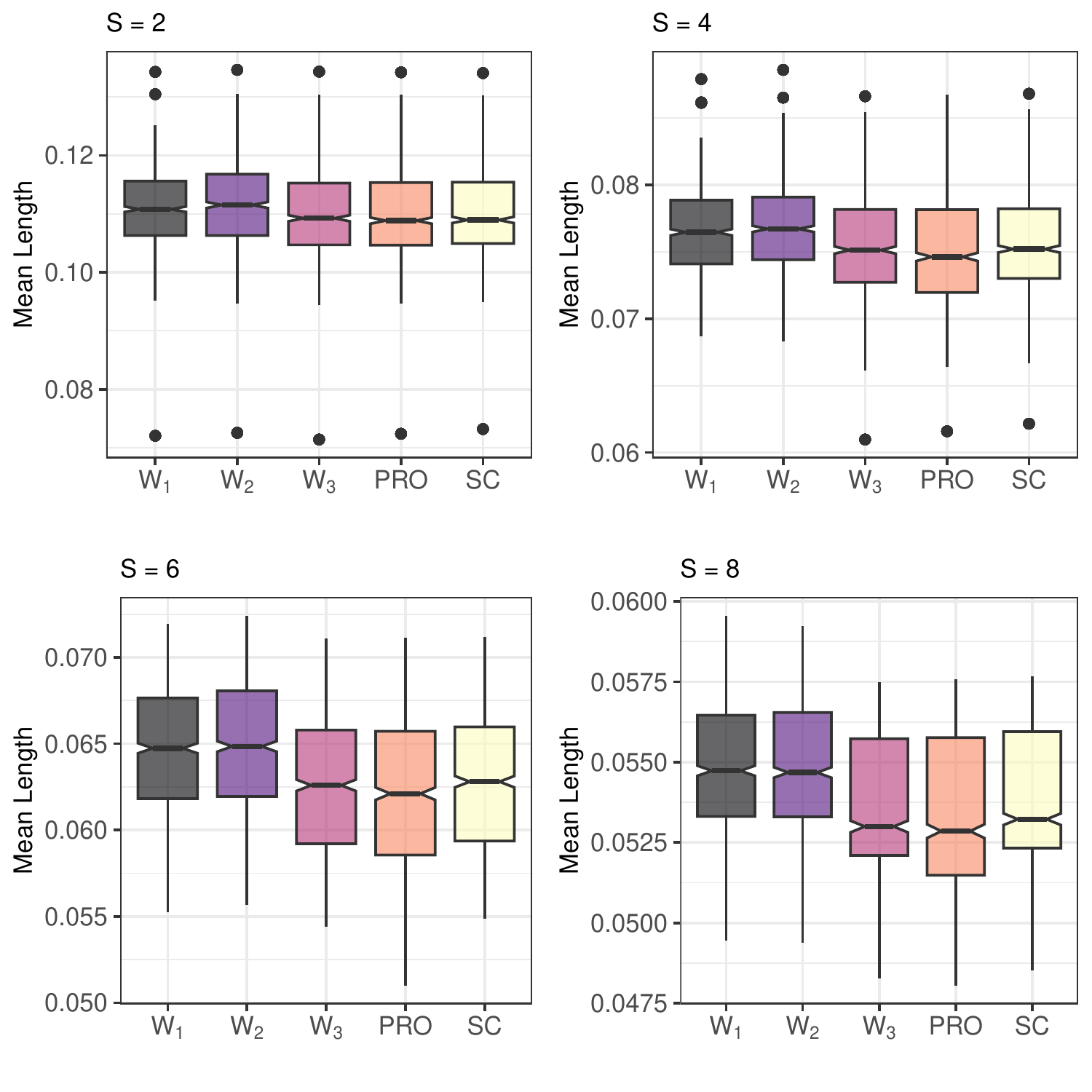}

\end{figure}

Finally, coverage probabilities and lengths of the five proposed confidence intervals under different settings of sample size, $d_s$, $\pi_{s1}$, and $\gamma_s$ are studied. Unbalanced studies are considered in order to distinguish the sample-based and the uniform Wald-type CIs. In each combination, a random sample under the null hypothesis $H_{00}: d_1=d_2=...=d_S=d_0$ is generated, where $d_0$ is a known quantity. Coverage probabilities are calculated as the proportion of the constructed CIs that cover the true risk difference $d_0$ among 10,000 replicates. Mean lengths of different types of CIs can be obtained easily. Notations of parameter configurations are presented in \hyperref[tab:ciconf2]{Table \ref*{tab:ciconf2}} to \hyperref[tab:ciconf8]{Table \ref*{tab:ciconf8}}. Denote $W_1$, $W_2$, $W_3$, PRO, and SC as sample-based Wald-type CIs, Wald-type CIs with uniform weights, constrained Wald-type CIs, profile likelihood CIs, and score CIs, respectively. The results are presented in \hyperref[tab:ci2]{Table \ref*{tab:ci2}} to \hyperref[tab:ci8]{Table \ref*{tab:ci8}}. Moreover, we generate 1,000 random sets of ($d_0, \pi_{s1}, \gamma_s$) for each combination of stratum number $S$ and sample size ($n_{s1}, m_{s1}$) while constraining $p_{bsg0}$, $p_{bsg1}$, $p_{bsg2}$, $p_{usg0}$, $p_{usg1} \in$ (0.1, 1). Sample size configurations are the same as the settings for CIs. \hyperref[fig:ciprob1]{Figure \ref*{fig:ciprob1}} to \hyperref[fig:cilen3]{Figure \ref*{fig:cilen3}} display notched box plots of empirical coverage probabilities and mean lengths for all types of CIs based on the 1,000 sets of parameters. In general, all the proposed CIs yield similar mean lengths and coverage probabilities that are close to the nominal level 95$\%$ under the pre-specified parameter arrangements. Under random parameter settings, all five methods produce satisfactory coverage probabilities and similar mean lengths when the sample size is large. When the sample size is small, constrained Wald-type CI yields deflated coverage probabilities. Other four methods have similar coverage probabilities while the profile likelihood CI generates slightly shorter mean lengths, therefore, it is recommended. 

\begin{singlespace}
\begin{table}[H]
\captionsetup[subtable]{justification=centering}
\caption{Parameter configurations for CIs (8 strata)}
\label{tab:ciconf8}
\begin{subtable}{1\linewidth}
\centering
\begin{tabular}[t]{l|c|c}
\toprule
  Notation   & $(n_{11}, n_{21}, n_{31}, n_{41}, n_{51}, n_{61}, n_{71}, n_{81})$  & $(m_{11}, m_{21}, m_{31}, m_{41}, m_{51}, m_{61}, m_{71}, m_{81})$  \\ \hline
$E^{C, 8}_{N,1}$ & (35, 35, 35, 35, 40, 40, 40, 40)   & (20, 20, 20, 20, 25, 25, 25, 25)     \\ 
$E^{C, 8}_{N,2}$ & (60, 60, 60, 60, 80, 80, 80, 80) & (50, 50, 50, 50, 40, 40, 40, 40) \\ 
$E^{C, 8}_{N,3}$ & (90, 90, 90, 90, 100, 100, 100, 100) &  (80, 80, 80, 80, 90, 90, 90, 90) \\ 

\bottomrule
\end{tabular}                                                

\caption{\footnotesize Note: $n_{s1}=n_{s2}$, $m_{s1}=m_{s2}$, $S=1, ..., 8.$}

\end{subtable}
\bigskip
\\
\begin{subtable}{1\linewidth}
\centering
\begin{tabular}[t]{l|c}
\toprule
  Notation   & $d_1=d_2=...=d_8=d_0$     \\ \hline
$E^{C, 8}_{d,1}$ & 0            \\ 
$E^{C, 8}_{d,2}$ & 0.1     \\ 
$E^{C, 8}_{d,3}$ & 0.15 \\ 

\bottomrule
\end{tabular}
\caption{\footnotesize }

\end{subtable}

\bigskip

\begin{subtable}{1\linewidth}
\centering
\begin{tabular}[t]{l|c}
\toprule
 Notation  & $(\gamma_1, \gamma_2, \gamma_3, \gamma_4, \gamma_5, \gamma_6, \gamma_7 \gamma_8)$     \\ \hline
$E^{C, 8}_{\gamma,1}$ & (0.4, 0.4, 0.5, 0.5, 0.6, 0.6, 0.7, 0.7)            \\

\bottomrule
\end{tabular}
\caption{\footnotesize }
\end{subtable}
\bigskip
\\
\begin{subtable}{1\linewidth}
\centering
\begin{tabular}[t]{l|c}
\toprule
 Notation  & $(\pi_{11}, \pi_{21}, \pi_{31}, \pi_{41}, \pi_{51}, \pi_{61}, \pi_{71}, \pi_{81})$     \\ \hline
$E^{C, 8}_{\pi,1}$ & (0.45, 0.45, 0.5, 0.5, 0.45, 0.45, 0.5, 0.5)            \\ 
$E^{C, 8}_{\pi,2}$ & (0.45, 0.5, 0.45, 0.5, 0.45, 0.5, 0.45, 0.5)  \\

\bottomrule
\end{tabular}
\caption{\footnotesize }
\end{subtable}
\end{table}
\end{singlespace}

\begin{singlespace}
\begin{center}
\addtolength{\tabcolsep}{-0.15em}
\begin{longtable}[H]{c c c c | c c c c c | c c c c c }
\caption{The empirical coverage probabilities (percent) and mean lengths of various CIs at $\alpha$=0.05 based on 10,000 replicates (2 strata) } \label{tab:ci2} \\

\hline

  &  &  &  & \multicolumn{5}{c|}{Coverage probability} & \multicolumn{5}{c}{Mean length}  \\ \cline{5-14}
N &  $d$ & $\gamma$ & $\pi$ & $W_1$ & $W_2$ & $W_3$ & PRO & SC & $W_1$ & $W_2$ & $W_3$ & PRO & SC\\ \hline

\endfirsthead

\multicolumn{14}{l}%
{{ \tablename\ \thetable{}: (Continued)}} \\
\hline

  &  &  &  & \multicolumn{5}{c|}{Coverage probability} & \multicolumn{5}{c}{Mean length}  \\ \cline{5-14}
N &  $d$ & $\gamma$ & $\pi$ & $W_1$ & $W_2$ & $W_3$ & PRO & SC & $W_1$ & $W_2$ & $W_3$ & PRO & SC\\ \hline

\endhead

\hline \multicolumn{14}{c}{{Continued on next page}} \\ \hline
\endfoot

\hline \hline
\endlastfoot

        $E^{C, 2}_{N,1}$ & $E^{C, 2}_{d,1}$ & $E^{C, 2}_{\gamma,1}$ & $E^{C, 2}_{\pi,1}$ & 94.49 & 94.71 & 94.66 & 94.81 & 94.97 & 0.175 & 0.174 & 0.173 & 0.174 & 0.174 \\ 
        ~ & ~ & ~ & $E^{C, 2}_{\pi,2}$ & 94.61 & 94.67 & 94.16 & 94.67 & 95.09 & 0.161 & 0.160 & 0.159 & 0.161 & 0.162 \\ 
        ~ & ~ & $E^{C, 2}_{\gamma,2}$ & $E^{C, 2}_{\pi,1}$ & 94.58 & 94.60 & 94.57 & 94.76 & 94.91 & 0.206 & 0.205 & 0.205 & 0.205 & 0.205 \\ 
        ~ & ~ & ~ & $E^{C, 2}_{\pi,2}$ & 94.42 & 94.30 & 94.24 & 94.57 & 94.78 & 0.200 & 0.198 & 0.198 & 0.199 & 0.199 \\ 
        ~ & $E^{C, 2}_{d,2}$ & $E^{C, 2}_{\gamma,1}$ & $E^{C, 2}_{\pi,1}$ & 94.44 & 94.30 & 94.35 & 94.56 & 94.57 & 0.181 & 0.180 & 0.180 & 0.180 & 0.180 \\ 
        ~ & ~ & ~ & $E^{C, 2}_{\pi,2}$ & 94.89 & 94.87 & 94.68 & 94.84 & 94.99 & 0.173 & 0.172 & 0.171 & 0.172 & 0.172 \\ 
        ~ & ~ & $E^{C, 2}_{\gamma,2}$ & $E^{C, 2}_{\pi,1}$ & 95.21 & 95.19 & 95.11 & 95.31 & 95.34 & 0.208 & 0.207 & 0.207 & 0.207 & 0.206 \\ 
        ~ & ~ & ~ & $E^{C, 2}_{\pi,2}$ & 94.27 & 94.34 & 94.30 & 94.50 & 94.60 & 0.205 & 0.204 & 0.204 & 0.204 & 0.203 \\
\end{longtable}
\end{center}
\end{singlespace}

\begin{singlespace}
\begin{center}
\addtolength{\tabcolsep}{-0.15em}
\begin{longtable}[H]{c c c c | c c c c c | c c c c c }
\caption{The empirical coverage probabilities (percent) and mean lengths of various CIs at $\alpha$=0.05 based on 10,000 replicates (4 strata) } \label{tab:ci4} \\

\hline

  &  &  &  & \multicolumn{5}{c|}{Coverage probability} & \multicolumn{5}{c}{Mean length}  \\ \cline{5-14}
N &  $d$ & $\gamma$ & $\pi$ & $W_1$ & $W_2$ & $W_3$ & PRO & SC & $W_1$ & $W_2$ & $W_3$ & PRO & SC\\ \hline

\endfirsthead

\multicolumn{14}{l}%
{{ \tablename\ \thetable{}: (Continued)}} \\
\hline

  &  &  &  & \multicolumn{5}{c|}{Coverage probability} & \multicolumn{5}{c}{Mean length}  \\ \cline{5-14}
N &  $d$ & $\gamma$ & $\pi$ & $W_1$ & $W_2$ & $W_3$ & PRO & SC & $W_1$ & $W_2$ & $W_3$ & PRO & SC\\ \hline

\endhead

\hline \multicolumn{14}{c}{{Continued on next page}} \\ \hline
\endfoot

\hline \hline
\endlastfoot

        $E^{C, 4}_{N,1}$ & $E^{C, 4}_{d,1}$ & $E^{C, 4}_{\gamma,1}$ & $E^{C, 4}_{\pi,1}$ & 94.46 & 94.56 & 94.30 & 94.51 & 94.79 & 0.119 & 0.119 & 0.118 & 0.119 & 0.120 \\ 
        ~ & ~ & ~ & $E^{C, 4}_{\pi,2}$ & 94.89 & 94.82 & 94.45 & 94.85 & 95.15 & 0.119 & 0.118 & 0.117 & 0.118 & 0.119 \\ 
        ~ & ~ & $E^{C, 4}_{\gamma,2}$ & $E^{C, 4}_{\pi,1}$ & 94.66 & 94.75 & 94.63 & 94.79 & 94.93 & 0.133 & 0.132 & 0.131 & 0.131 & 0.132 \\ 
        ~ & ~ & ~ & $E^{C, 4}_{\pi,2}$ & 94.65 & 94.57 & 94.45 & 94.69 & 94.83 & 0.133 & 0.131 & 0.130 & 0.131 & 0.131 \\ 
        ~ & $E^{C, 4}_{d,2}$ & $E^{C, 4}_{\gamma,1}$ & $E^{C, 4}_{\pi,1}$ & 94.82 & 94.85 & 94.74 & 94.96 & 95.07 & 0.125 & 0.125 & 0.125 & 0.125 & 0.125 \\ 
        ~ & ~ & ~ & $E^{C, 4}_{\pi,2}$ & 94.84 & 94.94 & 94.91 & 95.02 & 95.15 & 0.125 & 0.125 & 0.124 & 0.125 & 0.125 \\ 
        ~ & ~ & $E^{C, 4}_{\gamma,2}$ & $E^{C, 4}_{\pi,1}$ & 94.61 & 94.64 & 94.65 & 94.78 & 94.87 & 0.137 & 0.136 & 0.135 & 0.135 & 0.135 \\ 
        ~ & ~ & ~ & $E^{C, 4}_{\pi,2}$ & 95.13 & 95.19 & 95.17 & 95.28 & 95.40 & 0.137 & 0.135 & 0.135 & 0.135 & 0.135 \\
\end{longtable}
\end{center}
\end{singlespace}

\begin{singlespace}
\begin{center}
\addtolength{\tabcolsep}{-0.15em}
\begin{longtable}[H]{c c c c | c c c c c | c c c c c }
\caption{The empirical coverage probabilities (percent) and mean lengths of various CIs at $\alpha$=0.05 based on 10,000 replicates (6 strata) } \label{tab:ci6} \\

\hline

  &  &  &  & \multicolumn{5}{c|}{Coverage probability} & \multicolumn{5}{c}{Mean length}  \\ \cline{5-14}
N &  $d$ & $\gamma$ & $\pi$ & $W_1$ & $W_2$ & $W_3$ & PRO & SC & $W_1$ & $W_2$ & $W_3$ & PRO & SC\\ \hline

\endfirsthead

\multicolumn{14}{l}%
{{ \tablename\ \thetable{}: (Continued)}} \\
\hline

  &  &  &  & \multicolumn{5}{c|}{Coverage probability} & \multicolumn{5}{c}{Mean length}  \\ \cline{5-14}
N &  $d$ & $\gamma$ & $\pi$ & $W_1$ & $W_2$ & $W_3$ & PRO & SC & $W_1$ & $W_2$ & $W_3$ & PRO & SC\\ \hline

\endhead

\hline \multicolumn{14}{c}{{Continued on next page}} \\ \hline
\endfoot

\hline \hline
\endlastfoot

       $E^{C, 6}_{N,1}$ & $E^{C, 6}_{d,1}$ & $E^{C, 6}_{\gamma,1}$ & $E^{C, 6}_{\pi,1}$ & 94.73 & 94.70 & 94.38 & 94.55 & 94.68 & 0.099 & 0.099 & 0.098 & 0.099 & 0.099 \\ 
        ~ & ~ & ~ & $E^{C, 6}_{\pi,2}$ & 94.67 & 94.67 & 94.31 & 94.56 & 94.83 & 0.098 & 0.098 & 0.096 & 0.097 & 0.098 \\ 
        ~ & ~ & $E^{C, 6}_{\gamma,2}$ & $E^{C, 6}_{\pi,1}$ & 94.34 & 94.36 & 94.44 & 94.55 & 94.70 & 0.108 & 0.107 & 0.107 & 0.107 & 0.107 \\ 
        ~ & ~ & ~ & $E^{C, 6}_{\pi,2}$ & 94.97 & 94.87 & 94.68 & 94.97 & 95.06 & 0.107 & 0.107 & 0.106 & 0.106 & 0.107 \\ 
        ~ & $E^{C, 6}_{d,2}$ & $E^{C, 6}_{\gamma,1}$ & $E^{C, 6}_{\pi,1}$ & 94.66 & 94.61 & 94.60 & 94.71 & 94.79 & 0.102 & 0.102 & 0.102 & 0.102 & 0.102 \\ 
        ~ & ~ & ~ & $E^{C, 6}_{\pi,2}$ & 94.72 & 94.67 & 94.60 & 94.72 & 94.83 & 0.102 & 0.101 & 0.101 & 0.101 & 0.101 \\ 
        ~ & ~ & $E^{C, 6}_{\gamma,2}$ & $E^{C, 6}_{\pi,1}$ & 95.08 & 95.13 & 95.21 & 95.26 & 95.31 & 0.109 & 0.109 & 0.109 & 0.109 & 0.109 \\ 
        ~ & ~ & ~ & $E^{C, 6}_{\pi,2}$ & 94.95 & 94.94 & 94.87 & 94.93 & 95.03 & 0.109 & 0.109 & 0.108 & 0.109 & 0.109 \\
\end{longtable}
\end{center}
\end{singlespace}

\begin{singlespace}
\begin{center}
\addtolength{\tabcolsep}{-0.15em}
\begin{longtable}[H]{c c c c | c c c c c | c c c c c }
\caption{The empirical coverage probabilities (percent) and mean lengths of various CIs at $\alpha$=0.05 based on 10,000 replicates (8 strata) } \label{tab:ci8} \\

\hline

  &  &  &  & \multicolumn{5}{c|}{Coverage probability} & \multicolumn{5}{c}{Mean length}  \\ \cline{5-14}
N &  $d$ & $\gamma$ & $\pi$ & $W_1$ & $W_2$ & $W_3$ & PRO & SC & $W_1$ & $W_2$ & $W_3$ & PRO & SC\\ \hline

\endfirsthead

\multicolumn{14}{l}%
{{ \tablename\ \thetable{}: (Continued)}} \\
\hline

  &  &  &  & \multicolumn{5}{c|}{Coverage probability} & \multicolumn{5}{c}{Mean length}  \\ \cline{5-14}
N &  $d$ & $\gamma$ & $\pi$ & $W_1$ & $W_2$ & $W_3$ & PRO & SC & $W_1$ & $W_2$ & $W_3$ & PRO & SC\\ \hline

\endhead

\hline \multicolumn{14}{c}{{Continued on next page}} \\ \hline
\endfoot

\hline \hline
\endlastfoot

        $E^{C, 8}_{N,1}$ & $E^{C, 8}_{d,1}$ & $E^{C, 8}_{\gamma,1}$ & $E^{C, 8}_{\pi,1}$ & 94.58 & 94.61 & 94.26 & 94.54 & 94.66 & 0.092 & 0.091 & 0.090 & 0.090 & 0.091 \\ 
        ~ & ~ & ~ & $E^{C, 8}_{\pi,2}$ & 94.74 & 94.72 & 94.56 & 94.75 & 94.99 & 0.091 & 0.090 & 0.089 & 0.090 & 0.090 \\ 
        ~ & $E^{C, 8}_{d,2}$ & $E^{C, 8}_{\gamma,1}$ & $E^{C, 8}_{\pi,1}$ & 94.41 & 94.31 & 94.28 & 94.35 & 94.51 & 0.094 & 0.093 & 0.093 & 0.093 & 0.093 \\ 
        ~ & ~ & ~ & $E^{C, 8}_{\pi,2}$ & 94.72 & 94.68 & 94.69 & 94.75 & 94.82 & 0.094 & 0.093 & 0.092 & 0.093 & 0.093 \\ 
        ~ & $E^{C, 8}_{d,3}$ & $E^{C, 8}_{\gamma,1}$ & $E^{C, 8}_{\pi,1}$ & 94.77 & 94.74 & 94.76 & 94.84 & 94.96 & 0.094 & 0.093 & 0.093 & 0.093 & 0.093 \\ 
        ~ & ~ & ~ & $E^{C, 8}_{\pi,2}$ & 94.79 & 94.64 & 94.67 & 94.80 & 94.91 & 0.094 & 0.093 & 0.093 & 0.093 & 0.093 \\ 
        $E^{C, 8}_{N,2}$ & $E^{C, 8}_{d,1}$ & $E^{C, 8}_{\gamma,1}$ & $E^{C, 8}_{\pi,1}$ & 94.77 & 94.68 & 94.73 & 94.89 & 94.91 & 0.067 & 0.066 & 0.066 & 0.066 & 0.066 \\ 
        ~ & ~ & ~ & $E^{C, 8}_{\pi,2}$ & 94.62 & 94.73 & 94.72 & 94.76 & 94.82 & 0.067 & 0.066 & 0.065 & 0.066 & 0.066 \\
\end{longtable}
\end{center}
\end{singlespace}

\begin{figure}[H]
        \caption{Notched box plots of mean lengths for sample case $E_{N, 1}^{C, 4}$, $E_{N, 1}^{C, 6}$, and $E_{N, 1}^{C, 8}$}
        \label{fig:cilen1}
     \centering

         \includegraphics[width=\textwidth]{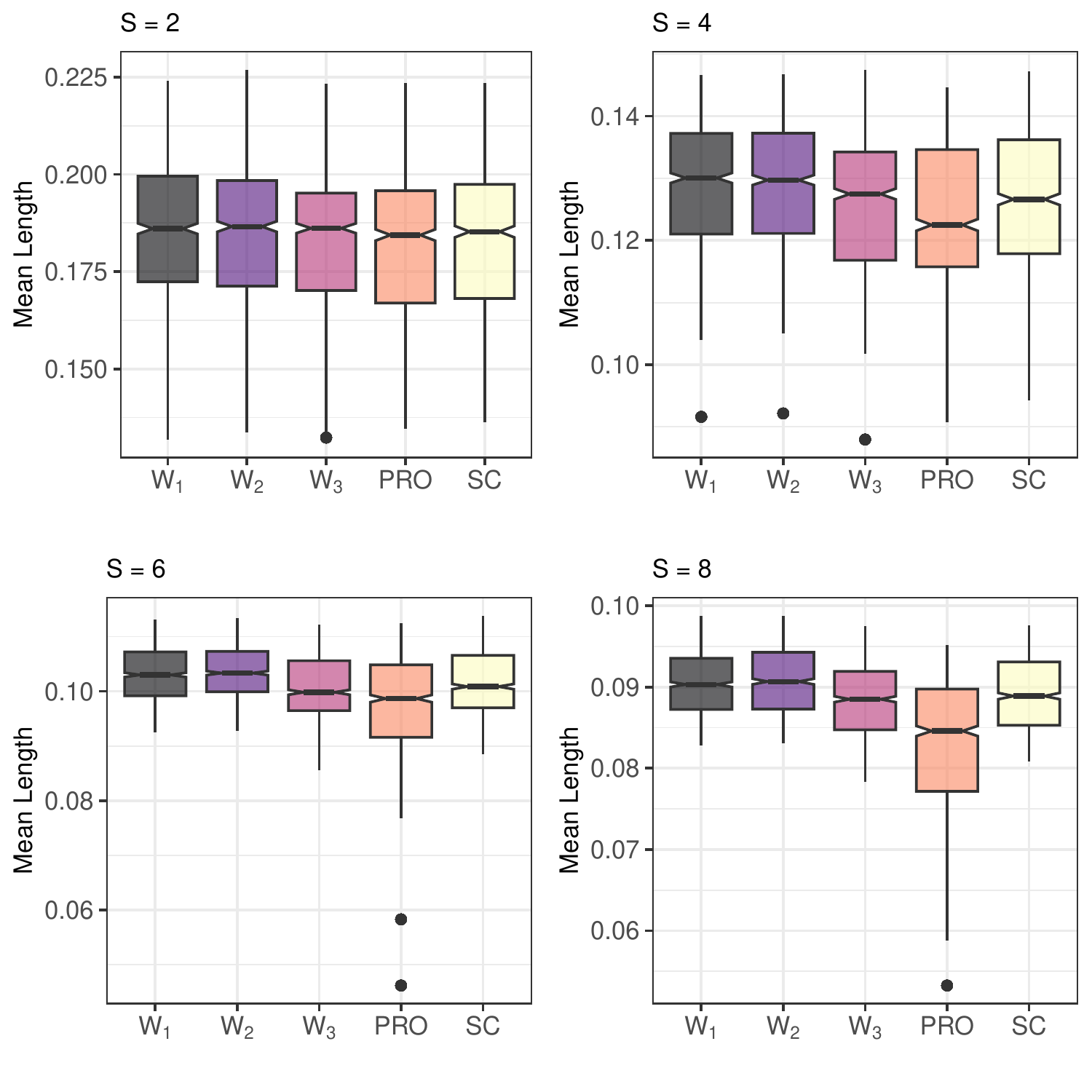}
  
\end{figure}

\begin{figure}[H]
        \caption{Notched box plots of empirical coverage probabilities for sample case $E_{N, 2}^{C, 4}$, $E_{N, 2}^{C, 6}$, and $E_{N, 2}^{C, 8}$}
        \label{fig:ciprob2}
     \centering
 
         \includegraphics[width=\textwidth]{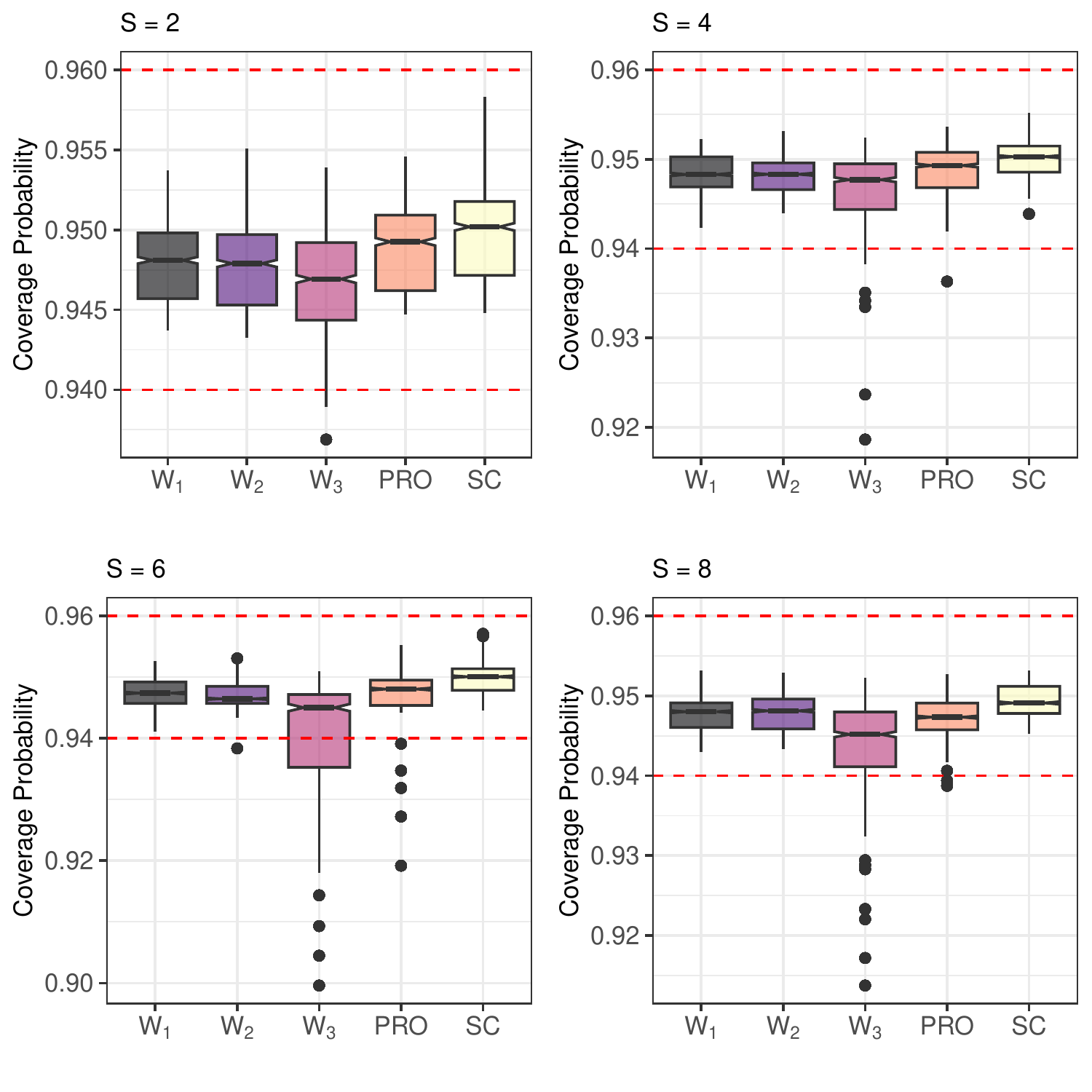}
  
\end{figure}

\begin{figure}[H]
        \caption{Notched box plots of mean lengths for sample case $E_{N, 2}^{C, 4}$, $E_{N, 2}^{C, 6}$, and $E_{N, 2}^{C, 8}$}
        \label{fig:cilen2}
     \centering
 
         \includegraphics[width=\textwidth]{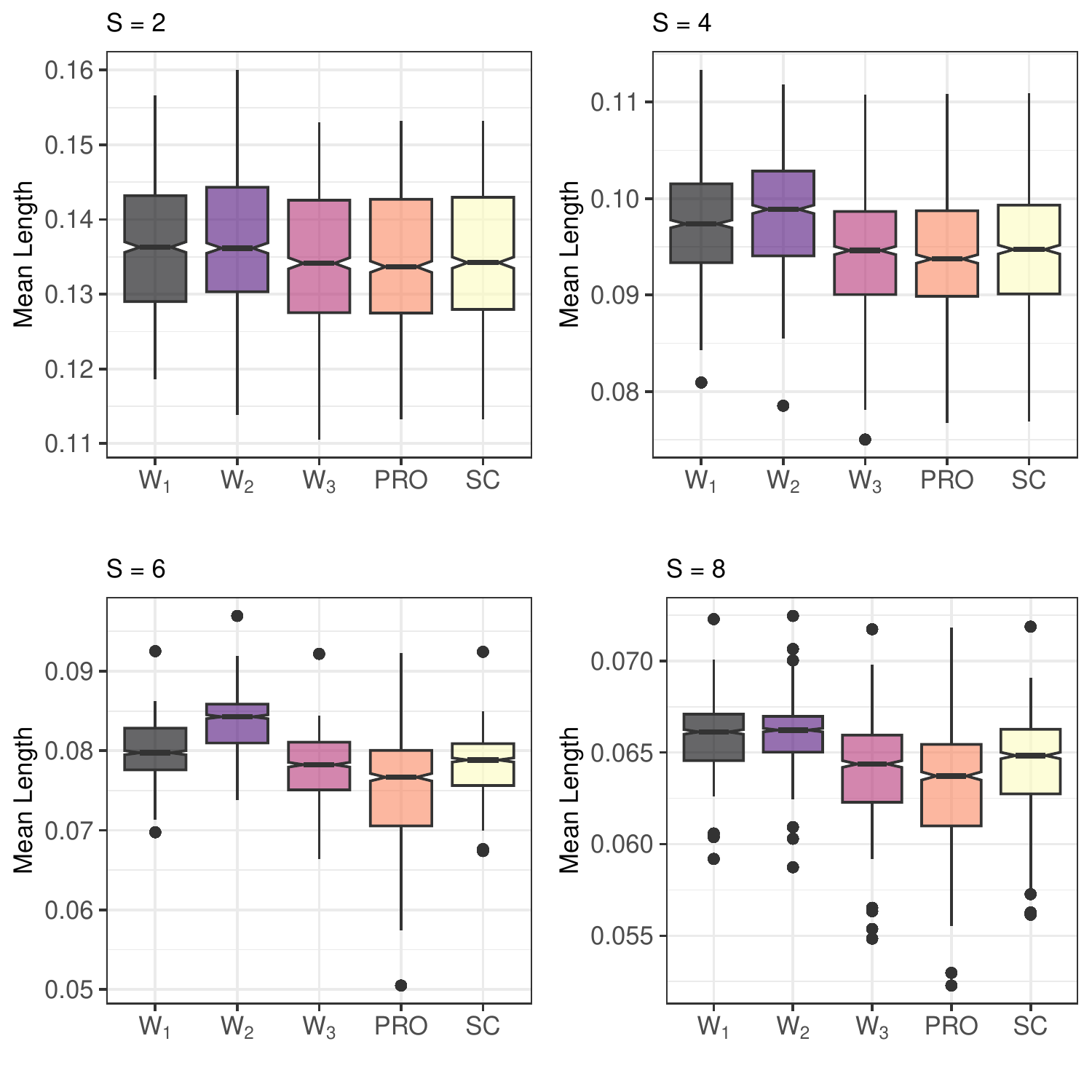}
  
\end{figure}

\begin{figure}[H]
        \caption{Notched box plots of empirical coverage probabilities for sample case $E_{N, 3}^{C, 4}$, $E_{N, 3}^{C, 6}$, and $E_{N, 3}^{C, 8}$}
        \label{fig:ciprob3}
     \centering

         \includegraphics[width=\textwidth]{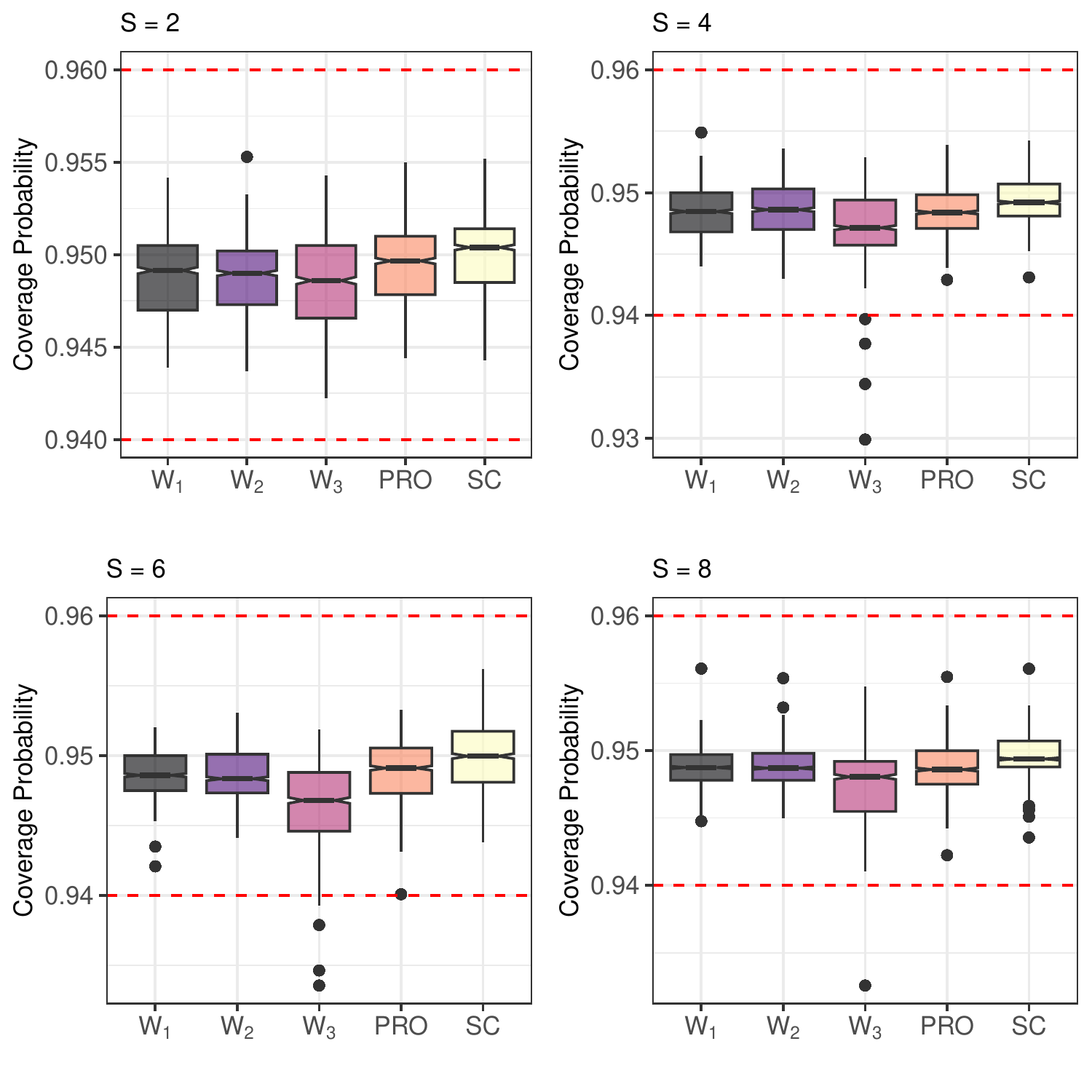}

\end{figure}

\section{Real Examples}
In this section, two real examples are provided to illustrate the proposed methods. The first one is a double-blind randomized clinical trial for studying acute otitis media with effusion (OME). A total of 214 children (293 ears) who underwent unilateral or bilateral tympanocentesis were randomly assigned to receive a 14-day course of either cefaclor or amoxicillin for the treatment of the disease (Mandel et al. \citeyearpar{mandel1982duration}). After 14 days there were 203 evaluable children left who did not experience repeat tympanocentesis, treatment change, or tympanic membrane perforations. The number of children at 14 days by age, the number of cured ears, and the treatment received is tabulated in \hyperref[tab:example]{Table \ref*{tab:example}}. It is of interest to explore if the difference in cure rates of the two treatments remains the same in different age strata. The cure rate difference is defined as the cure rate in the cefaclor group minus the cure rate in the amoxicillin group.

\begin{singlespace}
\begin{table}[H]
\centering
\caption{The number of children at 14 days by age, number of cured ears, and the treatment received (Example 1)}
\label{tab:example}
\begin{tabular}{c|c|ccc|cc}
\hline
\multirow{3}{*}{Age at Entry} & \multirow{3}{*}{Treatment} & \multicolumn{3}{c|}{Bilateral at entry}                        & \multicolumn{2}{c}{Unilateral at entry}          \\ \cline{3-7} 
&& \multicolumn{3}{c|}{No. of cured ears}& \multicolumn{2}{c}{No. of cured ears} \\ \cline{3-7} 
&& \multicolumn{1}{c|}{0}  & \multicolumn{1}{c|}{1} & 2  & \multicolumn{1}{c|}{0}         & 1       \\ \hline
\multirow{2}{*}{$<$2 years}     & Cefaclor                   & \multicolumn{1}{c|}{8}  & \multicolumn{1}{c|}{2} & 8  & \multicolumn{1}{c|}{9}         & 3       \\  
& Amoxicillin                & \multicolumn{1}{c|}{11}  & \multicolumn{1}{c|}{2} & 2 & \multicolumn{1}{c|}{10}         & 2      \\ \hline
\multirow{2}{*}{2-5 years}    & Cefaclor                   & \multicolumn{1}{c|}{6} & \multicolumn{1}{c|}{6} & 10  & \multicolumn{1}{c|}{7}        & 24       \\ 
& Amoxicillin                & \multicolumn{1}{c|}{3}  & \multicolumn{1}{c|}{1} & 5  & \multicolumn{1}{c|}{22}        & 14      \\ \hline
\multirow{2}{*}{$\ge$6 years} & Cefaclor                   & \multicolumn{1}{c|}{0}  & \multicolumn{1}{c|}{1} & 3  & \multicolumn{1}{c|}{8}        & 11       \\ 
& Amoxicillin                & \multicolumn{1}{c|}{1}  & \multicolumn{1}{c|}{0} & 6  & \multicolumn{1}{c|}{7}        & 11       \\ \hline
\end{tabular}
\end{table}
\end{singlespace}

\hyperref[tab:teststat]{Table \ref*{tab:teststat}} and \hyperref[tab:mle]{Table \ref*{tab:mle}} lay out the results of test statistics, p-values, unconstrained MLEs, and constrained MLEs. All test statistics are less than $\mathcal{X}_{2,0.95}^2=5.9915$ given the significance level of 5\%, suggesting there is insufficient evidence to reject the null hypothesis. Estimations of CIs of the difference of the cure rates presented in \hyperref[tab:cieg]{Table \ref*{tab:cieg}} demonstrate a higher cure rate in the cefaclor group across all age strata since all 95$\%$ CIs do not cover point zero.

\begin{singlespace}
\begin{table}[H]
\centering
\caption{Test statistics and p-values under $H_0$ (Example 1)}
\label{tab:teststat}
\begin{tabular}{l|ccc}
\hline
\multirow{2}{*}{Value} & \multicolumn{3}{c}{Test statistic}\\ \cline{2-4} 
& \multicolumn{1}{c|}{$T_{SC}$}& \multicolumn{1}{c|}{$T_{LR}$}& $T_W$                 \\ \hline
Statistic value& \multicolumn{1}{c|}{2.7487}& \multicolumn{1}{c|}{2.8475}& 5.8206\\ 
p-value& \multicolumn{1}{c|}{0.2530} & \multicolumn{1}{c|}{0.2408} & 0.0545 \\ \hline

\end{tabular}
\end{table}
\end{singlespace}

\begin{singlespace}
\begin{table}[H]
\centering
\caption{Unconstrained and constrained MLEs of $\pi$, $\gamma$, and $d$ (Example 1)}
\label{tab:mle}
\begin{tabular}{c|c|c|c|c|c|c|c}
\hline
\multirow{2}{*}{Age at Entry} & \multirow{2}{*}{Treatment} & \multicolumn{3}{c|}{Unconstrained MLEs}& \multicolumn{3}{c}{Constrained MLEs}\\ \cline{3-8} 
&& \multicolumn{1}{c|}{$\pi$}  & \multicolumn{1}{c|}{$\gamma$}                & $d$                      & \multicolumn{1}{c|}{$\pi$}  & \multicolumn{1}{c|}{$\gamma$}                & $d$                      \\ \hline
\multirow{2}{*}{$<$2 years}& Cefaclor    & 0.3958 & \multirow{2}{*}{0.8115} & \multirow{2}{*}{0.1939}  & 0.3847 & \multirow{2}{*}{0.8104} & \multirow{6}{*}{0.1742} \\
& Amoxicillin & 0.2018 & & & 0.2105 & & \\ \cline{1-7}
\multirow{2}{*}{2-5 years}    & Cefaclor    & 0.6881 & \multirow{2}{*}{0.8321} & \multirow{2}{*}{0.2536}  & 0.6565 & \multirow{2}{*}{0.8208} &\\
& Amoxicillin & 0.4346 &&& 0.4823 &&\\ \cline{1-7}
\multirow{2}{*}{$\ge$6 years} & Cefaclor    & 0.6522 & \multirow{2}{*}{0.9184} & \multirow{2}{*}{-0.0198} & 0.7424 & \multirow{2}{*}{0.9120} &\\
& Amoxicillin & 0.6720 &&& 0.5682 &&   \\ \hline                                 

\end{tabular}
\end{table}
\end{singlespace}

\begin{singlespace}
\begin{table}[H]
\centering
\caption{Estimations of confidence intervals (Example 1)}
\label{tab:cieg}

\begin{tabular}{c|c|c}
\hline
      & CI bounds                 & Width              \\ \hline
$W_1$ & (0.0504,0.2940) &  0.2436 \\ 
$W_2$ & (0.0155,0.2696) &  0.2542 \\ 
$W_3$ & (0.0526,0.2957) &  0.2431\\ 
PRO   & (0.0502,0.2953) &  0.2451 \\ 
SC    & (0.0486,0.2954) &  0.2468 \\ \hline
\end{tabular}

\end{table}
\end{singlespace}

The second example is an observational study conducted at the First Affiliated Hospital of Xiamen University in 2023 and is used for demonstration purpose only. Sixty subjects with myopia were recruited to receive Orthokeratology (Ortho-k), a non-surgical vision correction method that is designed to correct myopia using specialized contact lenses worn overnight to reshape the cornea temporarily. Custom lenses are created based on eye measurements. Patients wear them while sleeping, and they gently reshape the cornea to enhance vision. Wearing the lenses on both eyes or a single eye is based upon the patients' choice. It is important to notice that the outcomes achieved are not permanent, and lenses must be worn during the night. Otherwise, the eyes will eventually go back to their original shape. Benefits of Ortho-k include freedom from glasses during the day, potential myopia control in kids, reversibility if Ortho-k is discontinued, and a non-invasive alternative to surgery. Ortho-k candidacy varies, and regular eye check-ups are vital for success. One of the signs of good treatment is the axial length growth $< 0.3$ mm according to Rose et al . \citeyearpar{rose2021use}. Hence, we define an eye as having a "response" if $Axial - Axial0 < 0.3$ mm and as "without a response" if $Axial - Axial0 \geq 0.3$ mm. The brand S utilizes a lens design called corneal refractive therapy (CRT) while other brands rely on vision shaping treatment (VST). The VST lens design takes corneal curves into consideration, ensuring the the edge of the lens touches the edge of the cornea. In contrast, the CRT lens is characterized by congruent anterior and posterior surfaces and the edge of the lens does not touch the cornea according to Lu et al. \citeyearpar{lu2022comparison}. 

Our interest is if the difference in response rates of the two designs varies in different gender strata. The response rate difference is defined as the response rate of the CRT design minus the response rate of the VST design. Data summary is displayed in \hyperref[tab:example2]{Table \ref*{tab:example2}} and test statistics, p-values, unconstrained MLEs, and parameter estimates are presented in \hyperref[tab:teststat2]{Table \ref*{tab:teststat2}} and \hyperref[tab:mle2]{Table \ref*{tab:mle2}}. All test statistics are greater than $\mathcal{X}_{1,0.95}^2=3.84$, suggesting a rejection of the null hypothesis at the significance level of 5\%. Therefore, the downstream analyses may include subgroup analysis testing the equivalence of the two response rates for each gender category as suggested by Chen et al. \citeyearpar{chen2022further}.

\begin{singlespace}
\begin{table}[H]
\centering
\caption{The number of subjects by design, number of eyes with response, and gender after treatment (Example 2)}
\label{tab:example2}
\begin{tabular}{c|c|ccc|cc}
\hline
\multirow{3}{*}{Gender} & \multirow{3}{*}{Design} & \multicolumn{3}{c|}{Bilateral}                        & \multicolumn{2}{c}{Unilateral}          \\ \cline{3-7} 
&& \multicolumn{3}{c|}{No. of eyes with response}& \multicolumn{2}{c}{No. of eyes with response} \\ \cline{3-7} 
&& \multicolumn{1}{c|}{0}  & \multicolumn{1}{c|}{1} & 2  & \multicolumn{1}{c|}{0}         & 1       \\ \hline
\multirow{2}{*}{Female}     & VST                   & \multicolumn{1}{c|}{9}  & \multicolumn{1}{c|}{3} & 7  & \multicolumn{1}{c|}{2}         & 1       \\  
& CRT                & \multicolumn{1}{c|}{7}  & \multicolumn{1}{c|}{0} & 0 & \multicolumn{1}{c|}{0}         & 0      \\ \hline
\multirow{2}{*}{Male}    & VST                   & \multicolumn{1}{c|}{11} & \multicolumn{1}{c|}{4} & 3  & \multicolumn{1}{c|}{1}        & 2       \\ 
& CRT                & \multicolumn{1}{c|}{6}  & \multicolumn{1}{c|}{2} & 2  & \multicolumn{1}{c|}{0}        & 0      \\ \hline

\end{tabular}
\end{table}
\end{singlespace}

\begin{singlespace}
\begin{table}[H]
\centering
\caption{Test statistics and p-values under $H_0$ (Example 2)}
\label{tab:teststat2}
\begin{tabular}{l|ccc}
\hline
\multirow{2}{*}{Value} & \multicolumn{3}{c}{Test statistic}\\ \cline{2-4} 
& \multicolumn{1}{c|}{$T_{SC}$}& \multicolumn{1}{c|}{$T_{LR}$}& $T_W$                 \\ \hline
Statistic value& \multicolumn{1}{c|}{4.1229}& \multicolumn{1}{c|}{5.1650}&  4.7647\\ 
p-value& \multicolumn{1}{c|}{0.0423} & \multicolumn{1}{c|}{0.0230} & 0.0290 \\ \hline

\end{tabular}
\end{table}
\end{singlespace}

\begin{singlespace}
\begin{table}[H]
\centering
\caption{Unconstrained and constrained MLEs of $\pi$, $\gamma$, and $d$ (Example 2)}
\label{tab:mle2}
\begin{tabular}{c|c|c|c|c|c|c|c}
\hline
\multirow{2}{*}{Gender} & \multirow{2}{*}{Design} & \multicolumn{3}{c|}{Unconstrained MLEs}& \multicolumn{3}{c}{Constrained MLEs}\\ \cline{3-8} 
&& \multicolumn{1}{c|}{$\pi$}  & \multicolumn{1}{c|}{$\gamma$}                & $d$                      & \multicolumn{1}{c|}{$\pi$}  & \multicolumn{1}{c|}{$\gamma$}                & $d$                      \\ \hline
\multirow{2}{*}{Female}& VST    & 0.4340 & \multirow{2}{*}{0.8189 } & \multirow{2}{*}{-0.4340}  & 0.3480 & \multirow{2}{*}{0.8036} & \multirow{4}{*}{ -0.2060} \\
& CRT & 0.0000 & & & 0.1420 & & \\ \cline{1-7}
\multirow{2}{*}{Male}    & VST    & 0.3226 & \multirow{2}{*}{0.6468} & \multirow{2}{*}{-0.0270}  & 0.3984 & \multirow{2}{*}{0.6721} &\\
& CRT & 0.2956 &&& 0.1924 &&\\ \cline{1-8}

\end{tabular}
\end{table}
\end{singlespace}

\section{Conclusions}
In this article, we proposed three test statistics, score $T_{SC}$, maximum likelihood ratio test $T_{LR}$, and Wald-type $T_W$ for the purpose of investigating homogeneity of risk differences among multiple strata under Dallal's model. In addition, five confidence intervals are constructed to make statistical inferences on a common risk difference when the homogeneity assumption holds. The asymptotic features of these tests and CIs were investigated.

Simulation results show that the score test controls type I errors and have reasonable powers, regardless of the number of stratum, sample sizes, and parameter configurations. When the sample size is large, three test methods produce similar results. When the sample size is small, the Wald-type test generates inflated type I errors and becomes liberal, while the likelihood-ratio test generates larger proportion of type I errors that lay outside $(0.04,0.06)$. Hence, the score test is highly recommended. For confidence interval constructions, five approaches are proposed and the profile likelihood method is recommended based on its acceptable coverage probabilities and relatively shorter interval widths.

Potential future works include exact methods for data with a small sample size in order to improve performances, and building simultaneous confidence intervals if the homogeneity of risk differences is invalid. In case of certain studies where proportion ratios are the parameters of interest instead, it is worth extending the current framework for such applications. Several researches have investigated this topic under models other than Dallal's model (Zhuang et al. \citeyearpar{zhuang2019confidence} and Xue \& Ma \citeyearpar{xue2020interval}). 

A user-friendly online calculator is accessible through the following link: \url{https://www.acsu.buffalo.edu/~cxma/Test_and_CIsForRiskDifferenceDarralModelStrafified.htm}. With this calculator, readers have the ability to conduct hypothesis tests and obtain model estimations using either their own data or simulated data based on user-specified parameters.

\bibliography{ref}

\begin{thebibliography}{}

\bibitem[Chen et~al., 2022]{chen2022further}
Chen, Y., Li, Z., and Ma, C. (2022).
\newblock Further study on testing the equality of response rates under dallal’s model.
\newblock {\em Statistics and Its Interface}, 15(1):115--126.

\bibitem[Da~Fonseca, 2007]{da2007eigenvalues}
Da~Fonseca, C. (2007).
\newblock On the eigenvalues of some tridiagonal matrices.
\newblock {\em Journal of Computational and Applied Mathematics}, 200(1):283--286.

\bibitem[Dallal, 1988]{dallal1988paired}
Dallal, G.~E. (1988).
\newblock Paired bernoulli trials.
\newblock {\em Biometrics}, pages 253--257.

\bibitem[Donner, 1989]{donner1989statistical}
Donner, A. (1989).
\newblock Statistical methods in ophthalmology: an adjusted chi-square approach.
\newblock {\em Biometrics}, pages 605--611.

\bibitem[Donner and Banting, 1988]{donner1988analysis}
Donner, A. and Banting, D. (1988).
\newblock Analysis of site-specific data in dental studies.
\newblock {\em Journal of Dental Research}, 67(11):1392--1395.

\bibitem[K{\i}l{\i}{\c{c}}, 2008]{kilicc2008explicit}
K{\i}l{\i}{\c{c}}, E. (2008).
\newblock Explicit formula for the inverse of a tridiagonal matrix by backward continued fractions.
\newblock {\em Applied Mathematics and Computation}, 197(1):345--357.

\bibitem[Li et~al., 2020]{li2020statistical}
Li, Z., Ma, C., and Ai, M. (2020).
\newblock Statistical tests under dallal’s model: Asymptotic and exact methods.
\newblock {\em Plos one}, 15(11):e0242722.

\bibitem[Li et~al., 2023]{li2023testing}
Li, Z., Ma, C., and Mou, K. (2023).
\newblock Testing the common risk difference of proportions for stratified uni-and bilateral correlated data.
\newblock {\em Statistica Neerlandica}.

\bibitem[Liu et~al., 2017]{liu2017exact}
Liu, X., Shan, G., Tian, L., and Ma, C.-X. (2017).
\newblock Exact methods for testing homogeneity of proportions for multiple groups of paired binary data.
\newblock {\em Communications in Statistics-Simulation and Computation}, 46(8):6074--6082.

\bibitem[Lu et~al., 2022]{lu2022comparison}
Lu, W., Ning, R., Diao, K., Ding, Y., Chen, R., Zhou, L., Lian, Y., McAlinden, C., Sanders, F.~W., Xia, F., et~al. (2022).
\newblock Comparison of two main orthokeratology lens designs in efficacy and safety for myopia control.
\newblock {\em Frontiers in Medicine}, 9:798314.

\bibitem[Ma et~al., 2015]{ma2015homogeneity}
Ma, C., Shan, G., and Liu, S. (2015).
\newblock Homogeneity test for correlated binary data.
\newblock {\em PloS one}, 10(4):e0124337.

\bibitem[Ma and Liu, 2017]{ma2017testing}
Ma, C.-X. and Liu, S. (2017).
\newblock Testing equality of proportions for correlated binary data in ophthalmologic studies.
\newblock {\em Journal of biopharmaceutical statistics}, 27(4):611--619.

\bibitem[Ma and Wang, 2022]{ma2022testing}
Ma, C.-X. and Wang, H. (2022).
\newblock Testing the equality of proportions for combined unilateral and bilateral data under equal intraclass correlation model.
\newblock {\em Statistics in Biopharmaceutical Research}, pages 1--10.

\bibitem[Ma and Wang, 2021]{ma2021testing}
Ma, C.-X. and Wang, K. (2021).
\newblock Testing the homogeneity of proportions for combined unilateral and bilateral data.
\newblock {\em Journal of Biopharmaceutical Statistics}, 31(5):686--704.

\bibitem[Mallik, 2001]{mallik2001inverse}
Mallik, R.~K. (2001).
\newblock The inverse of a tridiagonal matrix.
\newblock {\em Linear Algebra and its Applications}, 325(1-3):109--139.

\bibitem[Mandel et~al., 1982]{mandel1982duration}
Mandel, E.~M., Bluestone, C.~D., Rockette, H.~E., BLATTER, M.~M., Reisinger, K.~S., Wucher, F.~P., and Harper, J. (1982).
\newblock Duration of effusion after antibiotic treatment for acute otitis media: comparison of cefaclor and amoxicillin.
\newblock {\em The Pediatric Infectious Disease Journal}, 1(5):310--316.

\bibitem[Pei et~al., 2008]{pei2008testing}
Pei, Y., Tang, M.-L., and Guo, J. (2008).
\newblock Testing the equality of two proportions for combined unilateral and bilateral data.
\newblock {\em Communications in Statistics—Simulation and Computation{\textregistered}}, 37(8):1515--1529.

\bibitem[Pei et~al., 2012]{pei2012confidence}
Pei, Y., Tang, M.-L., Wong, W.-K., and Guo, J. (2012).
\newblock Confidence intervals for correlated proportion differences from paired data in a two-arm randomised clinical trial.
\newblock {\em Statistical Methods in Medical Research}, 21(2):167--187.

\bibitem[Qiu et~al., 2019a]{qiu2019tests}
Qiu, S.-F., Guo, L.-X., Zou, G., and Yu, D. (2019a).
\newblock Tests for homogeneity of risk differences in stratified design with correlated bilateral data.
\newblock {\em Journal of Applied Statistics}.

\bibitem[Qiu et~al., 2021]{qiu2021confidence}
Qiu, S.-F., Liu, Q.-S., and Ge, Y. (2021).
\newblock Confidence intervals of proportion differences for stratified combined unilateral and bilateral data.
\newblock {\em Communications in Statistics-Simulation and Computation}, pages 1--24.

\bibitem[Qiu et~al., 2019b]{qiu2019construction}
Qiu, S.-F., Poon, W.-Y., Tang, M.-L., and Tao, J.-R. (2019b).
\newblock Construction of confidence intervals for the risk differences in stratified design with correlated bilateral data.
\newblock {\em Journal of Biopharmaceutical Statistics}, 29(3):446--467.

\bibitem[Qiu and Tao, 2022]{qiu2022confidence}
Qiu, S.-F. and Tao, J.-R. (2022).
\newblock Confidence intervals for assessing equivalence of two treatments with combined unilateral and bilateral data.
\newblock {\em Journal of Applied Statistics}, 49(13):3414--3435.

\bibitem[Rao, 1948]{rao1948large}
Rao, C.~R. (1948).
\newblock Large sample tests of statistical hypotheses concerning several parameters with applications to problems of estimation.
\newblock In {\em Mathematical Proceedings of the Cambridge Philosophical Society}, volume~44, pages 50--57. Cambridge University Press.

\bibitem[Rose et~al., 2021]{rose2021use}
Rose, L.~V., Schulz, A.~M., and Graham, S.~L. (2021).
\newblock Use baseline axial length measurements in myopic patients to predict the control of myopia with and without atropine 0.01\%.
\newblock {\em Plos one}, 16(7):e0254061.

\bibitem[Rosner, 1982]{rosner1982statistical}
Rosner, B. (1982).
\newblock Statistical methods in ophthalmology: an adjustment for the intraclass correlation between eyes.
\newblock {\em Biometrics}, pages 105--114.

\bibitem[Shan and Ma, 2014]{shan2014exact}
Shan, G. and Ma, C. (2014).
\newblock Exact methods for testing the equality of proportions for binary clustered data from otolaryngologic studies.
\newblock {\em Statistics in Biopharmaceutical Research}, 6(1):115--122.

\bibitem[Shen and Ma, 2018]{shen2018testing}
Shen, X. and Ma, C.-X. (2018).
\newblock Testing homogeneity of difference of two proportions for stratified correlated paired binary data.
\newblock {\em Journal of Applied Statistics}, 45(8):1410--1425.

\bibitem[Shen et~al., 2019]{shen2019common}
Shen, X., Ma, C.-X., Yuen, K.~C., and Tian, G.-L. (2019).
\newblock Common risk difference test and interval estimation of risk difference for stratified bilateral correlated data.
\newblock {\em Statistical Methods in Medical Research}, 28(8):2418--2438.

\bibitem[Sun et~al., 2022]{sun2022risk}
Sun, S., Li, Z., Ai, M., and Jiang, H. (2022).
\newblock Risk difference tests for stratified binary data under dallal’s model.
\newblock {\em Statistical Methods in Medical Research}, 31(6):1135--1156.

\bibitem[Tang et~al., 2006]{tang2006statistical}
Tang, M.-L., Tang, N.-S., and Rosner, B. (2006).
\newblock Statistical inference for correlated data in ophthalmologic studies.
\newblock {\em Statistics in Medicine}, 25(16):2771--2783.

\bibitem[Tang et~al., 2011]{tang2011asymptotic}
Tang, N.-S., Qiu, S.-F., Tang, M.-L., and Pei, Y.-B. (2011).
\newblock Asymptotic confidence interval construction for proportion difference in medical studies with bilateral data.
\newblock {\em Statistical Methods in Medical Research}, 20(3):233--259.

\bibitem[Tang et~al., 2008]{tang2008testing}
Tang, N.-S., Tang, M.-L., and Qiu, S.-F. (2008).
\newblock Testing the equality of proportions for correlated otolaryngologic data.
\newblock {\em Computational Statistics \& Data Analysis}, 52(7):3719--3729.

\bibitem[Usmani, 1994]{usmani1994inversion}
Usmani, R.~A. (1994).
\newblock Inversion of a tridiagonal jacobi matrix.
\newblock {\em Linear Algebra and its Applications}, 212(213):413--414.

\bibitem[Wald, 1943]{wald1943tests}
Wald, A. (1943).
\newblock Tests of statistical hypotheses concerning several parameters when the number of observations is large.
\newblock {\em Transactions of the American Mathematical society}, 54(3):426--482.

\bibitem[Wilks, 1938]{wilks1938large}
Wilks, S.~S. (1938).
\newblock The large-sample distribution of the likelihood ratio for testing composite hypotheses.
\newblock {\em The annals of mathematical statistics}, 9(1):60--62.

\bibitem[Xue and Ma, 2020]{xue2020interval}
Xue, Y. and Ma, C.-X. (2020).
\newblock Interval estimation of proportion ratios for stratified bilateral correlated binary data.
\newblock {\em Statistical methods in medical research}, 29(7):1987--2014.

\bibitem[Yang et~al., 2021]{yang2021simultaneous}
Yang, Z., Tian, G.-L., Liu, X., and Ma, C.-X. (2021).
\newblock Simultaneous confidence interval construction for many-to-one comparisons of proportion differences based on correlated paired data.
\newblock {\em Journal of Applied Statistics}, 48(8):1442--1456.

\bibitem[Zhang and Ying, 2018]{zhang2018statistical}
Zhang, H.~G. and Ying, G.-s. (2018).
\newblock Statistical approaches in published ophthalmic clinical science papers: a comparison to statistical practice two decades ago.
\newblock {\em British Journal of Ophthalmology}, 102(9):1188--1191.

\bibitem[Zhuang et~al., 2019]{zhuang2019confidence}
Zhuang, T., Tian, G.-L., and Ma, C.-X. (2019).
\newblock Confidence intervals for proportion ratios of stratified correlated bilateral data.
\newblock {\em Journal of Biopharmaceutical Statistics}, 29(1):203--225.

\end{thebibliography}

\end{document}